\documentclass[10pt,twocolumn]{IEEEtran}
\makeatletter
\def\@endtheorem{\endtrivlist}
\makeatother

\usepackage{latexsym}
\usepackage{amssymb}
\usepackage{amsmath}
\usepackage{graphics}

\newtheorem{theorem}{Theorem}
\newtheorem{definition}[theorem]{Definition}
\newtheorem{lemma}[theorem]{Lemma}
\newtheorem{example}[theorem]{Example}

\newcommand{\bs}{\boldsymbol}
\newcommand{\mb}{\mathbb}
\newcommand{\mc}{\mathcal}
\newcommand{\blank}{\;\;}

\renewcommand{\l}{\langle}
\renewcommand{\r}{\rangle}
\newcommand{\ua}{\rangle}
\DeclareMathOperator*{\argmax}{arg\,max}

\begin{document}

\title{Worst-Case Services and State-Based Scheduling}
\author{Yike Xu and Mark S. Andersland
\thanks{Date of this version: \today. {\it  (Corresponding author: Yike Xu.)}}
\thanks{Y. Xu is a freelancer in Shanghai, China (e-mail: yike\_xu@hotmail.com).}
\thanks{M. S. Andersland is with the Department of Electrical and Computer Engineering, University of Iowa, Iowa City, IA 52242 USA (e-mail: mark-andersland@uiowa.edu).}}
\date{}
\maketitle

\begin{abstract}

In this paper, we shed new light on a classical scheduling problem: given a slot-timed, constant-capacity server, what short-run scheduling decisions must be made to provide long-run service guarantees to competing flows of unit-sized tasks? We model each flow's long-run guarantee as a worst-case service that maps each queued arrival vector recording the flow's cumulative task arrivals, including those initially queued, to a worst-case acceptable departure vector lower-bounding its cumulative served tasks. We show that these maps are states that can be updated as tasks arrive and are served, introduce state-based scheduling, find the schedulability condition necessary and sufficient to maintain all flows' long-run guarantees, and use this condition to identify all short-run scheduling decisions that preserve schedulability. Our framework is general but computationally complex. To reduce complexity, we consider three specializations. First, we show that when satisfactory short-run scheduling decisions exist, at least one can be efficiently identified by maximizing the server's capacity slack, a generalization of the earliest-deadline-first rule. Second, we show that a special class of worst-case services, min-plus services, can be efficiently specified and updated using properties of the min-plus algebra. Finally, we show that efficiency can be further improved by restricting attention to a min-plus service subclass, dual-curve services. This last specialization turns out to be a dynamic extension of service curves that maintains all essential features of our framework while approaching near practical viability.

\end{abstract}


\section{Introduction}\label{S:introduction}

A slot-timed, constant-capacity server, shared by multiple flows of unit-sized tasks, is a common resource allocation model. Within this model, given the server's limited capacity, a classical scheduling problem can be formulated, namely that of deciding which tasks to serve and which to defer to meet competing flow service requests. Generally, service is a long-run concept in that service guarantees cover, if not entireties, significant portions of flow lifetimes. In contrast, scheduling is a short-run concept in that scheduling decisions must be made slot-by-slot as tasks arrive and are served. The challenge is to determine how to maintain these long-run service guarantees using short-run scheduling decisions. In this paper, we shed new light on this classical problem by providing novel answers to two key questions: What to guarantee? and How to guarantee it?

\subsection{A General Framework}\label{SS:framework}

What to guarantee? This is foremost a question of service specification because we can only guarantee what we can specify. In \cite{Cruz:1991A, Cruz:1991B, Parekh:1993, Parekh:1994}, cumulative curves were introduced to characterize long-run flow traffic. They become cumulative vectors in slot-timed systems. In particular, arrival vectors can be used to record a flow's cumulative task arrivals, and departure vectors, its cumulative served tasks. We extend the definition of arrival vectors to include initially queued tasks and define a worst-case service to be a map from each such queued arrival vector to a worst-case acceptable departure vector. While this definition encompasses a wide variety of guarantees, and leaves open endless intricacies, it is precisely this generality that underlies our framework's generality.

How to guarantee it? This is a question of methodology and our answer is state-based scheduling. Our motivating insight is that, fundamentally, the worst-case service guaranteed to each flow is a state that can be updated as tasks arrive and are served. The key to scheduling is then finding the schedulability condition on the flows' aggregate state necessary and sufficient to ensure that all flows' long-run guarantees can be met. Once this condition is found, it can be used, on the one hand, to admit or deny new service requests, and on the other, to identify all feasible schedules, that is, all short-run scheduling decisions that can be made without endangering any long-run guarantee, because all such decisions must preserve schedulability. In this way, state-based scheduling allows us to systematically identify all scheduling policies that can simultaneously guarantee all flows their respective services. This is quite a contrast to the traditional approach according to which scheduling policies are first proposed and only thereafter are their capabilities for guaranteeing services examined, verified, and, if possible and necessary, refined.

To find the schedulability condition, we introduce the concept of the spectrum of a worst-case service. During any slot interval, the least capacity that must be reserved to guarantee a worst-case service is specified by a spectral value. If all guarantees are to be maintained, the total capacity to be reserved, that is, the sum of the spectral values of all worst-case services, cannot exceed the server's capacity during the given slot interval. This turns out to be the schedulability condition that we seek. Using this condition, we then identify all feasible schedules. The principal constraint imposed on these schedules is determined by a baseline function that specifies the least number of tasks that must be served from any given subset of flows if, during every slot interval, the sum of all spectral values is to remain less than or equal to the available capacity. As the baseline function is supermodular, we use polymatroid theory to refine our feasible schedule characterization. In particular, we show that when the total service is fixed, the feasible schedules form a permutohedron, a special polytope from polymatroid theory.

\subsection{Three Specializations}\label{SS:specailizations}

A downside of our framework's generality is its complexity. This complexity is two-fold. On the one hand, to fully exploit the flexibility of selecting any feasible schedule, we must fully determine a feasible permutohedron, which is combinatorially difficult. On the other hand, worst-case services, as full-blown maps between cumulative vectors, are challenging to specify and update. To address these difficulties, but also for their own merits, we consider three specializations: max-slack schedules, min-plus services, and dual-curve services.

Max-slack schedules maximize the server's capacity slack, that is, they leave maximum room for the server to admit new service requests. According to the schedulability condition, this is achieved by {\it simultaneously} minimizing {\it all} sums of {\it all} spectral values during {\it all} slot intervals. This, with a little reflection, suggests that, as we will prove, the max-slack schedule is feasible if a feasible schedule exists, and that its identification is independent of the shape of the feasible permutohedron that contains it. Aggregating flows into classes such that, intra-class, flows are max-slack scheduled, enables intermediate tradeoffs of flexibility and efficiency, because when the total service is fixed, the feasible inter-class schedules still form a permutohedron, but of a lower dimension. When all flows' worst-case services allow static deadlines to be assigned to all tasks as they arrive, the max-slack schedule reduces to the well-known earliest-deadline-first (EDF) schedule.

Min-plus services are of interest because, using properties of the min-plus algebra, they can be completely identified by their spectra. To specify and update these services, instead of an uncountably infinite full-blown map between cumulative vectors, we need only specify and update a countably infinite spectral matrix. Among all worst-case services that share the same spectrum, there turns out to be one min-plus service that is maximum in that it maps each queued arrival vector to the maximum worst-case acceptable departure vector. Therefore, every schedulable set of non-min-plus services is dominated by a schedulable set of min-plus services. Thus, by upgrading the former to the latter, we can improve the services, simplify their specification and update, and yet, preserve schedulability.

The efficiency of min-plus services can be further improved by restricting attention to a min-plus service subclass, dual-curve services. This specialization, while maintaining all essential features of our general framework, approaches near practical viability. To specify and update dual-curve services, instead of a spectral matrix, we need only specify and update a pair of cumulative vectors, one static and one dynamic. We call them dual-curve, as opposed to dual-vector, services to highlight their connection to service curves. First suggested in \cite{Parekh:1993, Parekh:1994}, service curves were fully developed in \cite{Cruz:1995}. Each, according to \cite{Cruz:1995}, can be specified by a static cumulative vector. Adding a dynamic vector, yields the dynamic extension, a dual-curve service. This extension allows dual-curve services to be updated as tasks arrive and are served, which is essential for state-based scheduling.

The rest of the paper is organized as follows. In Section~\ref{S:serverModel}, we introduce our service model. In Section~\ref{S:WCS}, we define worst-case services and their spectra. In Sections~\ref{S:schedulability} and \ref{S:feasible}, we introduce state-based scheduling, find the schedulability condition, and then use this condition to identify all feasible schedules. In Sections~\ref{S:MSS}, \ref{S:MPS}, and \ref{S:DVS}, we introduce, respectively, max-slack schedules, min-plus services, and dual-curve services. In Section~\ref{S:remarks}, we conclude with remarks on possible extensions. The paper also includes an appendix in which two lemmas from polymatroid theory are proved for completeness.

\section{The Service Model}\label{S:serverModel}

\begin{figure}[t]
\centering \scalebox{0.972}{\includegraphics{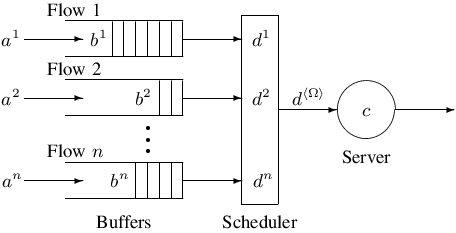}} \caption{The service model, composed of a buffer system, a scheduler and a constant-capacity server.} \label{F:serverModel}
\end{figure}

Our service model is discrete in that time is slotted and all tasks are of unit size. As illustrated in Fig.~\ref{F:serverModel}, the tasks arrive from $n$ distinct flows indexed by $\Omega = \{1,2,\ldots,n\}$. Each flow's tasks are buffered separately, either physically or virtually. All tasks, however, share a single $c$-task-per-slot server, access to which is controlled by a scheduler.

For each flow, indexed by $\omega\in\Omega$, at the beginning of slot~$t$, $a^\omega$ tasks arrive and are immediately queued behind the $b^\omega$ tasks left unserved after slot $t-1$. At this instant, there are
\begin{equation}\label{E:qomega}
q^\omega := a^\omega+b^\omega
\end{equation}
tasks queued in flow $\omega$'s buffer. During slot~$t$, the scheduler determines $d^\omega$, the number of these tasks to serve. As tasks cannot be served before they arrive,
\begin{equation}\label{E:causalityA}
d^\omega \leq q^\omega.
\end{equation}
Within each flow, the service order is first come, first served, so that, at the end of slot~$t$, the first $d^\omega$ tasks queued in flow $\omega$'s buffer depart and are served, leaving
\begin{equation}\label{E:bUpdate}
\dot{b}^\omega = q^\omega-d^\omega
\end{equation}
tasks unserved after slot~$t$. As in (\ref{E:qomega})-(\ref{E:bUpdate}), unless noted otherwise, all variables are implicitly indexed by current slot~$t$. To index $t+1$, we add a dot, as in $\dot{b}^\omega$. To index $t+2$, we add two dots, and so on.

The scheduler's choices are constrained. Denoting the ensemble of flow variables, $[x^1,x^2,\ldots,x^n]$, by $x^{[\Omega]}$, and the sum, $\sum_{\omega \in \Omega} x^\omega$, by $x^{\l\Omega\r}$, it follows that the selected schedule, $d^{[\Omega]}$, must satisfy a {\it \textbf{causality constraint}},
\begin{equation}\label{E:causalityB}
d^{[\Omega]} \leq q^{[\Omega]},
\end{equation}
which restates (\ref{E:causalityA}) in ensemble form, and a {\it \textbf{capacity constraint}},
\begin{equation}\label{E:capacity}
d^{\l\Omega\r} \leq c,
\end{equation}
which requires that the number of tasks served not exceed the server's capacity. We call a $d^{[\Omega]}$ that satisfies both (\ref{E:causalityB}) and (\ref{E:capacity}) a {\it \textbf{valid schedule}}. The scheduler may only select valid schedules.

\section{Worst-Case Services}\label{S:WCS}

In this section, we use cumulative vectors to define worst-case services and analyze their performance bounds. When the scheduler is causal, we show that these services are states that can be updated as tasks arrive and are served. We also introduce the spectrum of a worst-case service as a characterization of the capacities that must be reserved to guarantee it. As a worst-case service is updatable, so is its spectrum.

\subsection{Definition}\label{SS:WCSDef}

A {\it \textbf{cumulative vector}} is a semi-infinite, non-decreasing vector that starts with $0$, with elements in $\mb{N}^+  := \mb{N} \cup \{\infty\}$, where $\mb{N}$ denotes the set of natural numbers. Let $\mb{U}$ be the set of all cumulative vectors. Then $\bs{x}=[x_j]_{j \in \mb{N}}=[x_0,x_1,x_2,\ldots] \in \mb{U}$ if $x_0=0$, and for all $j \in \mb{N}$, $x_j \in \mb{N}^+$ and $x_j \leq x_{j+1}$. Important subsets of $\mb{U}$ include
\begin{equation}\label{E:U>x}
\mb{U} \ua x := \{\bs{x}\in\mb{U} | x_1 \geq x\},
\end{equation}
and
\begin{equation}\label{E:Ux}
\mb{U}|x := \{\bs{x}\in\mb{U} | x_1=x\}.
\end{equation}

For each flow, we use an {\it \textbf{arrival vector}}, $\bs{a}=[a_j]_{j\in\mb{N}} \in \mb{U}$, to record its cumulative task arrivals, and a {\it \textbf{departure vector}}, $\bs{d}=[d_j]_{j\in\mb{N}} \in \mb{U}$, to record its cumulative served tasks. In particular, for all $j>0$, $a_j$ and $d_j$ count the tasks that, respectively, arrive and are served during interval $[t, t+j)$, that is, from slot~$t$ to $t+j-1$. Here, as we are referencing a generic flow, we suppress its index. Note that $\bs{a}$ and $\bs{d}$ completely characterize flow traffic in that they specify when each task arrives and is served. To be precise, let
\begin{equation}\label{E:tau}
\tau_h(\bs{x}):= \max \{j\in\mb{N}^+|x_j < h\} \blank\blank\forall\, h>0.
\end{equation}
Then the $h$th task in $\bs{a}$ arrives in slot $t+\tau_h(\bs{a})$, while the $h$th task in $\bs{d}$ is served in slot $t+\tau_h(\bs{d})$. According to (\ref{E:tau}), $\tau_h(\bs{x}) = \infty$ if $h > x_\infty := \lim_{j \to \infty} x_j$, where the existence of $x_\infty$ in $\mb{N}^+$ is guaranteed because $\bs{x}$ is non-decreasing. For instance, if $h > d_\infty$, $\tau_h(\bs{d}) = \infty$, implying that the $h$th task, even if it exists, is never served.

Although it is up to the scheduler to determine the specific relation between $\bs{a}$ and $\bs{d}$, as tasks cannot be served before they arrive, the number of tasks served during any slot interval cannot exceed the number of arrivals during the same interval plus the number of previously unserved tasks. That is to say,
\begin{equation}\label{E:arrival}
\forall\, j\in\mb{N}, \blank\blank d_j \leq q_j := \left\{
\begin{IEEEeqnarraybox}[][c]{ll}
0       & \blank\blank\text{if } j=0\\
a_j + b & \blank\blank\text{if } j>0
\end{IEEEeqnarraybox}\right. ~,
\end{equation}
or in vector form,
\begin{equation}\label{E:vecArrival}
\bs{d} \leq \bs{q} := \bs{a}+b\bs\delta,
\end{equation}
where $\bs\delta=[\delta_j]_{j \in \mb{N}} := [0, 1, 1, \ldots] \in \mb{U}$, that is,
\begin{equation}\label{E:delta}
\delta_j := \left\{
\begin{IEEEeqnarraybox}[][c]{ll}
0 & \blank\blank\text{if } j=0\\
1 & \blank\blank\text{if } j>0
\end{IEEEeqnarraybox}\right. ~.
\end{equation}
Clearly, (\ref{E:vecArrival}) is the vector extension of (\ref{E:causalityA}) and (\ref{E:qomega}). By definition, $\bs{q} \in \mb{U} \ua b$. As $b$ is entirely fixed by the flow's past, $\bs{q}$ can be viewed as a bijective function of $\bs{a}$ mapping $\mb{U}$ to $\mb{U} \ua b$. Compared to $\bs{a}$, it is as if the $b$ tasks left unserved in the buffer are miscounted by $\bs{q}$ as new arrivals. So we call $\bs{q}$ the {\it \textbf{queued arrival vector}}.

A scheduler, through its scheduling decisions, maps each $\bs{q}$ to some $\bs{d}$. A natural way to specify a service is then in terms of a map from each $\bs{q}$ to a worst-case acceptable $\bs{d}$.

\begin{definition}\label{D:WCS}
{\it For a flow with $b$ tasks left unserved in its buffer, $\bs\psi: \mb{U} \ua b \rightarrow \mb{U}$ is a \textbf{worst-case service} if
\begin{equation}\label{E:WCSDef}
\bs\psi(\bs{q}) \leq \bs{q} \blank\blank\forall\, \bs{q} \in \mb{U} \ua b.
\end{equation}
The flow is said to be guaranteed worst-case service $\bs\psi$ if
\begin{equation}\label{E:WCSDef+}
\bs{d} \geq \bs\psi(\bs{q}) \blank\blank\forall\, \bs{q} \in \mb{U} \ua b.
\end{equation}
}
\end{definition}

\begin{figure}[t]
\centering \scalebox{0.875}{\includegraphics{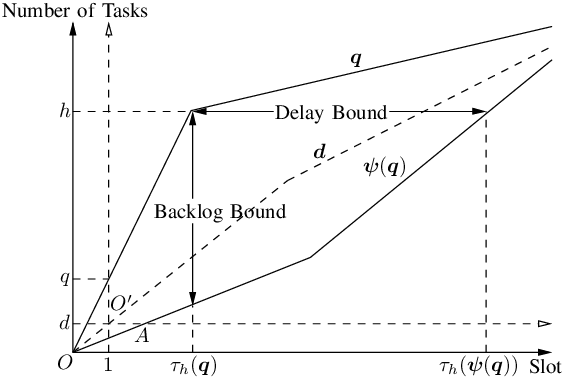}} \caption{Performance bounds and the update rule for a worst-case service.}\label{F:Bounds}
\end{figure}

Subject to (\ref{E:WCSDef}), a worst-case service $\bs\psi$ is simply a map between all queued arrival vectors and their corresponding worst-case departure vectors. Since $\bs\psi$ is conditioned on $b$, whenever we refer to $\bs\psi$, we implicitly refer to the pair, $(\bs\psi, b)$. As illustrated in Fig.~\ref{F:Bounds}, to guarantee $\bs\psi$, $\bs{d}$ must lie between $\bs{q}$ and $\bs\psi(\bs{q})$. This is impossible unless (\ref{E:WCSDef}) holds because otherwise, (\ref{E:vecArrival}) and (\ref{E:WCSDef+}) would contradict each other.

To specify a worst-case service, we need only identify an acceptable $\bs\psi(\bs{q})$ for each $\bs{q}\in\mb{U} \ua b$. This is not practical in general because $\mb{U} \ua b$ is uncountably infinite. Nonetheless, the theoretical possibility of specifying services so broadly itself makes it possible for us to frame the question of ``What to guarantee?" in a most general way.

\subsubsection*{\textup{[}Performance Bounds\textup{]}} In \cite{Cruz:1991A, Cruz:1991B, Parekh:1993, Parekh:1994}, cumulative curve models were shown to be useful tools for performance analysis. The same methods can be applied here given our cumulative vector models. Consider first flow backlogs. For all $j > 0$, the number of tasks left unserved after slot $t+j-1$ is
\begin{equation}\label{E:bj}
b_j := q_j-d_j,
\end{equation}
according to which, $b_1 = \dot{b}$, $b_2 = \ddot{b}$, and so on. As illustrated in Fig.~\ref{F:Bounds}, if the flow is guaranteed $\bs\psi$, this backlog is bounded by
\begin{equation}\label{E:distV}
b_j \leq q_j-\psi_j(\bs{q}) \leq \max_{j > 0} (q_j-\psi_j(\bs{q})).
\end{equation}

Consider next task delays. Recall that, for all $h > 0$, the $h$th task in $\bs{d}$ is served in slot $t+\tau_h(\bs{d})$. Notice that this service corresponds to the $h$th task in $\bs{q}$, not $\bs{a}$, because the $b$ tasks left unserved after slot $t-1$ have to be served before any new arrival. One subtlety here is that there is not enough information to determine when the $b$ tasks arrived. However, if we disregard delays experienced prior to slot~$t$ and treat the $b$ tasks as if they are new arrivals, the $h$th task in $\bs{d}$ can be viewed as arriving in slot $t+\tau_h(\bs{q})$, and its delay is
\begin{equation}\label{E:thetah}
\theta_h := \tau_h(\bs{d})-\tau_h(\bs{q}).
\end{equation}
Then, as illustrated in Fig.~\ref{F:Bounds}, if the flow is guaranteed $\bs\psi$, this delay is bounded by
\begin{equation}\label{E:distH}
\theta_h \leq \tau_h(\bs\psi(\bs{q}))-\tau_h(\bs{q}) \leq \max_{h>0} (\tau_h(\bs\psi(\bs{q}))-\tau_h(\bs{q})).
\end{equation}

Note that the bounds in (\ref{E:distV}) and (\ref{E:distH}) are tied to a specific $\bs{q}$. Confining $\bs{q}$ to a subset of $\mb{U} \ua b$ so that the worst case of the worst-case bounds can be determined, or endowing $\mb{U} \ua b$ with a probability measure so that distributions on the bounds can be derived, extends the bounds to cases when $\bs{q}$ is uncertain. Examples of the former approach can be found in \cite{Cruz:1991A, Cruz:1991B, Parekh:1993, Parekh:1994}, while an example of the latter approach can be found in \cite{Sidi:1993}, where tasks arrive according to a {\it Poisson} process and are served according to a leaky-bucket scheme.

Not only can we bound the performance guaranteed by given worst-case services, we can also design worst-case services to guarantee given performance bounds.

\begin{example}\label{Ex:CBS}
{\it Given $\bar{b} \in \mb{N}$, let
\begin{equation}\label{E:CBS}
\bs\psi(\bs{q}) = (\bs{q}-\bar{b}\bs\delta)^+ \blank\blank\forall\, \bs{q} \in \mb{U} \ua b,
\end{equation}
where $x^+$ denotes $\max \{x, 0\}$. We call $\bs\psi$ a \textbf{uniform-backlog service}, because guaranteeing it bounds all flow backlogs by $\bar{b}$.}
\end{example}

\begin{example}\label{Ex:CDS}
{\it Given $\bar\theta\in\mb{N}$, let
\begin{equation}\label{E:CDS}
\bs\psi(\bs{q}) = \mc{R}^{\bar\theta}\bs{q} \blank\blank\forall\, \bs{q} \in \mb{U} \ua b,
\end{equation}
where $\mc{R}: \mb{U} \rightarrow \mb{U}|0$ is the right-shift operator defined by
\begin{equation}\label{E:RShift}
[\mc{R}\bs{x}]_{j+1}:= x_j \blank\blank\forall\, j\in\mb{N}.
\end{equation}
We call $\bs\psi$ a {\it \textbf{uniform-delay service}}, because guaranteeing it bounds all task delays by $\bar\theta$.}
\end{example}

Using (\ref{E:distV}) and (\ref{E:distH}), we can even design worst-case services to fine-tune the backlog bounds slot-by-slot and the delay bounds task-by-task, although we will not do so here.

\subsection{The Causal Scheduler}\label{SS:causal}

To find $\bs\psi(\bs{q})$, we need to know all of $\bs{q}$. But this is not practical for the scheduler. In this paper, as in practice, we assume the scheduler to be {\it \textbf{causal}} in the sense that it cannot foresee future arrivals. So scheduling decisions must be made when $\bs{q}$ is only partially known. But how can we guarantee $\bs\psi(\bs{q})$ when $\bs{q}$ is only partially known?

Let $\bs{x} \stackrel{j}{=} \bs{y}$ denote the relation that $x_i=y_i$ for all $i \leq j$. Then, as long as $\bs{q} \stackrel{j}{=} \bs{q'}$, the scheduler cannot distinguish between $\bs{q}$ and $\bs{q'}$ before slot $t+j$. So it must ensure that
\[
d_j \geq \max\{\psi_j(\bs{q}), \psi_j(\bs{q'})\},
\]
to ensure that worst-case service $\bs\psi$ can be guaranteed no matter whether $\bs{q}$ or $\bs{q'}$ is realized. Applying this logic repeatedly, the scheduler must ensure that
\[
d_j \geq \max_{\bs{q'} \in \mb{U} \ua b, \bs{q'} \stackrel{j}{=} \bs{q}} \psi_j(\bs{q'}).
\]
Unlike $d_j$, this lower bound need not be non-decreasing with respect to $j$. But, since served tasks cannot be unserved, in guaranteeing $\bs\psi$, the scheduler is effectively guaranteeing a second service, $\bs\psi^\text{C}$, defined by
\begin{equation}\label{E:causalService}
\psi_j^\text{C}(\bs{q}) := \max_{\bs{q'} \in \mb{U} \ua b, i \leq j, \bs{q'} \stackrel{i}{=} \bs{q}} \psi_i(\bs{q'}) \blank\blank\forall\, \bs{q}\in\mb{U} \ua b, j\in\mb{N}.
\end{equation}
Clearly $\bs\psi^\text{C} \geq \bs\psi$, that is, $\bs\psi^\text{C}(\bs{q}) \geq \bs\psi(\bs{q})$ for all $\bs{q}\in\mb{U} \ua b$. Intuitively this is because the causal scheduler, always assuming the worst, may, in hindsight, over-guarantee some services.

\subsubsection*{\textup{[}Causal Services\textup{]}} In view of (\ref{E:causalService}), $\bs\psi^\text{C}$ is not just any service but one such that
\begin{equation}\label{E:causalService+}
\forall\, \bs{q}, \bs{q'} \in \mb{U} \ua b, j\in\mb{N},\blank\blank \bs\psi^\text{C}(\bs{q}) \stackrel{j}{=} \bs\psi^\text{C}(\bs{q'}) \text{ if } \bs{q} \stackrel{j}{=} \bs{q'}.
\end{equation}
That is to say, $\psi_j^\text{C}(\bs{q})$ depends on $q_1, q_2, \ldots, q_j$ alone. So we call $\bs\psi^\text{C}$ {\it \textbf{causal}}. For instance, it is easy to verify that both the uniform-backlog and uniform-delay services in Examples~\ref{Ex:CBS} and \ref{Ex:CDS} are causal. In fact, for the causal scheduler, it would not be at all restrictive to require that all worst-case services be causal, although we will not do so here because non-causality gives us the flexibility to specify each $\bs\psi(\bs{q})$ independently.

\begin{example}\label{Ex:STS}
{\it Given $\bar\theta \in \mb{N}$, let $b=0$ and, for all $\bs{q} \in \mb{U}$, let
\begin{equation}\label{E:STSDef}
\bs\psi(\bs{q})=\left\{
\begin{IEEEeqnarraybox}[][c]{ll}
\mc{R}^{i+\bar\theta}\bs\delta & \blank\blank\textup{if } \bs{q} = \mc{R}^i\bs\delta \textup{ for some } i\in\mb{N}\\
\bs{0}                         & \blank\blank\textup{otherwise}
\end{IEEEeqnarraybox}\right. ~,
\end{equation}
where $\bs{0}$ is the all-zero vector. We call $\psi$ a \textbf{single-task service}, because it guarantees that a task will be served, and that the task's delay will not exceed $\bar\theta$, if and only if the flow's traffic consists of a single task. But the scheduler can never be certain that flow traffic consists of a single task because there is always the possibility that additional tasks will arrive later. So single-task services are non-causal. In this case, using (\ref{E:causalService}) and (\ref{E:STSDef}), it can be verified that
\begin{equation}\label{E:STSCausal}
\bs\psi^\textup{C}(\bs{q})=\left\{
\begin{IEEEeqnarraybox}[][c]{ll}
\mc{R}^{i+\bar\theta}\bs\delta & \blank\blank\textup{if } \bs{q} \stackrel{i+\bar\theta}{=} \mc{R}^i\bs\delta \textup{ for some } i\in\mb{N}\\
\bs{0}                         & \blank\blank\textup{otherwise}
\end{IEEEeqnarraybox}\right. ~.
\end{equation}
Unlike the single-task service, its causal counterpart bounds the delay for the first task in both single and multi-task flows, but only if the second task arrives after the first task's service deadline.
}
\end{example}

\subsection{The Update Rule}\label{SS:WCSUpdate}

When $a$ tasks arrive in slot~$t$, the set of possible $\bs{a}$'s shrinks from $\mb{U}$ to $\mb{U}|a$, and consequently, due to (\ref{E:vecArrival}) and (\ref{E:qomega}), that of possible $\bs{q}$'s shrinks from $\mb{U} \ua b$ to $\mb{U}|q$. Since the scheduler is causal, it cannot foresee which $\bs{q} \in \mb{U}|q$ will be realized, so it must ensure that
\begin{equation}\label{E:WCSImmediate}
d \geq p := \max_{\bs{q} \in \mb{U}|q} \psi_1(\bs{q}),
\end{equation}
to ensure that $\bs{d} \geq \bs\psi(\bs{q})$ can be guaranteed no matter which $\bs{q}\in\mb{U}|q$ is realized.\footnote{Using (\ref{E:causalService}) and (\ref{E:WCSImmediate}), it is easy to verify that $p = \psi_1^\text{C}(\bs{q})$ for all $\bs{q} \in \mb{U}|q$, so $d \geq \psi_1^\text{C}(\bs{q})$ if $d \geq p$. As we will see, $\bs\psi$ can be updated to $\bs{\dot\psi}$, $\bs{\ddot\psi}$, and so on. Accordingly, it can be shown that $\bs\psi^\text{C}$ is guaranteed if $d \geq p$, $\dot{d} \geq \dot{p}$, $\ddot{d} \geq \ddot{p}$, and so on.}

The immediate portion of $\bs\psi$ to be met by $d$ is denoted by $p$ in (\ref{E:WCSImmediate}). But what about the remaining portion? As we will see, it turns out to be yet another worst-case service. Intuitively, as illustrated in Fig.~\ref{F:Bounds}, after $d$ tasks have been served, $\bs\psi(\bs{q})$ can be re-expressed in a translated coordinate frame in which the origin moves from $O$ to $O'$ at $(1,d)$ in the original frame. Discounting the immediate portion met by $d$, $\bs{\psi}(\bs{q})$ is truncated in this new frame. A new worst-case departure vector can then, roughly speaking, be constructed by splicing the line segment $\overline{O'A}$ to the truncated $\bs{\psi}(\bs{q})$, that is, by replacing $\overline{OA}$ with $\overline{O'A}$.

To formalize this intuition, observe first that since the counting process for $\bs{d}$ starts in slot~$t$ while that for $\bs{\dot{d}}$ starts in $t+1$, $d_{j+1}$ and $\dot{d}_j$ count, respectively, the number of tasks that are served during intervals $[t, t+j+1)$ and $[t+1, t+j+1)$. Therefore, when $d$ tasks are served in slot~$t$, for all $j\in\mb{N}$,
\begin{equation}\label{E:dUpdate}
d_{j+1}=\dot{d}_j+d.
\end{equation}
Using (\ref{E:RShift}), this can be rewritten in vector form as
\begin{equation}\label{E:vecdUpdate}
\bs{d} = \mc{R}\bs{\dot{d}}+d\bs\delta.
\end{equation}
Similarly, when $a$ tasks arrive in slot~$t$, $a_{j+1}=\dot{a}_j+a$, so using (\ref{E:arrival}), (\ref{E:qomega}), and (\ref{E:bUpdate}), we have
\begin{equation}\label{E:qUpdate++}
q_{j+1} = a_{j+1}+b = \dot{a}_j+q =\left\{
\begin{IEEEeqnarraybox}[][c]{ll}
q           & \blank\blank\text{if } j=0\\
\dot{q}_j+d & \blank\blank\text{if } j>0
\end{IEEEeqnarraybox}\right. ~.
\end{equation}
Using (\ref{E:RShift}) and (\ref{E:vecArrival}), this can be rewritten in vector form as
\begin{equation}\label{E:vecqUpdate}
\bs{q} = \mc{R}\bs{\dot{a}}+q\bs\delta = \mc{R}(\bs{\dot{q}}-\dot{b}\bs\delta)+q\bs\delta.
\end{equation}

Observe next that if $\bs\psi$ is guaranteed, for all $\bs{q} \in \mb{U}|q$, it must be guaranteed that $\bs{d} \geq \bs\psi(\bs{q})$. Then, as $\mc{R}\bs{\dot{d}} \in \mb{U}|0$, it is immediate from (\ref{E:vecdUpdate}) that
\begin{equation}\label{E:WCSUpdateA}
\bs{\dot{d}} \geq \mc{R}^{-1}(\bs\psi(\bs{q})-d\bs\delta)^+,
\end{equation}
where $\mc{R}^{-1}: \mb{U}|0 \rightarrow \mb{U}$ is the inverse of $\mc{R}$, implying that
\begin{equation}\label{E:LShift}
[\mc{R}^{-1}\bs{x}]_j= x_{j+1} \blank\blank\forall\, j\in\mb{N}.
\end{equation}
Notice that (\ref{E:vecqUpdate}) establishes that $\bs{q}$ is a bijective function of $\bs{\dot{q}}$, mapping $\mb{U}\ua\dot{b}$ to $\mb{U}|q$. This fact, together with (\ref{E:WCSUpdateA}), leads to the following update rule.

\begin{theorem}\label{T:WCSUpdate}
{\it When $b$ tasks are queued, $a$ tasks arrive, and $d$ tasks are served in slot~$t$, if both (\ref{E:causalityA}) and (\ref{E:WCSImmediate}) hold, that is, if $q \geq d \geq p$, the flow is guaranteed worst-case service $\bs\psi$ if and only if for all $\bs{\dot{q}} \in \mb{U}\ua\dot{b}$,
\begin{IEEEeqnarray}{rRl}
\bs{\dot{d}} \geq \bs{\dot\psi}(\bs{\dot{q}}) &:=& \mc{R}^{-1}(\bs\psi(\bs{q})-d\bs\delta)^+ \label{E:WCSUpdateB}\\
                                              &= & \mc{R}^{-1}(\bs\psi(\mc{R}(\bs{\dot{q}}-\dot{b}\bs\delta)+q\bs\delta)-d\bs\delta)^+, \label{E:WCSUpdateC}
\end{IEEEeqnarray}
where $\bs{\dot\psi}$ is a worst-case service for the flow in slot $t+1$.}
\end{theorem}

According to this theorem, $\bs\psi$ is a state variable that can be updated to $\bs{\dot\psi}$, the remaining portion of $\bs\psi$ to be guaranteed after slot~$t$. Recall that $\bs\psi$ is conditioned on $b$, so $\bs{\dot\psi}$ is conditioned on $\dot{b}$. According to (\ref{E:bUpdate}), $b$ is also a state variable that can be updated to $\dot{b}$. Therefore, whenever we update $\bs\psi$ to $\bs{\dot\psi}$, we implicitly update $(\bs\psi, b)$ to $(\bs{\dot\psi}, \dot{b})$ through (\ref{E:WCSUpdateB}) and (\ref{E:bUpdate}).

\begin{IEEEproof}[Proof of Theorem~\ref{T:WCSUpdate}]
The necessity of (\ref{E:WCSUpdateB}) follows directly from the derivation of (\ref{E:WCSUpdateA}). We need only show its sufficiency. As $d \geq p$, (\ref{E:WCSImmediate}) implies that $(\bs\psi(\bs{q})-d\bs\delta)^+ \in \mb{U}|0$ for all $\bs{q} \in \mb{U}|q$, justifying the use of $\mc{R}^{-1}$ in (\ref{E:WCSUpdateB}). Now, if (\ref{E:WCSUpdateB}) holds, for all $\bs{q} \in \mb{U}|q$, (\ref{E:vecdUpdate}) implies that
\[
\bs{d} = \mc{R}\bs{\dot{d}}+d\bs\delta \geq (\bs\psi(\bs{q})-d\bs\delta)^+ +d\bs\delta \geq \bs\psi(\bs{q}).
\]
So $\bs\psi$ is guaranteed.

It remains to show that $\bs{\dot\psi}$ is a worst-case service. For all $\bs{\dot{q}} \in \mb{U}\ua\dot{b}$ and $j>0$, using (\ref{E:WCSUpdateB}), (\ref{E:LShift}), (\ref{E:WCSDef}), and (\ref{E:qUpdate++}), we have
\[
\dot\psi_j(\bs{\dot{q}}) =    (\psi_{j+1}(\bs{q})-d)^+
                         \leq (q_{j+1}-d)^+
                         =    \dot{q}_j^+              = \dot{q}_j,
\]
where the final equality holds because, according to (\ref{E:bUpdate}), \mbox{$d \leq q$} guarantees that $\dot{b} \geq 0$, and thus $\dot{q}_j \geq 0$. It follows that $\bs{\dot\psi}(\bs{\dot{q}}) \leq \bs{\dot{q}}$. So, according to Definition~\ref{D:WCS}, $\bs{\dot\psi}$ is indeed a worst-case service.
\end{IEEEproof}

In later sections, we will consider various specializations of worst-case services that we would prefer persist from slot to slot. These considerations motivate the concept of update invariance. A class of worst-case services is {\it \textbf{update invariant}} if it is preserved by the update rule, that is, if $\bs\psi$ belongs to the class, so does $\bs{\dot\psi}$. Using (\ref{E:causalService+}) and (\ref{E:CBS}), it can be verified that both causal services, and the uniform-backlog services in Example~\ref{Ex:CBS}, are update invariant because, in both cases, if the defining property holds from slot~$t$ on, it must also hold from $t+1$ on. In contrast, the uniform-delay services in Example~\ref{Ex:CDS} are not update invariant, because the delay bound of any task left unserved after slot~$t$ has to be reduced by one.\footnote{This can be remedied by extending uniform-delay services to their dynamic closure. If there exists $\bs{r} \in \mb{U}$ with $\mc{R}^{\bar\theta} (b\bs\delta) \leq \bs{r} \leq b\bs\delta$ such that
\[
\bs\psi(\bs{q}) = \mc{R}^{\bar\theta}(\bs{q}-b\bs\delta) + \bs{r} \blank\blank\forall\, \bs{q} \in \mb{U} \ua b,
\]
we call $\bs\psi$ an {\it \textbf{extended uniform-delay service}}. It can be shown that these extended services are update invariant.} Neither are the single-task services in Example~\ref{Ex:STS}, because once a task has been served, these services' guarantees end.

\subsection{The Spectrum}\label{SS:signatures}

Full-blown maps between cumulative vectors are difficult to visualize. The concept of spectrum helps us distill what is essential. The key question is: for the causal scheduler to guarantee $\bs\psi$, what capacity must be reserved during each slot interval? According to (\ref{E:WCSDef+}), given $\bs{q} \in \mb{U} \ua b$, to guarantee $\bs\psi$, $d_j \geq \psi_j(\bs{q})$ tasks must be served during interval $[t,t+j)$. But according to (\ref{E:arrival}), only $d_i \leq q_i$ tasks can be served during $[t, t+i)$ because tasks cannot be served before they arrive. It follows that
\[
(d_j-d_i)^+ \geq (\psi_j(\bs{q})-q_i)^+,
\]
where $(d_j-d_i)^+$ counts the number of tasks that are served during $[t+i, t+j)$, and is $0$ by default if $i \geq j$. This observation, which holds for all $\bs{q} \in \mb{U} \ua b$, motivates the following definition.

\begin{definition}\label{D:spectrum}
{\it Given worst-case service $\bs\psi$, for all $i, j \in \mb{N}$, the \textbf{spectral value} of $\bs\psi$ over interval $[t+i, t+j)$ is
\begin{equation}\label{E:signature}
\lambda_{ij}(\bs\psi):= \max_{\bs{q} \in \mb{U} \ua b} (\psi_j(\bs{q})-q_i)^+.
\end{equation}
We call the collection of all such values the \textbf{spectrum} of $\bs\psi$.}
\end{definition}

By construction, $\lambda_{ij}(\bs\psi)$ is the least capacity that must be reserved during interval $[t+i, t+j)$ to ensure that $\bs{d} \geq \bs\psi(\bs{q})$ can be guaranteed no matter which $\bs{q}\in\mb{U}\ua b$ is realized. If \mbox{$\bs\psi \geq \bs{\psi'}$}, it is immediate that $\lambda_{ij}(\bs\psi) \geq \lambda_{ij}(\bs{\psi'})$. So, the better the guarantee, the greater the capacity that must be reserved. Our choice of the term, spectrum, is fundamentally arbitrary, but can be contextualized as follows. If $N$ is a normal matrix, that is, $N^\dagger N = N N^\dagger$, and $\lambda_{\max}$ is its maximum-magnitude eigenvalue, it is well known that
\[
|\lambda_{\max}| = \max_{\|\bs{x}\| \neq 0} (\|N\bs{x}\| \div \|\bs{x}\|).
\]
Comparing this expression to (\ref{E:signature}), it is apparent that $\lambda_{\max}$, $N$, and $\bs{x}$ loosely correspond to, respectively, $\lambda_{ij}(\bs\psi)$, $\bs\psi$, and $\bs{q}$ in (\ref{E:signature}). There is even a correspondence between operators $\div$ and $-$. In the so-called min-plus algebra to be introduced in Section~\ref{S:MPS}, $+$ replaces $\times$ in the standard algebra, so it is only natural to replace $\div$ by $-$.

Henceforth, we will denote $\lambda_{ij}(\bs\psi)$ by $\lambda_{ij}$ when no confusion can be introduced. For instance, if $\bs\psi$ is a uniform-backlog service, for all $i, j \in \mb{N}$, using (\ref{E:CBS}), we have
\begin{equation}\label{E:spectrumCBS}
\lambda_{ij}=\left\{
\begin{IEEEeqnarraybox}[][c]{ll}
\infty & \blank\blank\textup{if } j > i\\
0      & \blank\blank\textup{if } j \leq i
\end{IEEEeqnarraybox}\right. ~.
\end{equation}
If $\bs\psi$ is a uniform-delay service, using (\ref{E:CDS}), we have
\begin{equation}\label{E:spectrumCDS}
\lambda_{ij}=\left\{
\begin{IEEEeqnarraybox}[][c]{ll}
\infty & \blank\blank\textup{if } j-i > \bar\theta\\
0      & \blank\blank\textup{if } j-i \leq \bar\theta
\end{IEEEeqnarraybox}\right. ~.
\end{equation}
If $\bs\psi$ is a single-task service, using (\ref{E:STSDef}), we have
\begin{equation}\label{E:spectrumSTS}
\lambda_{ij}=\left\{
\begin{IEEEeqnarraybox}[][c]{ll}
1 & \blank\blank\textup{if } j-i > \bar\theta\\
0 & \blank\blank\textup{if } j-i \leq \bar\theta
\end{IEEEeqnarraybox}\right. ~.
\end{equation}
In the last case, using (\ref{E:STSCausal}), it is easy to verify that (\ref{E:spectrumSTS}) also applies to $\bs\psi$'s causal counterpart, $\bs\psi^\text{C}$. So the single-task service and its causal counterpart share the same spectrum. As we will see in Section~\ref{S:MPS}, this is not a coincidence.

The next theorem lists some basic properties of $\lambda_{ij}$. As will be shown in Section~\ref{SS:MPSSig}, these properties are not only necessary but also sufficient for a collection of $\lambda_{ij}$'s to be the spectrum of some worst-case service.

\begin{theorem}\label{T:SigProp}
{\it For all $i, j \in \mb{N}$,
\begin{equation}\label{E:SigPropA}
\lambda_{ij} = 0 \blank\textup{if } i \geq j,
\end{equation}
\begin{equation}\label{E:SigPropB}
\lambda_{ij} \leq \lambda_{i, j+1},
\end{equation}
\begin{equation}\label{E:SigPropC}
\lambda_{ij} \geq \lambda_{i+1, j},
\end{equation}
and
\begin{equation}\label{E:SigPropD}
\lambda_{ij} \leq (\lambda_{0j}-b\delta_i)^+.
\end{equation}}
\end{theorem}

\begin{IEEEproof}
First, if $i \geq j$, since $\psi_j(\bs{q}) \leq \psi_i(\bs{q}) \leq q_i$, according to (\ref{E:signature}), $\lambda_{ij} = 0$. Second, since $\psi_j(\bs{q}) \leq \psi_{j+1}(\bs{q})$ and $q_i \leq q_{i+1}$, according to (\ref{E:signature}), $\lambda_{ij} \leq \lambda_{i, j+1}$ and $\lambda_{ij} \geq \lambda_{i+1, j}$. Finally, since $\bs{q} \geq b\bs\delta$ for all $\bs{q}\in\mb{U}\ua b$, using (\ref{E:signature}), we have
\[
\lambda_{ij} \leq \max_{\bs{q} \in \mb{U} \ua b} (\psi_j(\bs{q})-b\delta_i)^+ = \left(\max_{\bs{q} \in \mb{U} \ua b} \psi_j(\bs{q})-b\delta_i\right)^+.\\
\]
But we also have
\[
\lambda_{0j} = \max_{\bs{q} \in \mb{U} \ua b} \psi_j(\bs{q}),
\]
so (\ref{E:SigPropD}) must be true.
\end{IEEEproof}

\subsection{Updating the Spectrum}\label{E:spectrumUpdate}

According to Theorem~\ref{T:WCSUpdate}, given $q \geq d \geq p$, we can update $\bs\psi$ to $\bs{\dot\psi}$. Denote $\lambda_{ij}(\bs{\dot\psi})$ by $\dot\lambda_{ij}$. Then, using (\ref{E:signature}), (\ref{E:WCSUpdateB}), and (\ref{E:LShift}), we have
\[
\begin{IEEEeqnarraybox}[][c]{rCl}
\dot\lambda_{ij} &=& \max_{\bs{\dot{q}} \in \mb{U}\ua \dot{b}} (\dot\psi_j(\bs{\dot{q}})-\dot{q}_i)^+\\
                 &=& \max_{\bs{q} \in \mb{U}|q} ([\mc{R}^{-1}(\bs\psi(\bs{q})-d\bs\delta)^+]_j-\dot{q}_i)^+ \\
                 &=& \max_{\bs{q} \in \mb{U}|q} (\psi_{j+1}(\bs{q})-d-\dot{q}_i)^+,
\end{IEEEeqnarraybox}
\]
where the second equality holds because, due to (\ref{E:vecqUpdate}), $\bs{\dot{q}} \in \mb{U}\ua \dot{b}$ is equivalent to $\bs{q} \in \mb{U}|q$. This implies that if $i=0$,
\begin{equation}\label{E:SigUpdProof+}
\dot\lambda_{0j} = \max_{\bs{q} \in \mb{U}|q} (\psi_{j+1}(\bs{q})-d)^+ = \left(\max_{\bs{q} \in \mb{U}|q} \psi_{j+1}(\bs{q})-d\right)^+,
\end{equation}
and if $i>0$, due to (\ref{E:qUpdate++}),
\begin{equation}\label{E:SigUpdProof++}
\dot\lambda_{ij} = \max_{\bs{q} \in \mb{U}|q} (\psi_{j+1}(\bs{q})-q_{i+1})^+.
\end{equation}
Let
\begin{equation}\label{E:CondSig}
\lambda_{ij}(\bs\psi|q) := \max_{\bs{q} \in \mb{U}|q} (\psi_j(\bs{q})-q_i)^+,
\end{equation}
and denote it by $\hat\lambda_{ij}$. Rewriting (\ref{E:SigUpdProof+}) and (\ref{E:SigUpdProof++}) in terms of $\hat\lambda_{ij}$, we obtain the next theorem.

\begin{theorem}\label{T:SigUpdate}
{\it Given $\bs{\dot\psi}$ in Theorem~\ref{T:WCSUpdate}, for all $i, j \in \mb{N}$,
\begin{equation}\label{E:SigUpdate}
\dot\lambda_{ij}=\left\{
\begin{IEEEeqnarraybox}[][c]{ll}
(\hat\lambda_{0, j+1}-d)^+ & \blank\blank\textup{if } i=0\\
\hat\lambda_{i+1, j+1}     & \blank\blank\textup{if } i>0
\end{IEEEeqnarraybox}\right. ~.
\end{equation}}
\end{theorem}

\subsubsection*{\textup{[}Conditional Spectra\textup{]}} According to Theorem~\ref{T:SigUpdate}, using $\hat\lambda_{ij}$, we can identify $\dot\lambda_{ij}$. Comparing (\ref{E:CondSig}) to (\ref{E:signature}), it is seen that the only difference is that $\bs{q}$'s range shrinks from $\mb{U} \ua b$ to $\mb{U} | q$. Accordingly, $\hat\lambda_{ij}$ is the least capacity that must be reserved during interval $[t+i, t+j)$ to ensure that $\bs{d} \geq \bs\psi(\bs{q})$ can be guaranteed no matter which $\bs{q}\in\mb{U}|q$ is realized. For this reason, we call $\hat\lambda_{ij}$ the {\it \textbf{conditional spectral value}} of $\bs\psi$ over interval \mbox{$[t+i, t+j)$}, and the collection of all such values, the {\it \textbf{conditional spectrum}} of $\bs\psi$. The next theorem lists some basic properties of $\hat\lambda_{ij}$.

\begin{theorem}\label{T:CondSigProp}
{\it For all $i, j \in \mb{N}$,
\begin{equation}\label{E:CondSigPropA}
\hat\lambda_{ij} = 0 \blank\textup{if } i \geq j,
\end{equation}
\begin{equation}\label{E:CondSigPropB}
\hat\lambda_{ij} \leq \hat\lambda_{i, j+1},
\end{equation}
\begin{equation}\label{E:CondSigPropC}
\hat\lambda_{ij} \geq \hat\lambda_{i+1, j},
\end{equation}
\begin{equation}\label{E:CondSigPropD}
\hat\lambda_{ij} \leq (\hat\lambda_{0j}-q\delta_i)^+,
\end{equation}
\begin{equation}\label{E:CondSigPropE}
\hat\lambda_{1j} = (\hat\lambda_{0j}-q)^+,
\end{equation}
\begin{equation}\label{E:p-lambda}
\hat\lambda_{01} = p,
\end{equation}
and
\begin{equation}\label{E:CondSigPropF}
\hat\lambda_{ij} \leq \lambda_{ij}.
\end{equation}}
\end{theorem}

\begin{IEEEproof}
First, as (\ref{E:CondSigPropA}), (\ref{E:CondSigPropB}), and (\ref{E:CondSigPropC}) are, respectively, the conditional counterparts of (\ref{E:SigPropA}), (\ref{E:SigPropB}), and (\ref{E:SigPropC}), they can be proved in exactly the same ways. Second, since $q_i \geq q\delta_i$ and $q_1 = q$ for all $\bs{q} \in \mb{U}|q$, using (\ref{E:CondSig}), we have
\[
\hat\lambda_{ij} \leq \max_{\bs{q} \in \mb{U}|q} (\psi_j(\bs{q})-q\delta_i)^+ = \left(\max_{\bs{q}\in\mb{U}|q} \psi_j(\bs{q})-q\delta_i\right)^+,
\]
and
\[
\hat\lambda_{1j} = \max_{\bs{q}\in\mb{U}|q}(\psi_j(\bs{q})-q_1)^+ = \left(\max_{\bs{q}\in\mb{U}|q} \psi_j(\bs{q})-q\right)^+.
\]
But we also have
\[
\hat\lambda_{0j} = \max_{\bs{q}\in\mb{U}|q} \psi_j(\bs{q}),
\]
so (\ref{E:CondSigPropD}) and (\ref{E:CondSigPropE}) must be true. Third, comparing the previous equation to (\ref{E:WCSImmediate}), we find that $\hat\lambda_{01} = p$. Finally, since $\mb{U}|q \subset \mb{U}\ua b$, comparing (\ref{E:CondSig}) to (\ref{E:signature}), it is immediate that $\hat\lambda_{ij} \leq \lambda_{ij}$.
\end{IEEEproof}

\subsubsection*{\textup{[}Interpreting Theorem~\ref{T:SigUpdate}\textup{]}} The spectral update relation in (\ref{E:SigUpdate}) has a straightforward interpretation. By definition, both $\dot\lambda_{ij}$ and $\hat\lambda_{i+1, j+1}$ specify the least capacity that must be reserved during interval $[t+i+1, t+j+1)$. The difference is that $\hat\lambda_{i+1, j+1}$ is computed relative to slot~$t$, {\it before} any task has been served, by letting $d_{i+1} = q_{i+1}$ for each $\bs{q} \in \mb{U} | q$, whereas $\dot\lambda_{ij}$ is computed relative to $t+1$, {\it after} the service of $d$ tasks, by letting $\dot{d}_i = \dot{q}_i$ for each $\bs{\dot{q}} \in \mb{U} \ua \dot{b}$. But, given any $d \leq q$, according to (\ref{E:vecqUpdate}), $\bs{\dot{q}} \in \mb{U} \ua \dot{b}$ is equivalent to $\bs{q} \in \mb{U} | q$, and moreover, if $i > 0$, according to (\ref{E:dUpdate}) and (\ref{E:qUpdate++}), $\dot{d}_i = \dot{q}_i$ is equivalent to $d_{i+1} = q_{i+1}$. So $\dot\lambda_{ij}=\hat\lambda_{i+1, j+1}$ in the case that $i>0$.

In the case that $i=0$, {\it pre}-service, the least capacity that must be reserved during interval $[t+1, t+j+1)$ is still $\hat\lambda_{1, j+1}$. It is computed by letting $d_1 = q_1$, that is, $d=q$. But, {\it post}-service, it cannot always remain $\hat\lambda_{1, j+1}$, because $d \leq q$ tasks have been actually served. According to (\ref{E:SigUpdate}), it now becomes $(\hat\lambda_{0, j+1}-d)^+$, the least capacity that must be reserved during $[t,t+j+1)$ minus the reserve released when the $d$ tasks are served. Consistent with this interpretation, given any $d \leq q$, according to (\ref{E:CondSigPropE}),
\[
\hat\lambda_{1, j+1} = (\hat\lambda_{0, j+1}-q)^+ \leq (\hat\lambda_{0, j+1}-d)^+,
\]
so it is immediate that
\begin{equation}\label{E:lambdaBound}
\dot\lambda_{0j} \geq \hat\lambda_{1, j+1} \blank\blank\forall\, j\in\mb{N},
\end{equation}
the lower bound of which is achieved when $d=q$.

\section{State-Based Scheduling}\label{S:schedulability}

When each flow in our service model is guaranteed a worst-case service, we call the model a {\it \textbf{worst-case system}}. For all $\omega\in\Omega$, let $\bs\psi^\omega$ denote the worst-case service guaranteed to flow $\omega$, and let $\bs\psi^{[\Omega]}$ denote the worst-case system. Since $\bs\psi^\omega$ is a state of flow $\omega$, $\bs\psi^{[\Omega]}$ can be viewed as the aggregate state for all flows in the system. Recall that $\bs\psi^\omega$ is conditioned on $b^\omega$, so $\bs\psi^{[\Omega]}$ is conditioned on $b^{[\Omega]}$. Accordingly, whenever we refer to $\bs\psi^{[\Omega]}$, we implicitly refer to $(\bs\psi^{[\Omega]}, b^{[\Omega]})$.

Given $\bs\psi^{[\Omega]}$, can the server guarantee all flows their respective worst-case services simultaneously? If yes, how? There is no problem if the server's capacity is infinite, because, according to the ensemble version of (\ref{E:WCSImmediate}), we can simply select $d^{[\Omega]} \geq p^{[\Omega]}$, $\dot{d}^{[\Omega]} \geq \dot{p}^{[\Omega]}$, $\ddot{d}^{[\Omega]} \geq \ddot{p}^{[\Omega]}$, and so on. But since the server's capacity is inevitably finite, we need to be more strategic. Given any $a^{[\Omega]}$, for $d^{[\Omega]} \geq p^{[\Omega]}$ to be possible, due to capacity constraint (\ref{E:capacity}), it is necessary that $p^{\l\Omega\r} \leq c$. But $d^{[\Omega]} \geq p^{[\Omega]}$ is not enough by itself, because to further ensure the possibility for $\dot{d}^{[\Omega]} \geq \dot{p}^{[\Omega]}$, $d^{[\Omega]}$ must also induce a $\bs{\dot\psi}^{\raisebox{-0.9mm}{\scriptsize$[\Omega]$}}$ such that given any $\dot{a}^{[\Omega]}$, $\dot{p}^{\l\Omega\r} \leq c$. Even this is not enough, because by the same logic, $d^{[\Omega]}$ must induce a $\bs{\dot\psi}^{\raisebox{-0.9mm}{\scriptsize$[\Omega]$}}$ such that given any $\dot{a}^{[\Omega]}$, there exists $\dot{d}^{[\Omega]} \geq \dot{p}^{[\Omega]}$ that induces a $\bs{\ddot\psi}^{\raisebox{-0.9mm}{\scriptsize$[\Omega]$}}$ such that given any $\ddot{a}^{[\Omega]}$, $\ddot{p}^{\l\Omega\r} \leq c$. This construction can be pursued indefinitely but quickly becomes unmanageable. Fortunately, there is a neat solution to this problem, the cornerstone of which lies in the idea of state-based scheduling.

In the remainder of this section, we outline how state-based scheduling is accomplished, and use the concept of spectrum to formulate the schedulability condition for worst-case systems. This condition will prove to be the key to state-based scheduling. We also explore how the schedulability condition can guide the design of multiplexing systems by introducing the concept of multiplexing gain.

\subsection{The General Idea}\label{E:SBSIdea}

Since the server's capacity is finite, no matter how smart the scheduler is, it will not be able to guarantee an arbitrary worst-case service. For instance, according to (\ref{E:CBS}) and (\ref{E:CDS}), it is impossible to guarantee a uniform-backlog service if $\bs{q} = (\bar{b}+c+1)\bs\delta$ or a uniform-delay service if $\bs{q} = (\bar\theta c + 1) \bs\delta$. What is needed is a condition to distinguish what can be guaranteed from what cannot. This condition must be sustainable because a $\bs\psi^{[\Omega]}$ that can be guaranteed must imply a $\bs{\dot\psi}^{\raisebox{-0.9mm}{\scriptsize$[\Omega]$}}$ that can be guaranteed, that implies a $\bs{\ddot\psi}^{\raisebox{-0.9mm}{\scriptsize$[\Omega]$}}$ that can be guaranteed, and so on, {\it ad infinitum}. This reasoning motivates the following definition.

\begin{definition}\label{D:proper}
{\it Given a condition \textup{P} on $\bs\psi^{[\Omega]}$, a \textbf{sustainer} of \textup{P} is a valid schedule, $d^{[\Omega]}$, that satisfies the ensemble version of (\ref{E:WCSImmediate}), $d^{[\Omega]} \geq p^{[\Omega]}$, and via (\ref{E:WCSUpdateB}), induces a $\bs{\dot\psi}^{\raisebox{-0.9mm}{\scriptsize$[\Omega]$}}$ that satisfies \textup{P} in the next slot. We call \textup{P} \textbf{sustainable} if a $\bs\psi^{[\Omega]}$'s satisfaction of \textup{P} implies that, given any $a^{[\Omega]}$, there exists at least one sustainer of \textup{P}.}
\end{definition}

According to this definition, once sustainable conditions are met, they can always be met. Sustainable conditions can be rather trivial. For instance, it is sustainable to require that $\bs\psi^\omega(\bs{q}^\omega) = \bs{0}$ for all $\omega \in \Omega$ and $\bs{q}^\omega \in \mb{U} | b^\omega$. But this is nothing but a disguised way of saying that there should be no guarantee to any flow at all. Sustainable conditions are also not unique. For instance, any condition that is sustainable for a server with capacity $c' \leq c$ must also be so for a server with capacity $c$. However, there does exist a unique, non-trivial, sustainable condition. To see this, let P1 and P2 both be sustainable. By definition, if $\bs\psi^{[\Omega]}$ satisfies ``P1 or P2", given any $a^{[\Omega]}$, there must exist at least one $d^{[\Omega]}$ that is a sustainer of ``P1 or P2". That is to say, ``P1 or P2" must also be sustainable. So the union of two sustainable conditions is sustainable. This observation leads to the next definition.

\begin{definition}\label{D:schedulability}
{\it The union of all sustainable conditions is the \textbf{schedulability condition}. We call $\bs\psi^{[\Omega]}$ \textbf{schedulable} if and only if it satisfies the schedulability condition, and call $d^{[\Omega]}$ a \textbf{feasible schedule} for $\bs\psi^{[\Omega]}$ if and only if it is a sustainer of the schedulability condition.}
\end{definition}

\begin{figure}[t]
\centering \scalebox{0.875}{\includegraphics{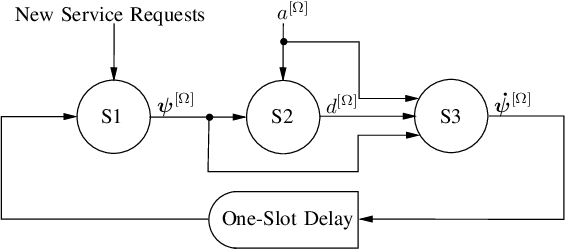}} \caption{The basic operating cycle of state-based scheduling.} \label{F:Diagram}
\end{figure}

The key to state-based scheduling is finding the schedulability condition. Once it is found, state-based scheduling works iteratively. Fig.~\ref{F:Diagram} illustrates the basic operating cycle:
\begin{itemize}
\item[S1] existing services, together with any newly admitted services, form a schedulable $\bs\psi^{[\Omega]}$;
\item[S2] given $\bs\psi^{[\Omega]}$ and $a^{[\Omega]}$, a feasible schedule, $d^{[\Omega]}$, is selected; and finally,
\item[S3] given $a^{[\Omega]}$ and $d^{[\Omega]}$, $\bs\psi^{[\Omega]}$ is updated to a schedulable $\bs{\dot\psi}^{\raisebox{-0.9mm}{\scriptsize$[\Omega]$}}$ which, after a one-slot delay, initializes the next cycle.
\end{itemize}
Here the role of the schedulability condition is two-fold. On the one hand, it is used to admit or deny new service requests to ensure that $\bs\psi^{[\Omega]}$, the current state, is schedulable. On the other hand, $d^{[\Omega]}$ can only be selected among feasible schedules to ensure that $\bs{\dot\psi}^{\raisebox{-0.9mm}{\scriptsize$[\Omega]$}}$, the next state, remains schedulable.

\begin{figure}[t]
\centering \scalebox{0.875}{\includegraphics{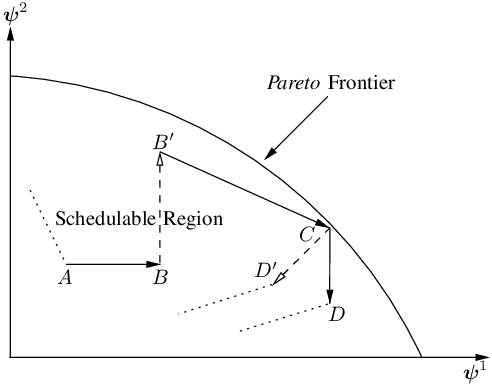}} \caption{A state-space path for a two-flow system.}
\label{F:StatePath}
\end{figure}

It is illuminating to trace a state-based scheduler's state evolution in its state space. In Fig.~\ref{F:StatePath}, a path from state $A$ to $D$ is illustrated for a two-flow system. Notice that the states visited never leave the schedulable region bounded by the {\it Pareto} frontier, on which the service guaranteed to one flow must be reduced to improve that guaranteed to the other. From this perspective, a state-based scheduler operates much like the state-based regulators used to regulate uncertain systems in control theory. In particular, $a^{[\Omega]}$ corresponds to the uncertain noise that perturbs the system; $d^{[\Omega]}$, the control signal applied to combat the noise and regulate the state; and the schedulability condition, the constraint on the state that the regulator aims to enforce.

One clear advantage of state-based scheduling is that it is fully dynamic. Subject to the schedulability condition, $\bs\psi^{[\Omega]}$ can be configured and reconfigured during each slot, that is, worst-case services can be negotiated and renegotiated on the fly. For instance, in Fig.~\ref{F:StatePath}, the jump from state $B$ to $B'$ reflects the admission of a new service request from flow $2$. Another advantage is that, as it systematically identifies all feasible schedules in each slot, state-based scheduling allows us to identify and pursue, not just one {\it ad hoc} policy, but all scheduling policies that can simultaneously guarantee all flows their respective worst-case services. Notice that although we do have the flexibility to select any feasible schedule during each slot, different selections will induce different future states. For instance, in Fig.~\ref{F:StatePath}, had a different feasible schedule been selected at $C$, the path would have diverged to $D'$ and thus have been entirely different from that point on.

\subsection{The Schedulability Condition}\label{SS:schedulability}

To fulfill the promise of state-based scheduling, we need to find the schedulability condition. Unfortunately, the condition's definition is not constructive. So, an intelligent guess must be made. This is where the concept of spectrum comes in. Recall, from Section~\ref{SS:signatures}, that the least capacity that must be reserved during interval $[t+i, t+j)$ to guarantee worst-case service $\bs\psi^\omega$ to flow $\omega$ is given by spectral value $\lambda_{ij}(\bs\psi^\omega)$. Intuitively, if all flows are to be simultaneously guaranteed their respective worst-case services, during every interval $[t+i, t+j)$, the total capacity that must be reserved, $\sum_{\omega\in\Omega} \lambda_{ij}(\bs\psi^\omega)$, cannot exceed the server's available capacity, $(j-i)^+c$. This reserve capacity restriction turns out to be the schedulability condition that we seek.

\begin{theorem}\label{T:schedulability}
{\it A worst-case system, $\bs\psi^{[\Omega]}$, is \textbf{schedulable} if and only if its spectrum system satisfies the following condition:
\begin{equation}\label{E:schedulability}
\lambda_{ij}^{\l\Omega\r} = \sum_{\omega\in\Omega} \lambda_{ij}(\bs\psi^\omega) \leq c_{ij} := (j-i)^+c \blank\blank\forall\, i,j\in\mb{N}.
\end{equation}
According to Definitions~\ref{D:proper} and \ref{D:schedulability}, this implies that:
\begin{itemize}
\item[I1] if $\bs\psi^{[\Omega]}$ is schedulable, given any $a^{[\Omega]}$, at least one feasible $d^{[\Omega]}$ exists; and
\item[I2] a valid $d^{[\Omega]}$ is \textbf{feasible} if and only if the spectrum system that it induces in the next slot satisfies (\ref{E:schedulability}), that is,
    \begin{equation}\label{E:schedulability+}
    \dot\lambda_{ij}^{\l\Omega\r} \leq \dot{c}_{ij} = c_{i+1, j+1} \blank\blank\forall\, i,j\in\mb{N},
    \end{equation}
    where, according to (\ref{E:SigUpdate}),
    \begin{equation}\label{E:dlSum}
    \!
    \dot\lambda_{ij}^{\l\Omega\r}(d^{[\Omega]}) = \left\{
    \begin{IEEEeqnarraybox}[][c]{ll}
    \textstyle \sum_{\omega\in\Omega} (\hat\lambda_{0, j+1}^\omega-d^\omega)^+ & \blank\blank\textup{if } i=0\\
    \hat\lambda_{i+1, j+1}^{\l\Omega\r}                                        & \blank\blank\textup{if } i>0
    \end{IEEEeqnarraybox}\right. ~.
    \end{equation}
\end{itemize}
}
\end{theorem}

As satisfying (\ref{E:schedulability}) is clearly necessary for $\bs\psi^{[\Omega]}$ to be schedulable, to prove this theorem, we need only show (\ref{E:schedulability}) to be sustainable because, according to Definition~\ref{D:schedulability}, the schedulability condition is the weakest among all sustainable conditions, and a necessary sustainable condition must be the weakest. Hence, proving Theorem~\ref{T:schedulability} reduces to establishing I1, which we leave for the next section.

Comparing I2's requirements for a schedule to be feasible to Definition~\ref{D:proper}'s requirements for a schedule to be a sustainer, it is immediate that the $d^{[\Omega]} \geq p^{[\Omega]}$ requirement is dropped in I2. This is because $d^{[\Omega]} \geq p^{[\Omega]}$ is implied by (\ref{E:schedulability+}). According to (\ref{E:schedulability+}), $\dot\lambda_{00}^{\l\Omega\r} \leq \dot{c}_{00} = 0$. So, according to (\ref{E:dlSum}) and (\ref{E:p-lambda}),
\[
\dot\lambda_{00}^{\l\Omega\r} = \sum_{\omega\in\Omega} (\hat\lambda_{01}^\omega-d^\omega)^+ = \sum_{\omega\in\Omega} (p^\omega-d^\omega)^+ \leq 0,
\]
implying that $d^{[\Omega]} \geq p^{[\Omega]}$.

With the $d^{[\Omega]} \geq p^{[\Omega]}$ requirement dropped, it may seem that, to check $d^{[\Omega]}$'s feasibility, we may need to compute $\dot\lambda_{ij}^{\l\Omega\r}$ when $d^{[\Omega]} \ngeq p^{[\Omega]}$, although, in this case, we cannot relate $\dot\lambda_{ij}^{\l\Omega\r}$ to $\bs{\dot\psi}^{\raisebox{-0.9mm}{\scriptsize$[\Omega]$}}$ because, according to Theorem~\ref{T:WCSUpdate}, it is required that $d^{[\Omega]} \geq p^{[\Omega]}$ to update $\bs\psi^{[\Omega]}$ to $\bs{\dot\psi}^{\raisebox{-0.9mm}{\scriptsize$[\Omega]$}}$. In fact, as $d^{[\Omega]} \ngeq p^{[\Omega]}$ implies that $\dot\lambda_{00}^{\l\Omega\r} > 0$, by Theorem~\ref{T:schedulability}, $d^{[\Omega]} \ngeq p^{[\Omega]}$ is never feasible, so, in all cases of interest, $\dot\lambda_{ij}^{\l\Omega\r}$'s link to $\bs{\dot\psi}^{\raisebox{-0.9mm}{\scriptsize$[\Omega]$}}$ holds.

\subsubsection*{\textup{[}Multiplexing Gain\textup{]}} Independent of its importance to state-based scheduling, the schedulability condition can be used as a normative formula to guide the design of multiplexing systems much as various information theoretic capacity formulas have been used to guide the design of communication systems. In particular, it distinguishes what can be guaranteed from what cannot. To see this, observe that (\ref{E:schedulability}) can be rewritten as
\begin{equation}\label{E:rhoSchedulability}
c \geq \rho(\Omega) := \max_{i<j} \frac{\lambda_{ij}^{\l\Omega\r}}{j-i}~.
\end{equation}
If $c < \rho(\Omega)$, no matter how smart the scheduler is, it is impossible to simultaneously guarantee all flows their respective worst-case services. Using (\ref{E:rhoSchedulability}), for instance, it is easy to see that uniform-backlog and uniform-delay services cannot be guaranteed when $c$ is finite because, according to (\ref{E:spectrumCBS}) and (\ref{E:spectrumCDS}),  in both cases $\lambda_{ij}$ can become infinite when $i$ and $j$ are finite.

Extending the definition of $\rho(\Omega)$ in (\ref{E:rhoSchedulability}) to
\begin{equation}\label{E:rho}
\rho(\Gamma):= \max_{i<j} \frac{\lambda_{ij}^{\l\Gamma\r}}{j-i} \blank\blank\forall\, \Gamma\subseteq\Omega,
\end{equation}
where $\lambda_{ij}^{\l\Gamma\r}$ denotes $\sum_{\omega\in\Gamma} \lambda_{ij}^\omega$, with $\lambda_{ij}^{\l\phi\r}:=0$, enables finer distinctions to be made. If all flows in $\Gamma$ are served by a single server, $\rho(\Gamma)$ is the least capacity required by this server to guarantee $\bs\psi^\omega$ for all $\omega\in\Gamma$.\footnote{As our traffic model is discrete, $\rho(\Gamma)$ should be rounded up to $\lceil\rho(\Gamma)\rceil$. But we will not do so here. Besides being a good approximation of $\lceil\rho(\Gamma)\rceil$, $\rho(\Gamma)$ is also more tractable analytically. Moreover, it can be shown that $\rho(\Gamma)$ is exactly the least capacity required if we slightly tweak our model such that a task can be fractionally served, although not counted as served until it has been wholly served.} If each flow is served by a separate server, then, $\sum_{\omega\in\Omega} \rho(\{\omega\})$ is the least total capacity required by these servers. Let
\begin{equation}\label{E:eta}
\eta:= \frac{\sum_{\omega\in\Omega} \rho(\{\omega\})}{\rho(\Omega)}~.
\end{equation}
Clearly, the larger the $\eta$, the greater capacity utilization that can be achieved through multiplexing. For this reason, we call $\eta$ the {\it \textbf{multiplexing gain}}.

Using $\Gamma+\Gamma'$ to denote $\Gamma\cup\Gamma'$, we have
\begin{equation}\label{E:rhoProp}
\rho(\Gamma+\Gamma') \leq \rho(\Gamma)+\rho(\Gamma') \blank\blank\forall\, \Gamma,\Gamma'\subseteq\Omega,
\end{equation}
because
\[
\begin{IEEEeqnarraybox}[][c]{rCl}
\max_{i<j} \frac{\lambda_{ij}^{\l\Gamma+\Gamma'\r}}{j-i} = \frac{\lambda_{i_*j_*}^{\l\Gamma+\Gamma'\r}}{j_*-i_*} & \leq & \frac{\lambda_{i_*j_*}^{\l\Gamma\r}}{j_*-i_*} + \frac{\lambda_{i_*j_*}^{\l\Gamma'\r}}{j_*-i_*}\\
                                                                                                                 & \leq & \max_{i<j} \frac{\lambda_{ij}^{\l\Gamma\r}}{j-i}+\max_{i<j} \frac{\lambda_{ij}^{\l\Gamma'\r}}{j-i}~,
\end{IEEEeqnarraybox}
\]
where $i_*$ and $j_*$ maximize $\frac{\lambda_{ij}^{\l\Gamma+\Gamma'\r}}{j-i}$. It follows from (\ref{E:eta}) and (\ref{E:rhoProp}) that $\eta \geq 1$. In particular, it is easy to see that $\eta=1$ if for all $\omega,\omega'\in\Omega$, $\lambda_{ij}^\omega \propto \lambda_{ij}^{\omega'}$, while $\eta$ increases as the correlation among the $\lambda_{ij}^\omega$'s decreases. Roughly speaking, the more diverse the guarantees in $\bs\psi^{[\Omega]}$, the larger the $\eta$.

\begin{example}\label{Ex:eta}
{\it A {\it \textbf{single-task system}} is a worst-case system in which every worst-case service is a single-task service, as defined in Example~\ref{Ex:STS}.\footnote{Unlike other specialized worst-case systems to be introduced later, a single-task system is not update invariant because its constituent single-task services are not update invariant. That is to say, a single-task system in slot~$t$ need not remain a single-task system in $t+1$.} So, if $\bs\psi^{[\Omega]}$ is a single-task system, for all $\omega\in\Omega$, $b^\omega = 0$; for all $\bs{q}^\omega \in \mb{U}$, according to (\ref{E:STSDef}),
\begin{equation}\label{E:STS}
\bs\psi^\omega(\bs{q}^\omega)=\left\{
\begin{IEEEeqnarraybox}[][c]{ll}
\mc{R}^{i+\bar\theta^\omega}\bs\delta & \blank\blank\textup{if } \bs{q}^\omega = \mc{R}^i\bs\delta \textup{ for some } i\in\mb{N}\\
\bs{0}                                & \blank\blank\textup{otherwise}
\end{IEEEeqnarraybox}\right. ~;
\end{equation}
and, for all $i,j\in\mb{N}$, according to (\ref{E:spectrumSTS}),
\begin{equation}\label{E:specSTS}
\lambda_{ij}^\omega=\left\{
\begin{IEEEeqnarraybox}[][c]{ll}
1 & \blank\blank\textup{if } j-i > \bar\theta^\omega\\
0 & \blank\blank\textup{if } j-i \leq \bar\theta^\omega
\end{IEEEeqnarraybox}\right. ~.
\end{equation}
Without loss of generality, assume that $\bar\theta^1 \leq \bar\theta^2 \leq \cdots \leq \bar\theta^n$. Then, using (\ref{E:rho}), it can be verified that
\begin{equation}\label{E:rhoSTS}
\rho(\{\omega\}) = \frac{1}{\bar\theta^\omega+1} \blank\textup{and}\blank \rho(\Omega) = \max_{\omega\in\Omega} \frac{\omega}{\bar\theta^\omega+1}~.
\end{equation}
According to (\ref{E:eta}), $\eta$ is now a function of the $\bar\theta^\omega$'s. The more diverse the $\bar\theta^\omega$'s become, the larger the $\eta$. For instance, if $\bar\theta^1 = \bar\theta^2 = \cdots = \bar\theta^n$, $\eta=1$, while if $\frac{1}{\bar\theta^1+1} = \frac{2}{\bar\theta^2+1} = \cdots = \frac{n}{\bar\theta^n+1}$, $\eta = 1+\frac{1}{2}+\cdots+\frac{1}{n}$.
}
\end{example}

As the number of flows grows, the complexity of many scheduling policies grows so fast that, even when $\eta$ is large enough to justify multiplexing, divide-and-conquer scheduling strategies remain attractive. To illustrate, let $\mc{P} \subseteq 2^\Omega$ be a {\it \textbf{partition}} of $\Omega$, that is, $\bigcup_{\Gamma\in\mc{P}} \Gamma = \Omega$ and $\Gamma\Gamma'=\phi$ for all distinct $\Gamma,\Gamma'\in\mc{P}$, where we use $\Gamma\Gamma'$ to denote $\Gamma\cap\Gamma'$. If, for each $\Gamma \in \mc{P}$, all flows in $\Gamma$ are served by a separate server, then, $\sum_{\Gamma\in\mc{P}} \rho(\Gamma)$ is the least total capacity required by these servers. Similar to $\eta$, let
\begin{equation}\label{E:etaP}
\eta^\mc{P}:= \frac{\sum_{\omega\in\Omega} \rho(\{\omega\})}{\sum_{\Gamma\in\mc{P}} \rho(\Gamma)}~.
\end{equation}
Using (\ref{E:rhoProp}) and (\ref{E:eta}), it is easy to verify that $1 \leq \eta^\mc{P} \leq \eta$. Now, if we can identify some $\mc{P}$ such that $\max_{\Gamma \in \mc{P}} |\Gamma|$ is sufficiently small and $\eta^\mc{P}$ is sufficiently close to $\eta$, we can reduce the scheduling complexity without sacrificing much of the multiplexing gain.

\begin{example}\label{Ex:eta+}
{\it Let $\bs\psi^{[\Omega]}$ be a single-task system. Consider the case that $n=3$ and $\frac{1}{\bar\theta^1+1} = \frac{2}{\bar\theta^2+1} = \frac{2}{\bar\theta^3+1}$. Using (\ref{E:rhoSTS}), we find that
\[
\rho(\{1\}) = 2\rho(\{2\}) = 2\rho(\{3\}) = \frac{2}{3} \rho(\{1, 2, 3\}) = \frac{1}{\bar\theta^1+1}~.
\]
Hence $\eta = \frac{4}{3}$. Using (\ref{E:rho}), we also find that
\[
\rho(\{1, 2\}) = \rho(\{1, 3\}) = \frac{1}{\bar\theta^1+1}~.
\]
So, if $\mc{P} = \{\{1, 2\}, \{3\}\}$ or $\{\{1, 3\}, \{2\}\}$, $\eta^\mc{P} = \frac{4}{3}$, and neither partition results in any loss of multiplexing gain.}
\end{example}

\section{Feasible Schedules}\label{S:feasible}

In this section, we show that, if $\bs\psi^{[\Omega]}$ is schedulable, feasible schedules always exist, and we identify all of them. To get a sense of what lies ahead, consider first the single-flow case. If $\bs\psi$ is schedulable, (\ref{E:schedulability}) requires that
\begin{equation}\label{E:schedulabilitySingle}
\lambda_{ij} \leq c_{ij} = (j-i)^+c \blank\blank\forall\, i,j\in\mb{N}.
\end{equation}
According to I2 of Theorem~\ref{T:schedulability}, we need to identify a $d$ that ensures that $\dot\lambda_{ij} \leq c_{i+1, j+1}$ for all $i,j\in\mb{N}$. If $\bs\psi$ is schedulable, this requirement is satisfied by default in the case that $i > 0$, because using (\ref{E:SigUpdate}), (\ref{E:CondSigPropF}), and (\ref{E:schedulabilitySingle}), we have
\[
\dot\lambda_{ij} = \hat\lambda_{i+1, j+1} \leq \lambda_{i+1, j+1} \leq c_{i+1, j+1}.
\]
So we need only focus on the case that $i = 0$. In this case, according to (\ref{E:SigUpdate}), $\dot\lambda_{0j} \leq c_{1, j+1}$ implies that
\[
c_{1, j+1} \geq \dot\lambda_{0j} = (\hat\lambda_{0, j+1}-d)^+  \geq \hat\lambda_{0, j+1}-d,
\]
which in turn implies that
\begin{equation}\label{E:alpha}
d \geq \alpha := \max_{j\in\mb{N}} (\hat\lambda_{0, j+1}-c_{1, j+1}).
\end{equation}
If $d \geq \alpha$, by reversing this reasoning, it is easy to verify that $\dot\lambda_{ij} \leq c_{i+1, j+1}$ for all $i,j\in\mb{N}$. Then, according to I2 of Theorem~\ref{T:schedulability}, $d \geq \alpha$ need only be valid to be feasible, so it is feasible if $d \leq \min \{q, c\}$. This is impossible unless $\alpha \leq q$ and $\alpha \leq c$. Fortunately, both inequalities hold when $\bs\psi$ is schedulable because, using (\ref{E:alpha}), (\ref{E:CondSigPropE}), (\ref{E:CondSigPropF}), and (\ref{E:schedulabilitySingle}), we have
\[
\begin{IEEEeqnarraybox}[][c]{rCl}
\alpha & =    & q + \max_{j\in\mb{N}} (\hat\lambda_{0, j+1}-q-c_{1, j+1}) \\
       & \leq & q + \max_{j\in\mb{N}} (\hat\lambda_{1, j+1}-c_{1, j+1})   \\
       & \leq & q + \max_{j\in\mb{N}} (\lambda_{1, j+1}-c_{1, j+1}) \leq q,
\end{IEEEeqnarraybox}
\]
and using (\ref{E:alpha}), (\ref{E:CondSigPropF}), and (\ref{E:schedulabilitySingle}), we also have
\[
\alpha \leq \max_{j\in\mb{N}} (\lambda_{0, j+1}-c_{1, j+1}) \leq \max_{j\in\mb{N}} (c_{0, j+1}-c_{1, j+1}) \leq c.
\]

Although the single-flow case is highly simplified, it nonetheless can serve as a skeleton for the multi-flow case. In a certain sense, what we describe next simply puts flesh on this skeleton. Of course, the multi-flow case does pose new technical challenges. In particular, $\alpha$, as defined in (\ref{E:alpha}), explodes from a scalar to a full-blown set function over $\Omega$. It is in dealing with this function that polymatroid theory comes to our attention.

In the remainder of this section, we first give a primer on permutohedra, a class of polytopes from polymatroid theory. We then introduce the baseline function and use it to show that the set of feasible schedules is a polytope that can be constructed from a series of permutohedral slices. We also explain how feasible schedules can be selected to enforce priority and fairness criteria.

\subsection{Supermodular Functions and Permutohedra}\label{SS:polymatroid}

Polymatroid theory was first developed in \cite{Edmonds:1970}. An extensive survey can be found in \cite{Schrijver:2003} (ch. 44-49). The standard introduction usually begins with submodular functions and polymatroids, but, tailored for our application, we instead begin with supermodular functions and permutohedra.\footnote{\label{F:duality}In Definition~\ref{D:supermodular}, if we replace $\leq$ by $\geq$ in (\ref{E:supermodular}), we obtain the definition for a {\it \textbf{submodular function}}. If $\chi$ is submodular, we call
\[
\mb{M}(\chi) := \{d^{[\Omega]} | d^{\l\Gamma\r} \leq \chi(\Gamma) ~\forall\, \Gamma\subseteq\Omega\}
\]
the {\it \textbf{polymatroid}} generated by $\chi$.}

\begin{definition}\label{D:supermodular}
{\it $\chi:2^\Omega \rightarrow \mb{N}$ is a \textbf{supermodular function} over $\Omega$ if $\chi(\phi) = 0$ and
\begin{equation}\label{E:supermodular}
\chi(\Gamma)+\chi(\Gamma') \leq \chi(\Gamma+\Gamma')+\chi(\Gamma\Gamma') \blank\blank\forall\, \Gamma,\Gamma'\subseteq\Omega.\footnote{\textup{Our definition of supermodularity is more restrictive than is standard. But no generality is lost. We restrict the range of $\chi$ to $\mb{N}$, instead of $\mb{R}$, because in our model traffic is discrete and non-negative. We require that $\chi(\phi) = 0$ because $\mb{P}(\chi)$ is empty when $\chi(\phi) > 0$.}}
\end{equation}
If $\chi$ is supermodular, we call
\begin{equation}\label{E:permutohedron}
\mb{P}(\chi) := \{d^{[\Omega]} | d^{\l\Omega\r} =  \chi(\Omega) \textup{ and } d^{\l\Gamma\r} \geq \chi(\Gamma) ~\forall\, \Gamma\subseteq\Omega\}\!
\end{equation}
the \textbf{permutohedron} generated by $\chi$.}
\end{definition}

By definition, $\mb{P}(\chi)$ is potentially an $(n-1)$-polytope. To investigate its faces, for all $\{\phi, \Omega\} \subseteq \mc{S} \subseteq 2^\Omega$, let
\begin{equation}\label{E:P-S}
\mb{P}_\mc{S}(\chi) := \{d^{[\Omega]} \in \mb{P}(\chi )| d^{\l\Gamma\r} =  \chi(\Gamma) ~\forall\, \Gamma \in \mc{S}\}.\footnote{\textup{If $\chi$ is supermodular, by definition, $d^{\l\phi\r}=0=\chi(\phi)$. If, additionally, $d^{[\Omega]}\in\mb{P}(\chi)$, according to (\ref{E:permutohedron}), $d^{\l\Omega\r} =  \chi(\Omega)$. Therefore, requiring $\{\phi, \Omega\} \subseteq \mc{S}$ is not restrictive. We do so here only for completeness.}}
\end{equation}
Each $\mc{S}$ identifies a potential face of $\mb{P}(\chi)$. There are $2^{2^n-2}$ such $\mc{S}$'s, but for most, $\mb{P}_\mc{S}(\chi)$ is actually empty. So when is it non-empty? To answer this question, we make use of the concept of chains. Given $\{\phi, \Omega\} \subseteq \mc{C} \subseteq 2^\Omega$, we call $\mc{C}$ a {\it \textbf{chain}} if for all distinct $\Gamma,\Gamma'\in\mc{C}$, $\Gamma \subset \Gamma'$ or $\Gamma' \subset \Gamma$. The next lemma, the proof of which can be found in the appendix, makes clear the usefulness of this concept.

\begin{lemma}\label{L:non-empty}
{\it For all $\{\phi, \Omega\} \subseteq \mc{S} \subseteq 2^\Omega$, if $\mb{P}_\mc{S}(\chi)$ is non-empty, there must exist a chain, $\mc{C}$, such that $\mb{P}_\mc{S}(\chi) = \mb{P}_\mc{C}(\chi)$.}
\end{lemma}

According to this lemma, each non-empty face of $\mb{P}(\chi)$ can be identified by a chain. To investigate these faces, we start with the simplest, $\mb{P}(\chi)$'s vertices.

\subsubsection*{\textup{[}Vertices\textup{]}} Given a chain, $\mc{C}$, if $|\mc{C}|=m+1$, $\mb{P}_\mc{C}(\chi)$ is a potential $(n-m)$-face of $\mb{P}(\chi)$. We call $\mc{C}$ {\it \textbf{complete}} if $m=n$. So each complete chain identifies a potential $0$-face, which is of particular interest because a $0$-face is a vertex. Complete chains, and thus potential vertices, can in turn be identified by permutations.

A {\it \textbf{permutation}} over $\Omega$ is a bijective map, $\pi: \Omega \rightarrow \{1,2,\ldots,n\} \in \Pi^\Omega$, where $\Pi^\Omega$ denotes the set of all such permutations. For all $0 \leq i \leq n$, let
\begin{equation}\label{E:GammapiA}
\Gamma_\pi^i := \{\omega\in\Omega | \pi(\omega) \leq i\}.
\end{equation}
Denoting $\pi^{-1}(i)$ by $\omega_\pi^i$, it is immediate that
\begin{equation}\label{E:GammapiB}
\Gamma_\pi^i = \left\{
\begin{IEEEeqnarraybox}[][c]{ll}
\phi                                                 & \blank\blank\text{if } i=0 \\
\{\omega_\pi^1, \omega_\pi^2, \ldots, \omega_\pi^i\} & \blank\blank\text{if } 1 \leq i \leq n
\end{IEEEeqnarraybox}\right. ~.
\end{equation}
On the one hand, this naturally leads to a complete chain,
\begin{equation}\label{E:Cpi}
\mc{C}_\pi:=\{\Gamma_\pi^i\}_{i=0}^n,
\end{equation}
because, by definition, $\phi = \Gamma_\pi^0 \subset \Gamma_\pi^1 \subset \cdots \subset \Gamma_\pi^n = \Omega$. On the other hand, let $v_\pi^{[\Omega]}(\chi)$ be the unique solution to the system of linear equations defined by
\begin{equation}\label{E:vertexSum}
v_\pi^{\l\Gamma_\pi^i\r}(\chi) = \chi(\Gamma_\pi^i) \blank\blank\forall\, 0 \leq i \leq n,
\end{equation}
so that
\begin{equation}\label{E:vertex}
v_\pi^{\omega_\pi^i}(\chi) = \chi(\Gamma_\pi^i)-\chi(\Gamma_\pi^{i-1}) \blank\blank\forall\, 1 \leq i \leq n.
\end{equation}
Then, according to (\ref{E:P-S}), (\ref{E:Cpi}), and (\ref{E:vertexSum}), $\mb{P}_{\mc{C}_\pi}(\chi) = \phi$ or $\{v_\pi^{[\Omega]}(\chi)\}$. It turns out that $\mb{P}_{\mc{C}_\pi}(\chi) = \{v_\pi^{[\Omega]}(\chi)\}$ due to the next lemma, the proof of which can also be found in the appendix.

\begin{lemma}\label{L:vertex}
{\it For all $\Gamma\subseteq\Omega$, $v_\pi^{\l\Gamma\r}(\chi) \geq \chi(\Gamma)$.}
\end{lemma}

This lemma guarantees that $v_\pi^{[\Omega]}(\chi)$ is indeed a vertex of $\mb{P}(\chi)$. So $\mb{P}(\chi)$ is completely determined by the $n!$ vertices indexed by $\Pi^\Omega$, which explains why such polytopes are called permutohedra.

\subsubsection*{\textup{[}Faces\textup{]}} Recall that each non-empty face of $\mb{P}(\chi)$ can be identified by a chain. Given an arbitrary chain, $\mc{C}$, let
\begin{equation}\label{E:PiC}
\Pi_\mc{C}^\Omega:= \{\pi\in\Pi^\Omega | \mc{C} \subseteq \mc{C}_\pi\}.
\end{equation}
Then, for all $\pi\in\Pi_\mc{C}^\Omega$, according to (\ref{E:P-S}), $\mb{P}_\mc{C_\pi}(\chi) \subseteq \mb{P}_\mc{C}(\chi)$. It follows that $v_\pi^{[\Omega]}(\chi)$ is a vertex of $\mb{P}_\mc{C}(\chi)$, so $\mb{P}_\mc{C}(\chi)$ is completely determined by the $|\Pi_\mc{C}^\Omega|$ vertices indexed by $\Pi_\mc{C}^\Omega$. Moreover, as a corollary, each chain does identify a non-empty face of $\mb{P}(\chi)$.

\begin{figure}[t]
\centering \scalebox{0.6}{\includegraphics{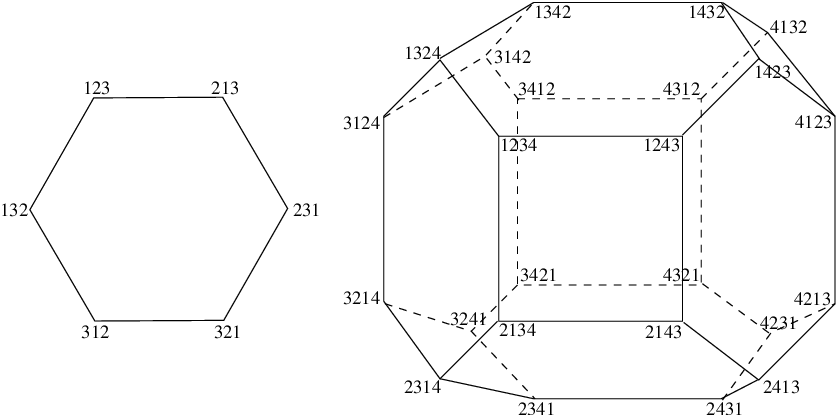}} \caption{Permutohedra of orders $n=3$ and $n=4$.} \label{F:Polymatroids}
\end{figure}

Two low-dimension permutohedra are illustrated in Fig.~\ref{F:Polymatroids}. In the case that $n=3$, the permutohedron is a hexagon with $6$ vertices and $6$ edges. In the case that $n=4$, it is a truncated octahedron with $24$ vertices, $36$ edges and $14$ facets, among which $6$ are rectangles and $8$ are hexagons. Notice how each vertex is indexed by a permutation and how these vertices are organized into faces according to their indexing permutations.

To explore further the topology of $\mb{P}(\chi)$'s face system, consider two arbitrary chains, $\mc{C}$ and $\mc{C}'$. On the one hand, although $\mc{C}+\mc{C}'$ may not be a chain, when it is, $\mb{P}_\mc{C}(\chi)$ and $\mb{P}_{\mc{C}'}(\chi)$ intersect at $\mb{P}_{\mc{C}+\mc{C}'}(\chi)$. On the other hand, $\mc{C}\mc{C}'$ is still a chain and $\mb{P}_{\mc{C}\mc{C}'}(\chi)$ is the minimum face to which both $\mb{P}_\mc{C}(\chi)$ and $\mb{P}_{\mc{C}'}(\chi)$ belong. As an application of this principle, consider the neighborhood of a vertex. Two vertices are neighbors if they belong to the same edge. An edge is a $1$-face, so if $v_\pi^{[\Omega]}(\chi)$ and $v_{\pi'}^{[\Omega]}(\chi)$ are neighbors, $|\mc{C}_\pi\mc{C}_{\pi'}|=n$. It is then easy to verify that there must exist $1 \leq i_* < n$ such that
\[
\forall\, 1 \leq i \leq n, \blank\blank
\pi'(\omega_\pi^i)=\left\{
\begin{IEEEeqnarraybox}[][c]{ll}
i_*+1 & \blank\blank\text{if } i=i_*\\
i_*   & \blank\blank\text{if } i=i_*+1\\
i     & \blank\blank\text{if } i \neq i_*, i_*+1
\end{IEEEeqnarraybox}\right. ~.
\]
That is to say, as illustrated in Fig.~\ref{F:Polymatroids}, the difference between neighboring indices, $\pi$ and $\pi'$, is no more than a simple transposition.

\subsection{The Feasible Polytope}\label{SS:polytope}

To guarantee a schedulable $\bs\psi^{[\Omega]}$ given any $a^{[\Omega]}$, we need to identify at least one feasible $d^{[\Omega]}$. According to I2 of Theorem~\ref{T:schedulability}, if $d^{[\Omega]}$ is feasible, it must ensure that $\dot\lambda_{ij}^{\l\Omega\r} \leq c_{i+1, j+1}$ for all $i,j \in \mb{N}$. In the case that $i=0$, this implies, according to (\ref{E:SigUpdate}), that for all $\Gamma\subseteq\Omega$,
\[
c_{1, j+1} \geq  \dot\lambda_{0j}^{\l\Omega\r} \geq \dot\lambda_{0j}^{\l\Gamma\r} = \sum_{\omega\in\Gamma} (\hat\lambda_{0, j+1}^\omega-d^\omega)^+ \geq \hat\lambda_{0, j+1}^{\l\Gamma\r}-d^{\l\Gamma\r},
\]
which in turn implies that
\begin{equation}\label{E:alpha+}
d^{\l\Gamma\r} \geq \alpha(\Gamma) := \max_{j \in \mb{N}} (\hat\lambda_{0, j+1}^{\l\Gamma\r}-c_{1, j+1}) \blank\blank\forall\, \Gamma\subseteq\Omega,
\end{equation}
the multi-flow extension of (\ref{E:alpha}).

This bound can be tightened by accounting for the conditional spectra of the flows omitted from $\Gamma$. Denote the set of these omitted flows by $\overline\Gamma := \Omega\setminus\Gamma$. According to (\ref{E:lambdaBound}), $\dot\lambda_{0j}^{\l\overline\Gamma\r} \geq \hat\lambda_{1, j+1}^{\l\overline\Gamma\r}$, the lower bound of which is achieved when $d^\omega = q^\omega$ for all $\omega\in\overline\Gamma$, that is, when the omitted flows' buffers are emptied. So, according to (\ref{E:SigUpdate}), $\dot\lambda_{0j}^{\l\Omega\r} \leq c_{1, j+1}$ implies that
\[
c_{1, j+1} \geq \dot\lambda_{0j}^{\l\Omega\r} = \dot\lambda_{0j}^{\l\Gamma\r}+\dot\lambda_{0j}^{\l\overline\Gamma\r} \geq \hat\lambda_{0, j+1}^{\l\Gamma\r}-d^{\l\Gamma\r} + \hat\lambda_{1, j+1}^{\l\overline\Gamma\r},
\]
which in turn implies that
\begin{equation}\label{E:beta}
d^{\l\Gamma\r} \geq \beta(\Gamma) := \max_{j \in \mb{N}} (\hat\lambda_{0, j+1}^{\l\Gamma\r} + \hat\lambda_{1, j+1}^{\l\overline\Gamma\r}-c_{1, j+1}) \blank\blank\forall\, \Gamma\subseteq\Omega.
\end{equation}

Comparing (\ref{E:beta}) to (\ref{E:alpha+}), it is immediate that $\beta \geq \alpha$, that is, $\beta(\Gamma) \geq \alpha(\Gamma)$ for all $\Gamma\subseteq\Omega$.\footnote{Although we will not do so here, it can be shown that
\[
\beta(\Gamma) = \max_{\Gamma' \subseteq \overline\Gamma} (\alpha(\Gamma+\Gamma')-q^{\l\Gamma'\r}) \blank\blank\forall\, \Gamma\subseteq\Omega,
\]
which directly relates $\alpha$ and $\beta$.} As (\ref{E:beta}) specifies the least number of tasks that must be served from any subset of flows to guarantee $\bs\psi^{[\Omega]}$, we call it the {\it \textbf{baseline constraint}} and $\beta$ the {\it \textbf{baseline function}}. Notice that $\beta$ implicitly depends on $a^{[\Omega]}$ because, according to (\ref{E:CondSig}), $\hat\lambda_{ij}$ depends on $q$, which in turn depends on $a$. The importance of $\beta$ is highlighted in the next theorem.

\begin{theorem}\label{T:polytope}
{\it If $\bs\psi^{[\Omega]}$ is schedulable, a valid schedule, $d^{[\Omega]}$, is feasible for $\bs\psi^{[\Omega]}$ if and only if it satisfies the baseline constraint, (\ref{E:beta}).}
\end{theorem}

\begin{IEEEproof}
The necessity of this condition follows directly from (\ref{E:beta})'s derivation. To establish its sufficiency, according to I2 of Theorem~\ref{T:schedulability}, we need only show that $d^{[\Omega]}$ ensures that $\dot\lambda_{ij}^{\l\Omega\r} \leq c_{i+1, j+1}$ for all $i,j \in \mb{N}$.

In the case that $i>0$, this is true by default because, if $\bs\psi^{[\Omega]}$ is schedulable, according to (\ref{E:dlSum}), (\ref{E:CondSigPropF}), and (\ref{E:schedulability}),
\[
\dot\lambda_{ij}^{\l\Omega\r} = \hat\lambda_{i+1, j+1}^{\l\Omega\r} \leq  \lambda_{i+1, j+1}^{\l\Omega\r} \leq c_{i+1, j+1}.
\]
In the case that $i=0$, on the one hand, using (\ref{E:dlSum}), and repeatedly applying the fact that
\[
\max\{{x, y\}}+z = \max\{x+z, y+z\},
\]
we have
\[
\dot\lambda_{0j}^{\l\Omega\r} = \sum_{\omega\in\Omega} \max\{\hat\lambda_{0, j+1}^\omega-d^\omega, 0\} = \max_{\Gamma\subseteq\Omega} (\hat\lambda_{0, j+1}^{\l\Gamma\r}-d^{\l\Gamma\r}).
\]
On the other hand, if (\ref{E:beta}) holds, for all $\Gamma\subseteq\Omega$, using (\ref{E:CondSig}), we also have
\[
d^{\l\Gamma\r} \geq \beta(\Gamma) \geq \hat\lambda_{0, j+1}^{\l\Gamma\r} + \hat\lambda_{1, j+1}^{\l\overline\Gamma\r}-c_{1, j+1} \geq \hat\lambda_{0, j+1}^{\l\Gamma\r}-c_{1, j+1}.
\]
It follows that
\[
\dot\lambda_{0j}^{\l\Omega\r} = \max_{\Gamma\subseteq\Omega} (\hat\lambda_{0, j+1}^{\l\Gamma\r}-d^{\l\Gamma\r}) \leq c_{1, j+1}.
\]
So $\dot\lambda_{ij}^{\l\Omega\r} \leq c_{i+1, j+1}$ for all $i,j \in \mb{N}$.
\end{IEEEproof}

\begin{figure}[t]
\centering \scalebox{0.875}{\includegraphics{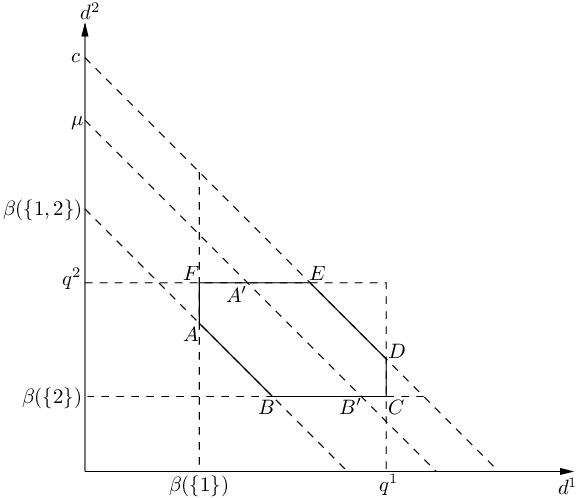}} \caption{The feasible polytope, a feasible permutohedron, and the baseline permutohedron in a two-flow case.}\label{F:FeasibleRegion}
\end{figure}

Let $\mb{F}$ denote the set of feasible schedules. If $\bs\psi^{[\Omega]}$ is schedulable, according to Theorem~\ref{T:polytope}, $\mb{F}$ is completely determined by three linear constraints, (\ref{E:causalityB}), (\ref{E:capacity}), and (\ref{E:beta}). So $\mb{F}$ is an $n$-polytope, which we call the {\it \textbf{feasible polytope}}. In the two-flow case illustrated in Fig.~\ref{F:FeasibleRegion}, $\mb{F}$ is the hexagon, $ABCDEF$, enclosed from above by the causality and capacity constraints, and from below, by the baseline constraints. In three-flow cases, $\mb{F}$ resembles a diamond. More generally, what is $\mb{F}$'s~structure? To answer this question, it is helpful to investigate $\beta$'s properties.

\subsection{The Properties of $\beta$}\label{SS:beta}

Alternative formulations of $\beta$ will prove useful in establishing $\beta$'s properties. To derive them, let $\bs{p} = [p_j]_{j \in \mb{N}} \in \mb{U}$, with
\begin{IEEEeqnarray}{rRl}
p_j &:=& \hat\lambda_{0j} - \hat\lambda_{1j} \label{E:p}\\
    & =& \hat\lambda_{0j} - \max\{\hat\lambda_{0j}-q, 0\} \IEEEnonumber\\
    & =& \min \{\hat\lambda_{0j}, q\} \label{E:p+}\\
    & =& \max_{\bs{q} \in \mb{U}|q} \min \{\psi_j(\bs{q}), q\}, \label{E:p++}
\end{IEEEeqnarray}
where the second equality follows from (\ref{E:CondSigPropE}), and the fourth, from (\ref{E:CondSig}). Intuitively (\ref{E:p++}) says that, to guarantee $\bs{d} \geq \bs\psi(\bs{q})$ no matter which $\bs{q} \in \mb{U}|q$ is realized, at least $p_j$ of the $q$ tasks must be served no later than slot $t+j-1$. According to (\ref{E:p+}), (\ref{E:CondSigPropA}), and (\ref{E:CondSigPropB}), $\bs{p}$ is a cumulative vector. It is also a natural extension of $p$ because, according to (\ref{E:p}), (\ref{E:CondSigPropA}), and (\ref{E:p-lambda}),
\begin{equation}\label{E:p1}
p_1 = \hat\lambda_{01} - \hat\lambda_{11} = \hat\lambda_{01} = p.
\end{equation}

It is immediate from (\ref{E:beta}) and (\ref{E:p}) that, for all $\Gamma\subseteq\Omega$,
\begin{IEEEeqnarray}{rCl}
\beta(\Gamma) &=& \max_{j \in \mb{N}} (p_{j+1}^{\l\Gamma\r} + \hat\lambda_{1, j+1}^{\l\Omega\r} - c_{1, j+1}) \label{E:beta+}\\
              &=& p_{j_\beta^\Gamma+1}^{\l\Gamma\r} + \hat\lambda_{1, j_\beta^\Gamma+1}^{\l\Omega\r} -c_{1, j_\beta^\Gamma+1}, \label{E:beta++}
\end{IEEEeqnarray}
where
\begin{equation}\label{E:jbeta}
j_\beta^\Gamma := \argmax_{j \in \mb{N}} (p_{j+1}^{\l\Gamma\r} + \hat\lambda_{1, j+1}^{\l\Omega\r} - c_{1, j+1}),
\end{equation}
with the understanding that, whenever comparisons result in ties, $j_\beta^\Gamma$ should be minimized.\footnote{Similar to (\ref{E:beta+}) and (\ref{E:beta++}), additional, easily derived, formulations of $\beta$ include
\[
\begin{IEEEeqnarraybox}[][c]{rCl}
\beta(\Gamma) &=& \max_{j \in \mb{N}} (\hat\lambda_{0, j+1}^{\l\Omega\r} - p_{j+1}^{\l\overline\Gamma\r}-c_{1, j+1})\\
              &=& \hat\lambda_{0, j_\beta^\Gamma+1}^{\l\Omega\r} - p_{j_\beta^\Gamma+1}^{\l\overline\Gamma\r}-c_{1, j_\beta^\Gamma+1}.
\end{IEEEeqnarraybox}
\]} This reformulation implies that (\ref{E:beta}) can also be expressed as
\begin{equation}\label{E:beta+++}
\beta(\Gamma) = \hat\lambda_{0, j_\beta^\Gamma+1}^{\l\Gamma\r} + \hat\lambda_{1, j_\beta^\Gamma+1}^{\l\overline\Gamma\r}-c_{1, j_\beta^\Gamma+1}.
\end{equation}

The next theorem is fundamental in that it links $\beta$, and thus the feasible polytope, to polymatroid theory.

\begin{theorem}\label{T:betaProp}
{\it If $\bs\psi^{[\Omega]}$ is schedulable, $\beta$ is a supermodular function over $\Omega$, that is, $\beta:2^\Omega \rightarrow \mb{N}$, $\beta(\phi) = 0$, and
\begin{equation}\label{E:betaProp}
\beta(\Gamma)+\beta(\Gamma') \leq \beta(\Gamma+\Gamma')+\beta(\Gamma\Gamma') \blank\blank\forall\, \Gamma,\Gamma'\subseteq\Omega.
\end{equation}}
\end{theorem}

\begin{IEEEproof}
By definition, $\beta$ can only take integer values, so we need only show it to be non-negative to restrict its range to $\mb{N}$. For all $\Gamma\subseteq\Omega$, using (\ref{E:beta}), (\ref{E:CondSig}), and (\ref{E:schedulability}), we have
\[
\beta(\Gamma) \geq (\hat\lambda_{0, j+1}^{\l\Gamma\r} + \hat\lambda_{1, j+1}^{\l\overline\Gamma\r}-c_{1, j+1})|_{j=0} = \hat\lambda_{01}^{\l\Gamma\r} + \hat\lambda_{11}^{\l\overline\Gamma\r} -c_{11} \geq 0.
\]
This also implies that $\beta(\phi) \geq 0$. But, if $\bs\psi^{[\Omega]}$ is schedulable, using (\ref{E:beta++}), (\ref{E:CondSigPropF}), and (\ref{E:schedulability}), we also have
\[
\beta(\phi) = \hat\lambda_{1, j_\beta^\phi+1}^{\l\Omega\r} - c_{1, j_\beta^\phi+1} \leq \lambda_{1, j_\beta^\phi+1}^{\l\Omega\r} - c_{1, j_\beta^\phi+1} \leq 0.
\]
So $\beta(\phi) = 0$.

To complete the proof, for all $\Gamma,\Gamma'\subseteq\Omega$, assume, without loss of generality, that $j_\beta^\Gamma \geq j_\beta^{\Gamma'}$. Then,
\[
\begin{IEEEeqnarraybox}[][c]{rCl}
p_{j_\beta^\Gamma+1}^{\l\Gamma\r} + p_{j_\beta^{\Gamma'}+1}^{\l\Gamma'\r} & =    & p_{j_\beta^\Gamma+1}^{\l\Gamma\r} + p_{j_\beta^{\Gamma'}+1}^{\l\overline\Gamma\Gamma'\r} + p_{j_\beta^{\Gamma'}+1}^{\l\Gamma\Gamma'\r}\\
                                                                          & \leq & p_{j_\beta^\Gamma+1}^{\l\Gamma\r} + p_{j_\beta^\Gamma+1}^{\l\overline\Gamma\Gamma'\r} + p_{j_\beta^{\Gamma'}+1}^{\l\Gamma\Gamma'\r}\\
                                                                          & = & p_{j_\beta^\Gamma+1}^{\l\Gamma+\Gamma'\r} + p_{j_\beta^{\Gamma'}+1}^{\l\Gamma\Gamma'\r}.
\end{IEEEeqnarraybox}
\]
So, using (\ref{E:beta++}) and (\ref{E:beta+}),
\[
\begin{IEEEeqnarraybox}[][c]{Cl}
     & \beta(\Gamma)+\beta(\Gamma')\\
=    & p_{j_\beta^\Gamma+1}^{\l\Gamma\r} \!+ \hat\lambda_{1, j_\beta^\Gamma+1}^{\l\Omega\r} \!- c_{1, j_\beta^\Gamma+1} \!+ p_{j_\beta^{\Gamma'}+1}^{\l\Gamma'\r} \!+ \hat\lambda_{1, j_\beta^{\Gamma'}+1}^{\l\Omega\r} \!- c_{1, j_\beta^{\Gamma'}+1}\\
\leq & p_{j_\beta^\Gamma+1}^{\l\Gamma+\Gamma'\r} \!+ \hat\lambda_{1, j_\beta^\Gamma+1}^{\l\Omega\r} \!- c_{1, j_\beta^\Gamma+1} \!+ p_{j_\beta^{\Gamma'}+1}^{\l\Gamma\Gamma'\r} \!+ \hat\lambda_{1, j_\beta^{\Gamma'}+1}^{\l\Omega\r} \!- c_{1, j_\beta^{\Gamma'}+1}\\
\leq & \beta(\Gamma+\Gamma')+\beta(\Gamma\Gamma').
\end{IEEEeqnarraybox}
\]
\end{IEEEproof}

The next theorem establishes two intuitive properties of $\beta$.

\begin{theorem}\label{T:betaProp+}
{\it For all $\Gamma,\Gamma'\subseteq\Omega$,
\begin{equation}\label{E:betaProp+}
\beta(\Gamma+\Gamma')-\beta(\Gamma') \leq q^{\l\Gamma\r} \blank\textup{if } \Gamma\Gamma'=\phi.
\end{equation}
If $\bs\psi^{[\Omega]}$ is schedulable,
\begin{equation}\label{E:betaProp++}
\beta(\Omega) \leq c.
\end{equation}}
\end{theorem}

The first property, (\ref{E:betaProp+}), says that, if $\Gamma\Gamma'=\phi$, $\beta(\Gamma+\Gamma')$ and $\beta(\Gamma')$, as lower bounds of $d^{\l\Gamma+\Gamma'\r}$ and $d^{\l\Gamma'\r}$, can be {\it jointly} tight without contradicting the causality constraint, (\ref{E:causalityB}). The second property, (\ref{E:betaProp++}), says that $\beta(\Omega)$, as a lower bound of $d^{\l\Omega\r}$, will not contradict the capacity constraint, (\ref{E:capacity}).

\begin{IEEEproof}[Proof of Theorem~\ref{T:betaProp+}]
On the one hand, according to (\ref{E:beta++}),
\[
\beta(\Gamma+\Gamma') = p_{j_\beta^{\Gamma+\Gamma'}+1}^{\l\Gamma+\Gamma'\r} + \hat\lambda_{1, j_\beta^{\Gamma+\Gamma'}+1}^{\l\Omega\r} - c_{1, j_\beta^{\Gamma+\Gamma'}+1}.
\]
On the other, according to (\ref{E:beta+}),
\[
\beta(\Gamma') \geq p_{j_\beta^{\Gamma+\Gamma'}+1}^{\l\Gamma'\r} + \hat\lambda_{1, j_\beta^{\Gamma+\Gamma'}+1}^{\l\Omega\r} - c_{1, j_\beta^{\Gamma+\Gamma'}+1}.
\]
Therefore, if $\Gamma\Gamma'=\phi$, using (\ref{E:p+}), we have
\[
\beta(\Gamma+\Gamma')-\beta(\Gamma') \leq p_{j_\beta^{\Gamma+\Gamma'}+1}^{\l\Gamma+\Gamma'\r}-p_{j_\beta^{\Gamma+\Gamma'}+1}^{\l\Gamma'\r} = p_{j_\beta^{\Gamma+\Gamma'}+1}^{\l\Gamma\r} \leq q^{\l\Gamma\r}.
\]
If $\bs\psi^{[\Omega]}$ is schedulable, using (\ref{E:beta+++}), (\ref{E:CondSigPropF}), and (\ref{E:schedulability}), we also have
\[
\begin{IEEEeqnarraybox}[][c]{rCrCl}
\beta(\Omega) = \hat\lambda_{0, j_\beta^\Omega+1}^{\l\Omega\r} - c_{1, j_\beta^\Omega+1} & \leq & \lambda_{0, j_\beta^\Omega+1}^{\l\Omega\r} -c_{1, j_\beta^\Omega+1} & & \\
                                                                                         & \leq & c_{0, j_\beta^\Omega+1} -c_{1, j_\beta^\Omega+1}                    &=& c.
\end{IEEEeqnarraybox}
\]
\end{IEEEproof}

According to (\ref{E:beta+}), to calculate $\beta(\Gamma)$, we need to find $j_\beta^\Gamma$. The following theorem suggests one systematic approach.

\begin{theorem}\label{T:jbeta}
{\it For all $\Gamma,\Gamma' \subseteq \Omega$, $j_\beta^{\Gamma'} \leq j_\beta^\Gamma$ if $\Gamma' \subseteq \Gamma$.}
\end{theorem}

To use this theorem to find all $j_\beta^\Gamma$'s, starting with $j_\beta^\Omega$, find all $j_\beta^\Gamma$'s with $|\Gamma| = n-1$, then all those with $|\Gamma| = n-2$, and so on. This ensures that all supersets of $\Gamma$ are visited before $\Gamma$ so that $j_\beta^\Gamma$ is bounded as tightly as possible. In particular, as we proceed, if we find $j_\beta^{\Gamma} = 0$, we can safely set $j_\beta^{\Gamma'} = 0$ for all $\Gamma' \subseteq \Gamma$.

\begin{IEEEproof}[Proof of Theorem~\ref{T:jbeta}]
If $\Gamma' \subseteq \Gamma$, for all $j > j_\beta^\Gamma$,
\[
\begin{IEEEeqnarraybox}[][c]{rCl}
p_{j+1}^{\l\Gamma'\r}-p_{j_\beta^\Gamma+1}^{\l\Gamma'\r} & \leq & p_{j+1}^{\l\Gamma'\r}-p_{j_\beta^\Gamma+1}^{\l\Gamma'\r} + p_{j+1}^{\l\Gamma\overline{\Gamma'}\r}-p_{j_\beta^\Gamma+1}^{\l\Gamma\overline{\Gamma'}\r}\\
                                                         & =    & p_{j+1}^{\l\Gamma\r}-p_{j_\beta^\Gamma+1}^{\l\Gamma\r}\\
                                                         & \leq & (\hat\lambda_{1, j_\beta^\Gamma+1}^{\l\Omega\r} - c_{1, j_\beta^\Gamma+1}) - (\hat\lambda_{1, j+1}^{\l\Omega\r} - c_{1, j+1}),
\end{IEEEeqnarraybox}
\]
where the final inequality holds because, according to (\ref{E:jbeta}),
\[
p_{j+1}^{\l\Gamma\r} + \hat\lambda_{1, j+1}^{\l\Omega\r} - c_{1, j+1} \leq p_{j_\beta^\Gamma+1}^{\l\Gamma\r} + \hat\lambda_{1, j_\beta^\Gamma+1}^{\l\Omega\r} - c_{1, j_\beta^\Gamma+1}.
\]
Then, for all $j > j_\beta^\Gamma$,
\[
p_{j+1}^{\l\Gamma'\r} + \hat\lambda_{1, j+1}^{\l\Omega\r} - c_{1, j+1} \leq p_{j_\beta^\Gamma+1}^{\l\Gamma'\r} + \hat\lambda_{1, j_\beta^\Gamma+1}^{\l\Omega\r} - c_{1, j_\beta^\Gamma+1}.
\]
But, according to (\ref{E:jbeta}),
\[
p_{j_\beta^{\Gamma'}+1}^{\l\Gamma'\r} + \hat\lambda_{1, j_\beta^{\Gamma'}+1}^{\l\Omega\r} - c_{1, j_\beta^{\Gamma'}+1} \geq p_{j_\beta^\Gamma+1}^{\l\Gamma'\r} + \hat\lambda_{1, j_\beta^\Gamma+1}^{\l\Omega\r} - c_{1, j_\beta^\Gamma+1}.
\]
So, if this inequality is strict, we have $j_\beta^{\Gamma'} < j_\beta^\Gamma$, while, if it holds with equality, we still have $j_\beta^{\Gamma'} \leq j_\beta^\Gamma$ because, according to (\ref{E:jbeta})'s tie-break rule, $j_\beta^{\Gamma'}$ is the minimum index achieving this equality.
\end{IEEEproof}

\subsection{Feasible Permutohedra}\label{SS:FP}

Because $\beta$ is supermodular, our hunch is that the feasible polytope, $\mb{F}$, should be related to some permutohedron. The problem is that $d^{\l\Omega\r}$ must remain constant in a permutohedron, which is not the case for $\mb{F}$. This observation motivates us to intersect $\mb{F}$ with the hyperplane,
\begin{equation}\label{E:Hmu}
\mb{H}_\mu := \{d^{[\Omega]} | d^{\l\Omega\r} = \mu\},
\end{equation}
as the intersection, $\mb{F}_\mu:=\mb{F}\cap\mb{H}_\mu$, turns out to be a permutohedron.

According to Theorem~\ref{T:polytope}, when $d^{[\Omega]} \in \mb{F}_\mu$, it must satisfy (\ref{E:causalityB}), (\ref{E:capacity}), and (\ref{E:beta}). Hence, for $\mb{F}_\mu$ to be non-empty, $\mu$ must~satisfy
\begin{equation}\label{E:muFeasible}
\beta(\Omega) \leq \mu \leq \min \{q^{\l\Omega\r}, c\}.
\end{equation}
We call such $\mu$ {\it \textbf{feasible}}. If $\bs\psi^{[\Omega]}$ is schedulable, it is immediate from (\ref{E:betaProp+}) and (\ref{E:betaProp++}) that $\beta(\Omega) \leq q^{\l\Omega\r}$ and $\beta(\Omega) \leq c$, so at least one feasible $\mu$ exists.

Now, for clarity, when $d^{[\Omega]} \in \mb{H}_\mu$, we denote it by $d_\mu^{[\Omega]}$ so that, by definition, $d_\mu^{\l\Omega\r} = \mu$. Then, when $d_\mu^{[\Omega]} \in \mb{F}_\mu$, since (\ref{E:beta}) and (\ref{E:causalityB}) imply that $d_\mu^{\l\Gamma\r} \geq \beta(\Gamma)$ and $d_\mu^{\l\Gamma\r}+q^{\l\overline\Gamma\r} \geq d_\mu^{\l\Omega\r} = \mu$,
\begin{equation}\label{E:betaMu}
d_\mu^{\l\Gamma\r} \geq \beta_\mu(\Gamma) := \max\{\beta(\Gamma), \mu-q^{\l\overline\Gamma\r}\} \blank\blank \forall\, \Gamma\subseteq\Omega.
\end{equation}
Also, if $\mu$ is feasible, according to (\ref{E:muFeasible}),
\begin{equation}\label{E:betaMuOmega}
\beta_\mu(\Omega) = \max\{\beta(\Omega), \mu\} = \mu.
\end{equation}
So, when $d_\mu^{[\Omega]} \in \mb{F}_\mu$, we have both
\begin{equation}\label{E:permuBetaMu}
d_\mu^{\l\Omega\r} = \mu = \beta_\mu(\Omega) \blank\text{and}\blank d_\mu^{\l\Gamma\r} \geq \beta_\mu(\Gamma) ~\forall\, \Gamma\subseteq\Omega.
\end{equation}
Comparing (\ref{E:permuBetaMu}) to (\ref{E:permutohedron}), it is immediate that $d_\mu^{[\Omega]} \in \mb{P}(\beta_\mu)$ if $\beta_\mu$ is supermodular. But is $\beta_\mu$ supermodular? Yes, according to the next theorem, it is.

\begin{theorem}\label{T:betaMuProp}
{\it If $\bs\psi^{[\Omega]}$ is schedulable and $\mu$ is feasible, $\beta_\mu$ is a supermodular function over $\Omega$, that is, $\beta_\mu:2^\Omega \rightarrow \mb{N}$, $\beta_\mu(\phi) = 0$, and
\begin{equation}\label{E:betaMuProp}
\beta_\mu(\Gamma)+\beta_\mu(\Gamma') \leq \beta_\mu(\Gamma+\Gamma')+\beta_\mu(\Gamma\Gamma') \blank\blank\forall\, \Gamma,\Gamma'\subseteq\Omega.
\end{equation}}
\end{theorem}

\begin{IEEEproof}
If $\bs\psi^{[\Omega]}$ is schedulable, according to Theorem~\ref{T:betaProp}, $\beta:2^\Omega \rightarrow \mb{N}$. So it is immediate from (\ref{E:betaMu}) that, not only can $\beta_\mu$ only take integer values but, for all $\Gamma\subseteq\Omega$, \mbox{$\beta_\mu(\Gamma) \geq \beta(\Gamma) \geq 0$}. Thus $\beta_\mu:2^\Omega \rightarrow \mb{N}$. Additionally, since $\beta(\phi) = 0$, and since, according to (\ref{E:muFeasible}), $\mu$'s feasibility implies that $\mu \leq q^{\l\Omega\r}$, according to (\ref{E:betaMu}),
\[
\beta_\mu(\phi) = \max\{\beta(\phi), \mu-q^{\l\Omega\r}\} = 0.
\]

To prove (\ref{E:betaMuProp}), we need only consider four cases:
\begin{itemize}
\item[C1] if $\beta_\mu(\Gamma) = \beta(\Gamma)$ and $\beta_\mu(\Gamma') = \beta(\Gamma')$, (\ref{E:betaMuProp}) follows directly from (\ref{E:betaProp});
\item[C2] if $\beta_\mu(\Gamma) = \mu-q^{\l\overline\Gamma\r}$ and $\beta_\mu(\Gamma') = \mu-q^{\l\overline{\Gamma'}\r}$,
\[
\begin{IEEEeqnarraybox}[][c]{rCl}
\beta_\mu(\Gamma)+\beta_\mu(\Gamma') & =    & \mu-q^{\l\overline\Gamma\r}+\mu-q^{\l\overline{\Gamma'}\r}\\
                                     & =    & \mu-q^{\l\overline{\Gamma+\Gamma'}\r}+\mu-q^{\l\overline{\Gamma\Gamma'}\r}\\
                                     & \leq & \beta_\mu(\Gamma+\Gamma')+\beta_\mu(\Gamma\Gamma');
\end{IEEEeqnarraybox}
\]
\item[C3] if $\beta_\mu(\Gamma) = \mu-q^{\l\overline\Gamma\r}$ and $\beta_\mu(\Gamma') = \beta(\Gamma')$,
\[
\begin{IEEEeqnarraybox}[][c]{rCl}
\beta_\mu(\Gamma)+\beta_\mu(\Gamma') & =    & \mu-q^{\l\overline\Gamma\r}+\beta(\Gamma')\\
                                     & \leq & \mu-q^{\l\overline\Gamma\r}+q^{\l\overline\Gamma\Gamma'\r}+\beta(\Gamma\Gamma')\\
                                     & =    & \mu-q^{\l\overline{\Gamma+\Gamma'}\r}+\beta(\Gamma\Gamma')\\
                                     & \leq & \beta_\mu(\Gamma+\Gamma')+\beta_\mu(\Gamma\Gamma'),
\end{IEEEeqnarraybox}
\]
where the first inequality holds because, according to (\ref{E:betaProp+}), $\beta(\Gamma') - \beta(\Gamma\Gamma') \leq q^{\l\overline\Gamma\Gamma'\r}$; and finally,
\item[C4] if $\beta_\mu(\Gamma) = \beta(\Gamma)$ and $\beta_\mu(\Gamma') = \mu-q^{\l\overline{\Gamma'}\r}$, (\ref{E:betaMuProp}) still holds as, aside from interchanging $\Gamma$ and $\Gamma'$, C4 is C3.
\end{itemize}
So, in all cases, (\ref{E:betaMuProp}) must also hold.
\end{IEEEproof}

Theorem~\ref{T:betaMuProp} ensures that $d_\mu^{[\Omega]} \in \mb{P}(\beta_\mu)$ when $d_\mu^{[\Omega]} \in \mb{F}_\mu$, and consequently, that $\mb{F}_\mu \subseteq \mb{P}(\beta_\mu)$. In fact, as shown by the next theorem, $\mb{F}_\mu$ is exactly $\mb{P}(\beta_\mu)$.

\begin{theorem}\label{T:FP}
{\it If $\bs\psi^{[\Omega]}$ is schedulable and $\mu$ is feasible,
\begin{equation}\label{E:Fmu}
\mb{F}_\mu = \mb{P}(\beta_\mu).
\end{equation}
}
\end{theorem}

Since $\mb{P}(\beta_\mu)$, the permutohedron generated by $\beta_\mu$, is non-empty, by proving this theorem, we establish I1 of Theorem~\ref{T:schedulability}, that is, we establish the existence of feasible schedules, and thus prove Theorem~\ref{T:schedulability}. We call $\mb{P}(\beta_\mu)$ the {\it \textbf{feasible permutohedron}} under $\mu$. Although each feasible permutohedron is but one slice of $\mb{F}$, by putting the slices together, we can assemble $\mb{F}$. When $n=2$, $\mb{P}(\beta_\mu)$ is a line segment. In Fig.~\ref{F:FeasibleRegion}, it is $\overline{A'B'}$. When $n=3$, $\mb{P}(\beta_\mu)$ is a hexagon. When $n=4$, it is a truncated octahedron. These cases are depicted in Fig.~\ref{F:Polymatroids}. Notice that $\mb{P}(\beta_\mu)$ can be highly degenerate. When, for instance, $\mu = q^{\l\Omega\r} \leq c$, $\mb{P}(\beta_\mu)$ shrinks to a single point, $q^{[\Omega]}$.\footnote{In general, $\mb{P}(\beta_\mu)$ is {\it \textbf{degenerate}} when not all of its $n!$ vertices are distinct. Although we will not do so here, it can be shown that, due to our service model's constant-capacity assumption, $\mb{P}(\beta_\mu)$ is almost always degenerate to a rather significant degree.}

\begin{IEEEproof}[Proof of Theorem~\ref{T:FP}]
As $d_\mu^{[\Omega]} \in \mb{F}_\mu$ implies (\ref{E:permuBetaMu}), which in turn implies that $\mb{F}_\mu \subseteq \mb{P}(\beta_\mu)$, we need only show that $\mb{P}(\beta_\mu)  \subseteq \mb{F}_\mu$. So, according to Theorem~\ref{T:polytope}, we need only show that (\ref{E:permuBetaMu}) implies (\ref{E:causalityB}), (\ref{E:capacity}), and (\ref{E:beta}). First, for all $\omega\in\Omega$, using (\ref{E:permuBetaMu}) and the definition of $\beta_\mu$ in (\ref{E:betaMu}), we have
\[
\begin{IEEEeqnarraybox}[][c]{rCl}
d_\mu^\omega = d_\mu^{\l\Omega\r} - d_\mu^{\l\overline{\{\omega\}}\r} & \leq & \mu-\beta_\mu(\overline{\{\omega\}})                   \\
                                                                      & =    & \mu-\max\{\beta(\overline{\{\omega\}}), \mu-q^\omega\} \leq q^\omega,
\end{IEEEeqnarraybox}
\]
so (\ref{E:causalityB}) holds. Second, since $\mu$ is feasible, according to (\ref{E:muFeasible}), $d_\mu^{\l\Omega\r} = \mu \leq c$, so (\ref{E:capacity}) holds. Finally, according to (\ref{E:permuBetaMu}) and the definition of $\beta_\mu$ in (\ref{E:betaMu}), $d_\mu^{\l\Gamma\r} \geq \beta_\mu(\Gamma) \geq \beta(\Gamma)$ for all $\Gamma\subseteq\Omega$, so (\ref{E:beta}) holds.
\end{IEEEproof}

\subsection{Selecting Feasible Schedules}\label{SS:how}

According to Theorem~\ref{T:FP}, one way to select a feasible schedule is to first select a feasible $\mu$, which fixes the total service, and then select $d_\mu^{[\Omega]}$ from $\mb{P}(\beta_\mu)$. To see how these selections can be used to, for instance, enforce a desired service priority or fairness objective, it is helpful to examine how service is distributed at $\mb{P}(\beta_\mu)$'s vertices.

Recall from Section~\ref{SS:polymatroid} that, for each permutation, \mbox{$\pi\in\Pi^\Omega$}, there exists a corresponding vertex of $\mb{P}(\beta_\mu)$, denoted by $v_\pi^{[\Omega]}(\beta_\mu)$. It is immediate from (\ref{E:vertexSum}) and (\ref{E:permuBetaMu}) that
\begin{equation}\label{E:priority}
v_\pi^{\l\Gamma_\pi^i\r}(\beta_\mu) = \beta_\mu(\Gamma_\pi^i) = \min_{d_\mu^{[\Omega]} \in \mb{P}(\beta_\mu)} d_\mu^{\l\Gamma_\pi^i\r} \blank\blank \forall\, 0 \leq i \leq n.
\end{equation}
This relation, according to (\ref{E:GammapiB}), implies that, when schedule $v_\pi^{[\Omega]}(\beta_\mu)$ is selected, first $d_\mu^{\omega_\pi^1}$ is minimized, then $d_\mu^{\omega_\pi^1}+d_\mu^{\omega_\pi^2}$ is minimized, then $d_\mu^{\omega_\pi^1}+d_\mu^{\omega_\pi^2}+d_\mu^{\omega_\pi^3}$ is minimized, and so on. Thus, a strict flow priority order is enforced. In particular, $\pi(\omega)$ can be viewed as a priority index. The larger the $\pi(\omega)$, the higher the priority that flow $\omega$ enjoys.

When, instead of enforcing priorities, the objective is ensuring fairness, priorities can be assigned equally by setting $d_\mu^{[\Omega]}$ equal to the vertex centroid of $\mb{P}(\beta_\mu)$,
\begin{equation}\label{E:fair}
v_\text{F}^{[\Omega]}(\beta_\mu):= \frac{1}{n!} \sum_{\pi\in\Pi^\Omega} v_\pi^{[\Omega]}(\beta_\mu).
\end{equation}
Intuitively, $v_\text{F}^{[\Omega]}(\beta_\mu)$ is fair in the sense that it gives all vertices, each enforcing a unique flow priority order, an equal weight.\footnote{In fact, $v_\text{F}^{[\Omega]}(\beta_\mu)$'s definition coincides with that of the {\it Shapley} value from cooperative game theory. More specifically, it is easy to verify that
\[
v_\text{F}^\omega(\beta_\mu) = \sum_{\Gamma\subseteq\overline{\{\omega\}}} \frac{|\Gamma|!(n-|\Gamma|-1)!}{n!} (\beta_\mu(\Gamma+\{\omega\})-\beta_\mu(\Gamma)) \blank\blank \forall\, \omega\in\Omega,
\]
which is the classical formulation of the {\it Shapley} value.
} Of course, (\ref{E:fair}) may not result in an integral point, in which case, it needs to be rounded.

\subsubsection*{\textup{[}The Baseline Permutohedron\textup{]}} An alternative approach to selecting feasible schedules is motivated by the following observation. In the case that $\mu = \beta(\Omega)$, using (\ref{E:betaMu}), it is easy to verify that $\beta_\mu = \beta$, that is, $\beta_\mu(\Gamma) = \beta(\Gamma)$ for all $\Gamma\subseteq\Omega$, because, according to (\ref{E:betaProp+}), $\beta(\Omega) - \beta(\Gamma) \leq q^{\l\overline\Gamma\r}$. So, in this case, $\mb{P}(\beta_\mu) = \mb{P}(\beta)$. We call $\mb{P}(\beta)$ the {\it \textbf{baseline permutohedron}}. It is the bottom face of $\mb{F}$ in the sense that $d^{\l\Omega\r}$ is minimized in $\mb{F}$ if $d^{[\Omega]} \in \mb{P}(\beta)$. In Fig.~\ref{F:FeasibleRegion}, $\mb{P}(\beta)$ is line segment $\overline{AB}$.

To use $\mb{P}(\beta)$ to select a feasible schedule, first select $d_*^{[\Omega]}$ from $\mb{P}(\beta)$, and then select $d^{[\Omega]}$ such that $d_*^{[\Omega]} \leq d^{[\Omega]} \leq q^{[\Omega]}$ and $d^{\l\Omega\r} \leq c$. Significantly, this second step requires nothing but a free allocation of the excess capacity, $c-\beta(\Omega)$, subject to the causality and capacity constraints. So, for instance, a generalized-processor-sharing (GPS) policy, as analyzed in \cite{Parekh:1993, Parekh:1994}, can be used to allocate this excess according to any preset ratios, or even ratios set dynamically by $d_*^{[\Omega]}$. The validity of this approach is assured by Theorem~\ref{T:polytope} and the next lemma.

\begin{lemma}\label{L:fundamental++}
{\it If $\bs\psi^{[\Omega]}$ is schedulable, $d^{[\Omega]}$ satisfies the baseline constraint, (\ref{E:beta}), if and only if there exists $d_*^{[\Omega]} \in \mb{P}(\beta)$ such that $d^{[\Omega]} \geq d_*^{[\Omega]}$.}
\end{lemma}

\begin{IEEEproof}
If there exists $d_*^{[\Omega]} \in \mb{P}(\beta)$ such that $d^{[\Omega]} \geq d_*^{[\Omega]}$, by definition, $d^{\l\Gamma\r} \geq d_*^{\l\Gamma\r} \geq \beta(\Gamma)$ for all $\Gamma \subseteq \Omega$. Therefore, the condition is sufficient. To establish its necessity, we will show that if $d^{\l\Gamma\r} \geq \beta(\Gamma)$ for all $\Gamma \subseteq \Omega$, there is always a direction along which $d^{[\Omega]}$ can descend towards $\mb{P}(\beta)$, while maintaining $d^{\l\Gamma\r} \geq \beta(\Gamma)$ for all $\Gamma \subseteq \Omega$, until it reaches $\mb{P}(\beta)$.

To begin, if $d^{\l\Gamma\r} = \beta(\Gamma)$ and $d^{\l\Gamma'\r} = \beta(\Gamma')$ for some $\Gamma,\Gamma'\subseteq\Omega$, according to (\ref{E:betaProp}),
\[
\begin{IEEEeqnarraybox}[][c]{rCl}
d^{\l\Gamma+\Gamma'\r}+d^{\l\Gamma\Gamma'\r} = d^{\l\Gamma\r}+d^{\l\Gamma'\r} &=& \beta(\Gamma)+\beta(\Gamma')\\
                                                                              & \leq & \beta(\Gamma+\Gamma')+\beta(\Gamma\Gamma').
\end{IEEEeqnarraybox}
\]
But, if $d^{\l\Gamma\r} \geq \beta(\Gamma)$ for all $\Gamma \subseteq \Omega$, this is impossible unless $d^{\l\Gamma+\Gamma'\r} = \beta(\Gamma+\Gamma')$ and $d^{\l\Gamma\Gamma'\r} = \beta(\Gamma\Gamma')$, so $d^{\l\Gamma\r} = \beta(\Gamma)$ and $d^{\l\Gamma'\r} = \beta(\Gamma')$ imply that $d^{\l\Gamma+\Gamma'\r} = \beta(\Gamma+\Gamma')$.

Next, let $\Gamma_*$ denote the union of all $\Gamma\subseteq\Omega$ such that $d^{\l\Gamma\r} = \beta(\Gamma)$. Then, applying the preceding result repeatedly, $d^{\l\Gamma_*\r} = \beta(\Gamma_*)$. If $d^{\l\Omega\r} > \beta(\Omega)$, clearly $\Gamma_* \subset \Omega$, so at least one $\omega_* \notin \Gamma_*$ must exist. By construction, $d^{\l\Gamma\r} > \beta(\Gamma)$ for all $\Gamma$ containing $\omega_*$, so $d^{\omega_*}$ can be reduced until $d^{\l\Gamma\r} = \beta(\Gamma)$ for some $\Gamma$ containing $\omega_*$. If, after this reduction, it remains the case that $d^{\l\Omega\r} > \beta(\Omega)$, we can use the same procedure to find another direction, $\omega_{**}$, for $d^{[\Omega]}$ to descend. This descent can continue, while maintaining $d^{\l\Gamma\r} \geq \beta(\Gamma)$ for all $\Gamma \subseteq \Omega$, until $d^{\l\Omega\r} = \beta(\Omega)$, which, by definition, guarantees that $d^{[\Omega]}$ reaches some $d_*^{[\Omega]} \in \mb{P}(\beta)$.
\end{IEEEproof}

\section{Max-Slack Schedules}\label{S:MSS}

So far we have demonstrated both the modeling power of worst-case services and the generality of our state-based approach to guaranteeing them. A downside of our framework's generality is its complexity. One source of this complexity is the fact that, given a feasible $\mu$, to fully exploit the flexibility of selecting any $d_\mu^{[\Omega]}$ from $\mb{P}(\beta_\mu)$, such as $v_\text{F}^{[\Omega]}(\beta_\mu)$ defined in (\ref{E:fair}), we must calculate all $2^n$ values of $\beta_\mu$, which, even for a small $n$, is daunting. In fact, much of this calculation can be avoided for some special feasible schedules.

For instance, if $\bs\psi^{[\Omega]}$ is schedulable, according to (\ref{E:CondSigPropF}) and (\ref{E:schedulability}), $\hat\lambda_{1, j_\beta^\Gamma+1}^{\l\Omega\r} \leq \lambda_{1, j_\beta^\Gamma+1}^{\l\Omega\r} \leq c_{1, j_\beta^\Gamma+1}$. So, according to (\ref{E:beta++}), if $j > j_\beta^\Gamma$, $p_j^{\l\Gamma\r} \geq p_{j_\beta^\Gamma+1}^{\l\Gamma\r} \geq \beta(\Gamma)$. But, according to Theorem~\ref{T:jbeta}, $j_\beta^\Gamma \leq j_\beta^\Omega$. So, if $j > j_\beta^\Omega$, $p_j^{\l\Gamma\r} \geq \beta(\Gamma)$ for all $\Gamma\subseteq\Omega$. It is also clear that $p_j^{[\Omega]}$ is valid if $p_j^{\l\Omega\r} \leq c$ because, according to (\ref{E:p+}), $p_j^{[\Omega]} \leq q^{[\Omega]}$. So, according to Theorem~\ref{T:polytope}, $p_j^{[\Omega]}$ is feasible if $j > j_\beta^\Omega$ and $ p_j^{\l\Omega\r} \leq c$.

This result can be improved. In particular, it can be shown that, if $p_j^{\l\Omega\r} \geq \beta(\Omega)$, $p_j^{\l\Gamma\r} \geq \beta(\Gamma)$ for all $\Gamma\subseteq\Omega$. As we have shown that $p_j^{\l\Gamma\r} \geq \beta(\Gamma)$ if $j > j_\beta^\Gamma$, we need only focus on the case that $j \leq j_\beta^\Gamma$. According to (\ref{E:beta+}),  $p_j^{\l\Omega\r} \geq \beta(\Omega)$ implies that
\[
p_j^{\l\Omega\r} \geq p_{j_\beta^\Gamma+1}^{\l\Omega\r} + \hat\lambda_{1, j_\beta^\Gamma+1}^{\l\Omega\r} -c_{1, j_\beta^\Gamma+1}.
\]
Hence, if $j \leq j_\beta^\Gamma$, according to (\ref{E:beta++}),
\[
p_j^{\l\Gamma\r} \geq p_j^{\l\Omega\r} - p_{j_\beta^\Gamma+1}^{\l\overline\Gamma\r} \geq p_{j_\beta^\Gamma+1}^{\l\Gamma\r} + \hat\lambda_{1, j_\beta^\Gamma+1}^{\l\Omega\r} -c_{1, j_\beta^\Gamma+1} = \beta(\Gamma).
\]
So $p_j^{[\Omega]}$ is feasible if $\beta(\Omega) \leq p_j^{\l\Omega\r} \leq c$. In this case, although we still need to calculate $\beta(\Omega)$ to ensure that $p_j^{\l\Omega\r} \geq \beta(\Omega)$, if we set $\mu = p_j^{\l\Omega\r}$, $p_j^{[\Omega]} \in \mb{P}(\beta_\mu)$. So no calculation of $\beta_\mu$ is required. However, such $p_j^{[\Omega]}$'s are disjoint points in $\mb{F}$ separated by irregular gaps. To fill these gaps, we introduce max-slack schedules.

In the remainder of this section, we first define max-slack schedules and investigate their properties. We then explore how scheduling max-slack scheduled classes of flows as opposed to individual flows, enables scheduling flexibility to be traded for scheduling efficiency. We also interpret max-slack schedules in terms of the scheduled tasks' worst-case deadlines. This leads us to compare max-slack schedules to EDF schedules, which in turn motivates the introduction of deadline-rigid services.

\subsection{Definition}\label{SS:MSSDef}

\begin{figure}[t]
\centering \scalebox{0.875}{\includegraphics{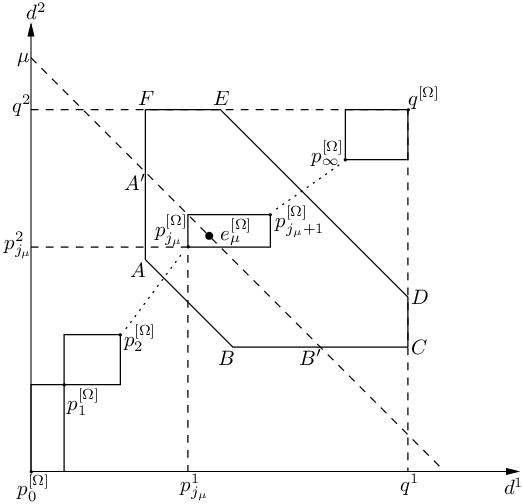}} \caption{In a two-flow case, $e_\mu^{[\Omega]}$ is selected from the intersection of $\mb{H}_\mu$ and the elevating staircase of rectangles, $[p_j^{[\Omega]}, p_{j+1}^{[\Omega]}]$, $j = 0, 1, 2, \ldots$.}\label{F:MaxSlack}
\end{figure}

Since $p_j^{[\Omega]}$ is feasible if $\beta(\Omega) \leq p_j^{\l\Omega\r} \leq c$, it is immediate that, if $\beta(\Omega) \leq p_j^{\l\Omega\r} \leq p_{j+1}^{\l\Omega\r} \leq c$, $d^{[\Omega]}$ is feasible if $d^{[\Omega]} \in [p_j^{[\Omega]}, p_{j+1}^{[\Omega]}]$, that is, if $p_j^{[\Omega]} \leq d^{[\Omega]} \leq p_{j+1}^{[\Omega]}$. So $[p_j^{[\Omega]}, p_{j+1}^{[\Omega]}]$, an $n$-dimension hypercuboid, must lie inside the feasible polytope, that is, $[p_j^{[\Omega]}, p_{j+1}^{[\Omega]}] \subseteq \mb{F}$. This is illustrated, for a two-flow case, in Fig.~\ref{F:MaxSlack}, where $[p_{j_\mu}^{[\Omega]}, p_{j_\mu+1}^{[\Omega]}] \subseteq \mb{F}$ is a rectangle, and $\mb{F}$ is the hexagon, $ABCDEF$. Moreover, when $\mb{H}_\mu$ intersects $[p_{j_\mu}^{[\Omega]}, p_{j_\mu+1}^{[\Omega]}]$, since $[p_{j_\mu}^{[\Omega]}, p_{j_\mu+1}^{[\Omega]}] \subseteq \mb{F}$, the intersection must lie inside $\mb{F}_\mu$, which, in Fig.~\ref{F:MaxSlack}, is line segment $\overline{A'B'}$. Clearly, a feasible schedule can be selected from this intersection. However, for the intersection to be possible, $j_\mu$ cannot be arbitrary but must depend on $\mu$. This motivates the following definition.

\begin{definition}\label{D:MSS}
{\it $e_\mu^{[\Omega]} \in \mb{H}_\mu$ is a \textbf{max-slack schedule} under~$\mu$~if
\begin{equation}\label{E:MSSDef}
e_\mu^{[\Omega]}\in\left\{
\begin{IEEEeqnarraybox}[][c]{ll}
[p_{j_\mu}^{[\Omega]}, p_{j_\mu+1}^{[\Omega]}] & \blank\blank\textup{if } j_\mu < \infty\\
{[}p_\infty^{[\Omega]}, q^{[\Omega]}]          & \blank\blank\textup{if } j_\mu = \infty
\end{IEEEeqnarraybox}\right. ~,
\end{equation}
where
\begin{equation}\label{E:jMu}
j_\mu := \tau_{\mu+1}(\bs{p}^{\l\Omega\r}).
\end{equation}}
\end{definition}

This definition says that $e_\mu^{[\Omega]}$ lies in the intersection of $\mb{H}_\mu$ and an elevating staircase of hypercuboids, at a location determined by $j_\mu$. To see why $j_\mu$ determines this location, notice that, if $\mu \leq q^{\l\Omega\r}$, using (\ref{E:jMu}) and (\ref{E:tau}), it is easy to verify that
\begin{equation}\label{E:jMu+}
\mu\in\left\{
\begin{IEEEeqnarraybox}[][c]{ll}
[p_{j_\mu}^{\l\Omega\r}, p_{j_\mu+1}^{\l\Omega\r}) & \blank\blank\text{if } j_\mu < \infty\\
{[}p_\infty^{\l\Omega\r}, q^{\l\Omega\r}]          & \blank\blank\text{if } j_\mu = \infty
\end{IEEEeqnarraybox}\right. ~.
\end{equation}
So, consistent with (\ref{E:MSSDef}), $\mb{H}_\mu$ intersects the staircase somewhere in $[p_{j_\mu}^{[\Omega]}, p_{j_\mu+1}^{[\Omega]}]$ if $j_\mu < \infty$, and somewhere in $[p_\infty^{[\Omega]}, q^{[\Omega]}]$ if $j_\mu = \infty$, where $[p_{j_\mu}^{[\Omega]}, p_{j_\mu+1}^{[\Omega]}]$ and $[p_\infty^{[\Omega]}, q^{[\Omega]}]$ are non-empty because, according to (\ref{E:p+}), $p_{j_\mu}^{[\Omega]} \leq p_{j_\mu+1}^{[\Omega]}$ and $p_\infty^{[\Omega]} \leq q^{[\Omega]}$. In particular, if $\mu = p_{j_\mu}^{\l\Omega\r}$, the intersection shrinks to a single point, $p_{j_\mu}^{[\Omega]}$, and, as $\mu$ increases, it moves away from $p_{j_\mu}^{[\Omega]}$ toward $p_{j_\mu+1}^{[\Omega]}$ if $j_\mu < \infty$, and toward $q^{[\Omega]}$ if $j_\mu = \infty$.

Comparing (\ref{E:jMu+}) to (\ref{E:MSSDef}), it is immediate that, since $\mu$, $\bs{p}^{[\Omega]}$, and $q^{[\Omega]}$ are all integral, there exists at least one way to split $\mu$ integrally into $e_\mu^{[\Omega]}$.\footnote{One subtlety: if $j_\mu < \infty$, according to (\ref{E:jMu+}), $\mu < p_{j_\mu+1}^{\l\Omega\r}$, so $e_\mu^{[\Omega]} \neq p_{j_\mu+1}^{[\Omega]}$. But, in this case, why does (\ref{E:MSSDef}) still require that $e_\mu^{[\Omega]} \leq p_{j_\mu+1}^{[\Omega]}$ instead of $e_\mu^{[\Omega]} < p_{j_\mu+1}^{[\Omega]}$? The reason is that, if $\mu < p_{j_\mu+1}^{\l\Omega\r}$, although $e_\mu^{[\Omega]} \neq p_{j_\mu+1}^{[\Omega]}$, it is still possible, and possibly even necessary, that $e_\mu^\omega = p_{j_\mu+1}^\omega$ for some, but not all, $\omega\in\Omega$. This possibility, however, is excluded when $e_\mu^{[\Omega]} < p_{j_\mu+1}^{[\Omega]}$, that is, when $e_\mu^\omega < p_{j_\mu+1}^\omega$ for all $\omega \in \Omega$.} That is to say, the intersection contains at least one integral point that can be selected as $e_\mu^{[\Omega]}$.\footnote{In the case that the intersection contains multiple integral points, any one of them can be selected as $e_\mu^{[\Omega]}$.} So, if $\mu \leq q^{\l\Omega\r}$, $e_\mu^{[\Omega]}$ exists. What if $\mu > q^{\l\Omega\r}$? In this case, $j_\mu=\infty$, but it is impossible for $e_\mu^{[\Omega]} \leq q^{[\Omega]}$. So, if $\mu > q^{\l\Omega\r}$, no $e_\mu^{[\Omega]}$ exists. This can be clearly seen in Fig.~\ref{F:MaxSlack} where, since the staircase ends at $q^{[\Omega]}$, no intersection is possible beyond $q^{[\Omega]}$.

\subsection{The Properties of $e_\mu^{[\Omega]}$}\label{SS:MSSProp}

From our argument leading to Definition~\ref{D:MSS}, it follows that, if $j_\mu < \infty$ and $\beta(\Omega) \leq p_{j_\mu}^{\l\Omega\r} \leq p_{j_\mu+1}^{\l\Omega\r} \leq c$, $[p_{j_\mu}^{[\Omega]}, p_{j_\mu+1}^{[\Omega]}] \subseteq \mb{F}$ and thus, according to (\ref{E:MSSDef}), $e_\mu^{[\Omega]} \in \mb{F}_\mu$. Additionally, according to (\ref{E:jMu+}), $p_{j_\mu}^{\l\Omega\r} \leq \mu < p_{j_\mu+1}^{\l\Omega\r}$. So, if $p_\infty^{\l\Omega\r} > c$, $e_\mu^{[\Omega]} \in \mb{F}_\mu$ as long as
\[
\min_{j\in\mb{N}, p_j^{\l\Omega\r} \geq \beta(\Omega)} p_j^{\l\Omega\r} \leq \mu < \max_{j\in\mb{N}, p_j^{\l\Omega\r} \leq c} p_j^{\l\Omega\r}.
\]
As any $\mu$ satisfying this condition necessarily satisfies (\ref{E:muFeasible}), we further conjecture that $e_\mu^{[\Omega]} \in \mb{F}_\mu$ if (\ref{E:muFeasible}) holds, that is, if $\mu$ is feasible. Then, according to Theorem~\ref{T:FP}, $e_\mu^{[\Omega]} \in \mb{F}_\mu$ if $\mb{F}_\mu$ is non-empty. This conjecture is confirmed by the next theorem.

\begin{theorem}\label{T:MSS}
{\it If $\mb{F}_\mu$ is non-empty, $e_\mu^{[\Omega]}$ exists, and $e_\mu^{[\Omega]} \in \mb{F}_\mu$.}
\end{theorem}

To prove this theorem, we first establish a key lemma.

\begin{lemma}\label{L:MSS}
{\it If $e_\mu^{[\Omega]}$ exists,
\begin{equation}\label{E:MSS}
\dot\lambda_{ij}^{\l\Omega\r}(e_\mu^{[\Omega]}) = \min_{d_\mu^{[\Omega]} \leq q^{[\Omega]}} \dot\lambda_{ij}^{\l\Omega\r}(d_\mu^{[\Omega]}) \blank\blank\forall\, i, j \in \mb{N},
\end{equation}
where $\dot\lambda_{ij}^{\l\Omega\r}$ is given by (\ref{E:dlSum}).}
\end{lemma}

\begin{IEEEproof}
If $e_\mu^{[\Omega]}$ exists, (\ref{E:MSSDef}) requires that $e_\mu^{[\Omega]} \leq q^{[\Omega]}$, so at least one $d_\mu^{[\Omega]} \leq q^{[\Omega]}$ exists. If $i>0$, it is immediate from (\ref{E:dlSum}) that (\ref{E:MSS}) holds by default because $\dot\lambda_{ij}^{\l\Omega\r}$ is constant with respect to $d_\mu^{[\Omega]}$. If $i = 0$, an alternative formulation of $\dot\lambda_{0j}$ is useful. Given any $d \leq q$, according to (\ref{E:lambdaBound}), $\dot\lambda_{0j} \geq \hat\lambda_{1, j+1}$. So, using (\ref{E:SigUpdate}) and (\ref{E:p}), we have
\[
\begin{IEEEeqnarraybox}{rCl}
\dot\lambda_{0j} &=& \hat\lambda_{1, j+1} + (\dot\lambda_{0j} - \hat\lambda_{1, j+1})^+\\
                 &=& \hat\lambda_{1, j+1} + (\hat\lambda_{0, j+1} - d - \hat\lambda_{1, j+1})^+\\
                 &=& \hat\lambda_{1, j+1} + (p_{j+1} - d)^+.
\end{IEEEeqnarraybox}
\]
Then, given any $d_\mu^{[\Omega]} \leq q^{[\Omega]}$,
\[
\dot\lambda_{0j}^{\l\Omega\r}(d_\mu^{[\Omega]}) \!=\! \hat\lambda_{1,j+1}^{\l\Omega\r} + \sum_{\omega\in\Omega} (p_{j+1}^\omega-d_\mu^\omega)^+ \!\geq\! \hat\lambda_{1,j+1}^{\l\Omega\r} + (p_{j+1}^{\l\Omega\r}-\mu)^+.
\]
This bound can be achieved for all $j\in\mb{N}$ simultaneously when $d_\mu^{[\Omega]} = e_\mu^{[\Omega]}$ because, if $j+1 \leq j_\mu$, according to (\ref{E:MSSDef}) and (\ref{E:jMu+}), $e_\mu^{[\Omega]} \geq p_{j+1}^{[\Omega]}$ and $\mu \geq p_{j+1}^{\l\Omega\r}$, implying that both sides equal $\hat\lambda_{1,j+1}^{\l\Omega\r}$, while, if $j+1 > j_\mu$, $e_\mu^{[\Omega]} \leq p_{j+1}^{[\Omega]}$ and $\mu < p_{j+1}^{\l\Omega\r}$, implying that both sides equal $\hat\lambda_{1,j+1}^{\l\Omega\r} + p_{j+1}^{\l\Omega\r}-\mu$. So (\ref{E:MSS}) holds.
\end{IEEEproof}

Lemma~\ref{L:MSS} is the key to understanding Theorem~\ref{T:MSS}. According to I2 of Theorem~\ref{T:schedulability}, a valid $d_\mu^{[\Omega]}$ is feasible if and only if $\dot\lambda_{ij}^{\l\Omega\r}(d_\mu^{[\Omega]}) \leq \dot{c}_{ij}$ for all $i,j \in \mb{N}$. Just as (\ref{E:schedulability}) can be rewritten as (\ref{E:rhoSchedulability}), this condition can be rewritten as
\begin{equation}\label{E:sealevel}
c \geq \dot\rho(\Omega|d_\mu^{[\Omega]}) := \max_{i < j} \frac{\dot\lambda_{ij}^{\l\Omega\r}(d_\mu^{[\Omega]})}{j-i}.
\end{equation}
It is immediate from (\ref{E:MSS}) that
\begin{equation}\label{E:sealevel+}
\dot\rho(\Omega|e_\mu^{[\Omega]}) = \min_{d_\mu^{[\Omega]} \leq q^{[\Omega]}} \dot\rho(\Omega|d_\mu^{[\Omega]}).
\end{equation}
Therefore, although the set of valid $d_\mu^{[\Omega]}$'s that satisfy (\ref{E:sealevel}), $\mb{F}_\mu$, grows or shrinks as $c$ increases or decreases, as long as it remains non-empty, $e_\mu^{[\Omega]}$ remains feasible. This explains why the shape of $\mb{F}_\mu$ is irrelevant in Theorem~\ref{T:MSS}.

\begin{IEEEproof}[Proof of Theorem~\ref{T:MSS}]
If $\mb{F}_\mu$ is non-empty, given any $d_\mu^{[\Omega]}\in\mb{F}_\mu$, as $d_\mu^{[\Omega]}$ is feasible, $d_\mu^{[\Omega]} \leq q^{[\Omega]}$, so $\mu = d_\mu^{\l\Omega\r} \leq q^{\l\Omega\r}$. Then, according to the discussion following (\ref{E:jMu+}), $e_\mu^{[\Omega]}$ exists.

By default, (\ref{E:MSSDef}) requires that $e_\mu^{[\Omega]} \leq q^{[\Omega]}$. Given any \mbox{$d_\mu^{[\Omega]}\in\mb{F}_\mu$}, as $d_\mu^{[\Omega]}$ is feasible, $e_\mu^{\l\Omega\r} = \mu = d_\mu^{\l\Omega\r} \leq c$. So $e_\mu^{[\Omega]}$ is valid. Also, as $d_\mu^{[\Omega]}$ is feasible, $\dot\lambda_{ij}^{\l\Omega\r}(d_\mu^{[\Omega]}) \leq \dot{c}_{ij}$ for all $i,j \in \mb{N}$. So (\ref{E:MSS}) guarantees that $\dot\lambda_{ij}^{\l\Omega\r}(e_\mu^{[\Omega]}) \leq \dot{c}_{ij}$ for all $i,j \in \mb{N}$. Then, according to I2 of Theorem~\ref{T:schedulability}, $e_\mu^{[\Omega]}$ is feasible, and $e_\mu^{[\Omega]} \in \mb{F}_\mu$.
\end{IEEEproof}

A corollary of Theorem~\ref{T:MSS} and Lemma~\ref{L:MSS} is that, if $\mb{F}_\mu$ is non-empty, $e_\mu^{[\Omega]}$ is not just any feasible schedule in $\mb{F}_\mu$ but one that minimizes $\dot\lambda_{ij}^{\l\Omega\r}$ for all $i,j \in \mb{N}$ simultaneously. According to Theorem~\ref{T:schedulability}, this implies that, in the next slot, $e_\mu^{[\Omega]}$ leaves maximum room for the server to admit new service requests. So $e_\mu^{[\Omega]}$  maximizes the server's capacity slack, which explains why we call it the max-slack schedule. If any feasible schedule other than $e_\mu^{[\Omega]}$ is selected from $\mb{F}_\mu$, there will be less capacity slack in the next slot, a cost that should be carefully weighed against any potential benefit of such a selection.

According to Theorems~\ref{T:MSS} and \ref{T:FP}, $e_\mu^{[\Omega]}$ is feasible if $\bs\psi^{[\Omega]}$ is schedulable and $\mu$ is feasible. According to (\ref{E:MSSDef}), $e_\mu^{[\Omega]}$ is confined to a hypercuboid that can be easily determined from $\mu$, $\bs{p}^{[\Omega]}$, and $q^{[\Omega]}$ alone. So, to construct a feasible $e_\mu^{[\Omega]}$, no calculation of $\beta_\mu$ is required. This is not surprising because $\beta_\mu$ determines the shape of $\mb{F}_\mu$ which, as already explained, is irrelevant. Of course, to ensure that $e_\mu^{[\Omega]}$ is feasible, according to (\ref{E:muFeasible}), we still need to calculate $\beta(\Omega)$ to ensure that $\mu$ is feasible. But even this calculation can be avoided if we let $\mu=\min\{c, q^{\l\Omega\r}\}$, that is, if we require that the server be {\it \textbf{work-conserving}} in the sense that it always serves as many tasks as possible.\footnote{Work-conservation is generally desirable. But non-work-conserving schedules may be preferred when they can serve other goals, such as: serving best-effort tasks that belong to no existing flow; controlling delay jitter; or in network settings, balancing buffer backlogs.}

\subsection{Per-Class Max-Slack Schedules}\label{SS:HMSS}

To fully exploit the flexibility of selecting any $d_\mu^{[\Omega]}$ from $\mb{P}(\beta_\mu)$, we must calculate all values of $\beta_\mu$. But, if we select $e_\mu^{[\Omega]}$ to avoid almost all calculation, we lose almost all scheduling flexibility. In many cases, what we would really prefer is to find a way to explore the full flexibility-efficiency continuum instead of being limited to either extreme.

To this end, we aggregate flows into classes. Let $\mc{P}$ denote the partition of $\Omega$ according to which flows are aggregated. Given any $\Gamma\in\mc{P}$, when $\Gamma$ is used as an index, as in $x^\Gamma$, we denote the ensemble vector of all $x^\Gamma$'s by $x^{[\mc{P}]}$, and the sum, $\sum_{\Gamma\in\mc{P}} x^\Gamma$, by $x^{\l\mc{P}\r}$. We call $\nu^{[\mc{P}]}$ an {\it \textbf{inter-class schedule}} if, for each $\Gamma\in\mc{P}$, $\nu^\Gamma$ is the total service allocated to flows in $\Gamma$ so that, flow-by-flow, schedules are restricted to
\begin{equation}\label{E:Hnu}
\mb{H}(\nu^{[\mc{P}]}) := \{d^{[\Omega]} | d^{\l\Gamma\r} = \nu^\Gamma ~\forall\, \Gamma\in\mc{P}\}.
\end{equation}
Let $\mu = \nu^{\l\mc{P}\r}$, and for clarity, denote $\nu^{[\mc{P}]}$ by $\nu_\mu^{[\mc{P}]}$. It is immediate that $\mb{H}(\nu_\mu^{[\mc{P}]}) \subseteq \mb{H}_\mu$. Additionally, let $\mb{F}(\nu_\mu^{[\mc{P}]}) := \mb{F} \cap \mb{H}(\nu_\mu^{[\mc{P}]})$. So $\mb{F}(\nu_\mu^{[\mc{P}]}) \subseteq \mb{F}_\mu$.

Roughly speaking, given $\nu_\mu^{[\mc{P}]}$, for each $\Gamma\in\mc{P}$, we can construct a max-slack schedule under $\nu_\mu^\Gamma$ for $\bs\psi^{[\Gamma]}$, where $\bs\psi^{[\Gamma]}$ denotes the subvector, and thus the subsystem, of $\bs\psi^{[\Omega]}$ indexed by $\Gamma$. By concatenating all such schedules for all $\bs\psi^{[\Gamma]}$'s, flows are, inter-class, scheduled according to $\nu_\mu^{[\mc{P}]}$ and, intra-class, max-slack scheduled. This motivates the following extensions of Definition~\ref{D:MSS} and Theorem~\ref{T:MSS}.

\begin{definition}\label{D:HMSS}
{\it $e^{[\Omega]}(\nu_\mu^{[\mc{P}]}) \in \mb{H}(\nu_\mu^{[\mc{P}]})$ is a \textbf{per-class max-slack schedule} under $\nu_\mu^{[\mc{P}]}$ if, for each $\Gamma\in\mc{P}$,
\begin{equation}\label{E:HMSSDef}
e^{[\Gamma]}(\nu_\mu^{[\mc{P}]}) \in\left\{
\begin{IEEEeqnarraybox}[][c]{ll}
[p_{j_\mu^\Gamma}^{[\Gamma]} , p_{j_\mu^\Gamma+1}^{[\Gamma]}] & \blank\blank\textup{if } j_\mu^\Gamma < \infty\\
{[}p_\infty^{[\Gamma]} , q^{[\Gamma]}]                        & \blank\blank\textup{if } j_\mu^\Gamma = \infty
\end{IEEEeqnarraybox}\right. ~,
\end{equation}
where
\begin{equation}\label{E:jT}
j_\mu^\Gamma := \tau_{\nu_\mu^\Gamma+1}(\bs{p}^{\l\Gamma\r}).
\end{equation}}
\end{definition}

\begin{theorem}\label{T:HMSS}
{\it If $\mb{F}(\nu_\mu^{[\mc{P}]})$ is non-empty, $e^{[\Omega]}(\nu_\mu^{[\mc{P}]})$ exists, and $e^{[\Omega]}(\nu_\mu^{[\mc{P}]}) \in \mb{F}(\nu_\mu^{[\mc{P}]})$.}
\end{theorem}

To prove this theorem, we need the following extension of Lemma~\ref{L:MSS}. Its proof is omitted because it replicates, almost verbatim, Lemma~\ref{L:MSS}'s proof.

\begin{lemma}\label{L:HMSS}
{\it If $e^{[\Omega]}(\nu_\mu^{[\mc{P}]})$ exists, for each $\Gamma\in\mc{P}$,
\begin{equation}\label{E:HMSS}
\dot\lambda_{ij}^{\l\Gamma\r}(e^{[\Gamma]}(\nu_\mu^{[\mc{P}]})) = \min_{d^{[\Gamma]} \leq q^{[\Gamma]}, d^{\l\Gamma\r} = \nu_\mu^\Gamma} \dot\lambda_{ij}^{\l\Gamma\r}(d^{[\Gamma]}) \blank\blank\forall\, i,j \in \mb{N},
\end{equation}
where, according to (\ref{E:SigUpdate}),
\begin{equation}\label{E:HMSSSum}
\dot\lambda_{ij}^{\l\Gamma\r}(d^{[\Gamma]}) = \left\{
\begin{IEEEeqnarraybox}[][c]{ll}
\textstyle \sum_{\omega\in\Gamma} (\hat\lambda_{0, j+1}^\omega-d^\omega)^+ & \blank\blank\textup{if } i=0\\
\hat\lambda_{i+1, j+1}^{\l\Gamma\r}                                        & \blank\blank\textup{if } i>0
\end{IEEEeqnarraybox}\right. ~.
\end{equation}}
\end{lemma}

Using (\ref{E:sealevel}) and (\ref{E:HMSS}), it is easy to verify that
\begin{equation}\label{E:sealevel++}
\dot\rho(\Omega|e^{[\Omega]}(\nu_\mu^{[\mc{P}]})) = \min_{d_\mu^{[\Omega]} \leq q^{[\Omega]}, d_\mu^{[\Omega]}\in\mb{H}(\nu_\mu^{[\mc{P}]})} \dot\rho(\Omega|d_\mu^{[\Omega]}),
\end{equation}
which extends (\ref{E:sealevel+}). Therefore, as long as the set of valid $d_\mu^{[\Omega]}$'s in $\mb{H}(\nu_\mu^{[\mc{P}]})$ that satisfy (\ref{E:sealevel}), $\mb{F}(\nu_\mu^{[\mc{P}]})$, is non-empty, $e^{[\Omega]}(\nu_\mu^{[\mc{P}]})$ is feasible.

\begin{IEEEproof}[Proof of Theorem~\ref{T:HMSS}]
If $\mb{F}(\nu_\mu^{[\mc{P}]})$ is non-empty, given any $d_\mu^{[\Omega]}\in\mb{F}(\nu_\mu^{[\mc{P}]})$, as $d_\mu^{[\Omega]}$ is feasible, $d_\mu^{[\Omega]} \leq q^{[\Omega]}$. So, for each $\Gamma\in\mc{P}$, $\nu_\mu^\Gamma = d_\mu^{\l\Gamma\r} \leq q^{\l\Gamma\r}$. Then, using (\ref{E:jT}) and (\ref{E:tau}), it is easy to verify that
\[
\nu_\mu^\Gamma\in\left\{
\begin{IEEEeqnarraybox}[][c]{ll}
[p_{j_\mu^\Gamma}^{\l\Gamma\r}, p_{j_\mu^\Gamma+1}^{\l\Gamma\r}) & \blank\blank\text{if } j_\mu^\Gamma < \infty\\
{[}p_\infty^{\l\Gamma\r}, q^{\l\Gamma\r}]                        & \blank\blank\text{if } j_\mu^\Gamma = \infty
\end{IEEEeqnarraybox}\right. ~.
\]
Comparing this to (\ref{E:HMSSDef}), since $\nu_\mu^\Gamma$, $\bs{p}^{[\Gamma]}$, and $q^{[\Gamma]}$ are all integral, there exists at least one way to split $\nu_\mu^\Gamma$ integrally into $e^{[\Gamma]}(\nu_\mu^{[\mc{P}]})$. So, as all of $e^{[\Omega]}(\nu_\mu^{[\mc{P}]})$'s subvectors exist, by concatenation, $e^{[\Omega]}(\nu_\mu^{[\mc{P}]})$ exists.

By default, (\ref{E:HMSSDef}) requires that $e^{[\Omega]}(\nu_\mu^{[\mc{P}]}) \leq q^{[\Omega]}$. Given any $d_\mu^{[\Omega]}\in\mb{F}(\nu_\mu^{[\mc{P}]})$, as $d_\mu^{[\Omega]}$ is feasible, $e^{\l\Omega\r}(\nu_\mu^{[\mc{P}]}) = \nu_\mu^{\l\mc{P}\r} = \mu = d_\mu^{\l\Omega\r} \leq c$. So $e^{[\Omega]}(\nu_\mu^{[\mc{P}]})$ is valid. Also, as $d_\mu^{[\Omega]}$ is feasible, $\dot\lambda_{ij}^{\l\Omega\r}(d_\mu^{[\Omega]}) \leq \dot{c}_{ij}$ for all $i,j \in \mb{N}$. So, by concatenation, (\ref{E:HMSS}) guarantees that $\dot\lambda_{ij}^{\l\Omega\r}(e^{[\Omega]}(\nu_\mu^{[\mc{P}]})) \leq \dot{c}_{ij}$ for all $i,j \in \mb{N}$. Then, according to I2 of Theorem~\ref{T:schedulability}, $e^{[\Omega]}(\nu_\mu^{[\mc{P}]})$ is feasible, and $e^{[\Omega]}(\nu_\mu^{[\mc{P}]}) \in \mb{F}(\nu_\mu^{[\mc{P}]})$.
\end{IEEEproof}

We call $\nu_\mu^{[\mc{P}]}$ {\it \textbf{feasible}} if $\mb{F}(\nu_\mu^{[\mc{P}]})$ is non-empty. According to Theorem~\ref{T:HMSS}, for every feasible $\nu_\mu^{[\mc{P}]}$, there exists a feasible per-class max-slack schedule. But how do we know if $\mb{F}(\nu_\mu^{[\mc{P}]})$ is non-empty?

\subsection{$\mb{F}(\nu_\mu^{[\mc{P}]})$'s Non-Emptiness Condition}\label{SS:Q3}

By definition, $\mb{F}(\nu_\mu^{[\mc{P}]}) \subseteq \mb{F}_\mu$, while, according to Theorem~\ref{T:FP}, $\mb{F}_\mu = \mb{P}(\beta_\mu)$. Therefore, if $\mb{F}(\nu_\mu^{[\mc{P}]})$ is non-empty, given any $d_\mu^{[\Omega]} \in \mb{F}(\nu_\mu^{[\mc{P}]})$, $d_\mu^{[\Omega]} \in \mb{P}(\beta_\mu)$. That is to say, $d_\mu^{[\Omega]}$ must satisfy (\ref{E:permuBetaMu}), so
\[
\nu_\mu^{\l\mc{P}\r} = d_\mu^{\l\Omega\r} = \mu = \beta_\mu(\Omega),
\]
and
\[
\nu_\mu^{\l\mc{S}\r} = \sum_{\Gamma\in\mc{S}} d_\mu^{\l\Gamma\r} = d_\mu^{\l\Sigma(\mc{S})\r} \geq \beta_\mu(\Sigma(\mc{S})) \blank\blank\forall\, \mc{S} \subseteq \mc{P},
\]
where $\nu_\mu^{\l\mc{S}\r}$ denotes $\sum_{\Gamma\in\mc{S}} \nu_\mu^\Gamma$ and $\Sigma(\mc{S})$ denotes $\bigcup_{\Gamma\in\mc{S}} \Gamma$. Now, let
\begin{equation}\label{E:betaP}
\beta_\mu^\mc{P}(\mc{S}) := \beta_\mu(\Sigma(\mc{S})) \blank\blank\forall\, \mc{S} \subseteq \mc{P}.
\end{equation}
Then,
\begin{equation}\label{E:permuBetaP}
\nu_\mu^{\l\mc{P}\r} = \mu = \beta_\mu^\mc{P}(\mc{P}) \blank\text{and}\blank \nu_\mu^{\l\mc{S}\r} \geq \beta_\mu^\mc{P}(\mc{S}) ~\forall\, \mc{S} \subseteq \mc{P}.
\end{equation}
By definition, $\beta_\mu^\mc{P}$ is the sampling of $\beta_\mu$ on the algebra generated by $\mc{P}$. It is easy to verify that, like $\beta_\mu$, $\beta_\mu^\mc{P}$ is supermodular. So, according to (\ref{E:permuBetaP}), a necessary condition for $\nu_\mu^{[\mc{P}]}$ to be feasible is that $\nu_\mu^{[\mc{P}]} \in \mb{P}(\beta_\mu^\mc{P})$. It turns out that this condition is also sufficient.

\subsubsection*{\textup{[}A Special Case\textup{]}} Consider first, the case that $\nu_\mu^{[\mc{P}]}$ is a vertex of $\mb{P}(\beta_\mu^\mc{P})$. Recall, from Section~\ref{SS:polymatroid}, that each vertex of $\mb{P}(\beta_\mu^\mc{P})$ can be identified by a permutation over $\mc{P}$. Let $m = |\mc{P}|$. Then, given a permutation, $\sigma: \mc{P} \rightarrow \{1, 2, \ldots, m\} \in \Pi^\mc{P}$, where $\Pi^\mc{P}$ denotes the set of all permutations over $\mc{P}$, the vertex identified by $\sigma$, $v_\sigma^{[\mc{P}]}(\beta_\mu^\mc{P})$, is the unique solution to the system of linear equations defined by
\begin{equation}\label{E:vertexSumExt}
v_\sigma^{\l\mc{S}_\sigma^j\r} (\beta_\mu^\mc{P}) = \beta_\mu^\mc{P}(\mc{S}_\sigma^j) \blank\blank\forall\, 0 \leq j \leq m,
\end{equation}
where
\begin{equation}\label{E:Spi}
\mc{S}_\sigma^j := \{\Gamma\in\mc{P} | \sigma(\Gamma) \leq j\}.
\end{equation}
The next theorem shows that $\mb{F}(v_\sigma^{[\mc{P}]}(\beta_\mu^\mc{P}))$, the set of feasible schedules that, inter-class, are scheduled according to $v_\sigma^{[\mc{P}]}(\beta_\mu^\mc{P})$, is a non-empty face of $\mb{P}(\beta_\mu)$.

\begin{theorem}\label{T:HMSSSpecial}
{\it If $\bs\psi^{[\Omega]}$ is schedulable and $\mu$ is feasible,
\begin{equation}\label{E:Fvertex}
\mb{F}(v_\sigma^{[\mc{P}]}(\beta_\mu^\mc{P})) = \mb{P}_{\mc{C}_\sigma^{\mc{P}}}(\beta_\mu) \blank\blank\forall\, \sigma \in \Pi^\mc{P},
\end{equation}
where $\mb{P}_{\mc{C}_\sigma^{\mc{P}}}(\beta_\mu)$ is the face of $\mb{P}(\beta_\mu)$ identified by
\begin{equation}\label{E:Csigma}
\mc{C}_\sigma^{\mc{P}}:= \{\Sigma(\mc{S}_\sigma^j)\}_{j=0}^m,
\end{equation}
which is a chain because, by definition, $\phi = \Sigma(\mc{S}_\sigma^0) \subset \Sigma(\mc{S}_\sigma^1) \subset \cdots \subset \Sigma(\mc{S}_\sigma^m) = \Omega$.
}
\end{theorem}

\begin{IEEEproof}
Using (\ref{E:Hnu}), (\ref{E:Spi}), (\ref{E:vertexSumExt}), (\ref{E:betaP}), and (\ref{E:Csigma}) consecutively, we have
\[
\begin{IEEEeqnarraybox}[][c]{rCl}
\mb{H}(v_\sigma^{[\mc{P}]}(\beta_\mu^\mc{P})) &=& \{d_\mu^{[\Omega]} | d_\mu^{\l\Gamma\r} = v_\sigma^\Gamma(\beta_\mu^\mc{P}) ~\forall\, \Gamma\in\mc{P}\}\\
&=& \{d_\mu^{[\Omega]} | d_\mu^{\l\Sigma(\mc{S}_\sigma^j)\r} = v_\sigma^{\l\mc{S}_\sigma^j\r}(\beta_\mu^\mc{P}) ~\forall\, 0 \leq j \leq m\}\\
&=& \{d_\mu^{[\Omega]} | d_\mu^{\l\Sigma(\mc{S}_\sigma^j)\r} = \beta_\mu^\mc{P}(\mc{S}_\sigma^j) ~\forall\, 0 \leq j \leq m\}\\
&=& \{d_\mu^{[\Omega]} | d_\mu^{\l\Sigma(\mc{S}_\sigma^j)\r} = \beta_\mu(\Sigma(\mc{S}_\sigma^j)) ~\forall\, 0 \leq j \leq m\}\\
&=& \{d_\mu^{[\Omega]} | d_\mu^{\l\Gamma\r} = \beta_\mu(\Gamma) ~\forall\, \Gamma\in\mc{C}_\sigma^\mc{P}\},
\end{IEEEeqnarraybox}
\]
where the second equality holds because the linear transformation involved is invertible. Then, $\mb{F}(v_\sigma^{[\mc{P}]}(\beta_\mu^\mc{P}))$'s definition, Theorem~\ref{T:FP}, and (\ref{E:P-S}) imply that
\[
\begin{IEEEeqnarraybox}[][c]{rCl}
\mb{F}(v_\sigma^{[\mc{P}]}(\beta_\mu^\mc{P})) &=& \mb{P}(\beta_\mu) \cap \mb{H}(v_\sigma^{[\mc{P}]}(\beta_\mu^\mc{P}))\\
                                              &=& \{d_\mu^{[\Omega]} \in \mb{P}(\beta_\mu)|d_\mu^{\l\Gamma\r} = \beta_\mu(\Gamma) ~\forall\, \Gamma\in\mc{C}_\sigma^\mc{P}\}\\
                                              &=& \mb{P}_{\mc{C}_\sigma^\mc{P}}(\beta_\mu).
\end{IEEEeqnarraybox}
\]
\end{IEEEproof}

\subsubsection*{\textup{[}The General Case\textup{]}}

Using convexity, Theorem~\ref{T:HMSSSpecial} can be generalized.

\begin{theorem}\label{T:HMSSCond}
{\it If $\bs\psi^{[\Omega]}$ is schedulable and $\mu$ is feasible, $\nu_\mu^{[\mc{P}]}$ is feasible if and only if $\nu_\mu^{[\mc{P}]} \in \mb{P}(\beta_\mu^\mc{P})$.}
\end{theorem}

\begin{IEEEproof}\footnote{\label{F:non-empty+}An alternative proof can be constructed using Frank's sandwich theorem \cite{Frank:1982}. See also, \cite{Schrijver:2003} (p. 799). Let
\[
\gamma_\mu(\Gamma) := \min_{\mc{S} \subseteq \mc{P}, \Sigma(\mc{S}) \supseteq \Gamma} \nu_\mu^{\l\mc{S}\r} \blank\blank\forall\, \Gamma \subseteq \Omega.
\]
It can be verified that $\gamma_\mu$ is a submodular function, that $\gamma_\mu \geq \beta_\mu$, and that $d_\mu^{[\Omega]} \in \mb{F}(\nu_\mu^{[\mc{P}]})$ if and only if $\gamma_\mu(\Gamma) \geq d_\mu^{\l\Gamma\r} \geq \beta_\mu(\Gamma)$ for all $\Gamma \subseteq \Omega$. According to Frank's theorem, these facts guarantee that $\mb{F}(\nu^{[\mc{P}]})$ is non-empty. In contrast, our treatment is self-contained, and reveals more structural information than just the non-emptiness of $\mb{F}(\nu_\mu^{[\mc{P}]})$.} The condition's necessity follows directly from (\ref{E:permuBetaP}), so we need only show its sufficiency. If $\nu_\mu^{[\mc{P}]} \in \mb{P}(\beta_\mu^\mc{P})$, it must be a convex combination of $\mb{P}(\beta_\mu^\mc{P})$'s vertices, that is, there must exist, for all $\sigma \in \Pi^\mc{P}$, $w_\sigma \in [0, 1]$ in $\mb{R}$ such that
\[
\sum_{\sigma\in\Pi^\mc{P}} w_\sigma = 1 \blank\text{and}\blank \nu_\mu^{[\mc{P}]} = \sum_{\sigma\in\Pi^\mc{P}} w_\sigma v_\sigma^{[\mc{P}]}(\beta_\mu^\mc{P}).\footnote{The weighting, $\{w_\sigma\}_{\sigma\in\Pi^\mc{P}}$, is not unique. For instance, it can be shown that it is always possible to set $w_\sigma = \frac{\xi_\sigma}{\sum_{\sigma'\in\Pi^\mc{P}} \xi_{\sigma'}}$, where, still letting $m = |\mc{P}|$,
\[
\xi_\sigma := \prod_{j=1}^{m-1} \frac{1}{\nu^{\l\mc{S}_\sigma^j\r}-v_\sigma^{\l\mc{S}_\sigma^j\r}(\beta_\mu^\mc{P})} = \prod_{j=1}^{m-1} \frac{1}{\nu^{\l\mc{S}_\sigma^j\r}-\beta_\mu^\mc{P}(\mc{S}_\sigma^j)}.
\]
}
\]
For each $\sigma \in \Pi^\mc{P}$, select any $d_\sigma^{[\Omega]} \in \mb{P}_{\mc{C}_\sigma^{\mc{P}}}(\beta_\mu)$. It is immediate from (\ref{E:Fvertex}) that $d_\sigma^{[\Omega]} \in \mb{F}_\mu$ and $d_\sigma^{[\Omega]} \in \mb{H}(v_\sigma^{[\mc{P}]}(\beta_\mu^\mc{P}))$. Construct
\[
d_\mu^{[\Omega]} = \sum_{\sigma\in\Pi^\mc{P}} w_\sigma d_\sigma^{[\Omega]}.
\]
On the one hand, since $d_\sigma^{[\Omega]} \in \mb{F}_\mu$, $d_\mu^{[\Omega]} \in \mb{F}_\mu$. On the other, since $d_\sigma^{[\Omega]} \in \mb{H}(v_\sigma^{[\mc{P}]}(\beta_\mu^\mc{P}))$, for all $\Gamma\in\mc{P}$, using (\ref{E:Hnu}), we have
\[
d_\mu^{\l\Gamma\r} = \sum_{\sigma\in\Pi^\mc{P}} w_\sigma d_\sigma^{\l\Gamma\r} = \sum_{\sigma\in\Pi^\mc{P}} w_\sigma v_\sigma^\Gamma(\beta_\mu^\mc{P}).
\]
But by $w_\sigma$'s definition, we also have
\[
\nu_\mu^\Gamma = \sum_{\sigma\in\Pi^\mc{P}} w_\sigma v_\sigma^\Gamma(\beta_\mu^\mc{P}).
\]
So $d_\mu^{\l\Gamma\r} = \nu_\mu^\Gamma$ for all $\Gamma\in\mc{P}$, that is, $d_\mu^{[\Omega]} \in \mb{H}(\nu_\mu^{[\mc{P}]})$. It follows that $d_\mu^{[\Omega]} \in \mb{F}(\nu_\mu^{[\mc{P}]})$, $\mb{F}(\nu_\mu^{[\mc{P}]})$ is non-empty, and $\nu_\mu^{[\mc{P}]}$ is feasible.
\end{IEEEproof}

One subtlety does need to be addressed. Close examination of Theorem~\ref{T:HMSSCond}'s proof reveals that what is proved is that $\mb{F}(\nu_\mu^{[\mc{P}]})$ is non-empty in $\mb{R}^n$, not that it contains any integral point. This, however, is not an obstacle to the joint application of Theorems~\ref{T:HMSS} and \ref{T:HMSSCond}. In fact, nowhere is the integrality of $\mb{F}(\nu_\mu^{[\mc{P}]})$ ever used in Theorem~\ref{T:HMSS}'s proof, including the omitted proof of Lemma~\ref{L:HMSS} that replicates that of Lemma~\ref{L:MSS} almost verbatim. So Theorem~\ref{T:HMSS} holds as long as $\mb{F}(\nu_\mu^{[\mc{P}]})$ is non-empty in $\mb{R}^n$.\footnote{In light of this observation, $\mb{F}(\nu_\mu^{[\mc{P}]})$'s non-emptiness in $\mb{R}^n$ is not only necessary but also sufficient for its integrality. If $\mb{F}(\nu_\mu^{[\mc{P}]})$ is non-empty in $\mb{R}^n$, Theorem~\ref{T:HMSS} guarantees that $e^{[\Omega]}(\nu_\mu^{[\mc{P}]}) \in \mb{F}(\nu_\mu^{[\mc{P}]})$. But, by definition, $e^{[\Omega]}(\nu_\mu^{[\mc{P}]})$ is integral. So $\mb{F}(\nu_\mu^{[\mc{P}]})$ contains at least one integral point. This result can also be established directly. For instance, if we apply Frank's sandwich theorem as suggested in footnote~\ref{F:non-empty+}, it follows directly from the discrete part of that theorem.}

\subsection{Selecting Per-Class Max-Slack Schedules}\label{SS:howHMSS}

According to Theorems~\ref{T:HMSS} and \ref{T:HMSSCond}, to select a per-class max-slack schedule, we can first select a feasible $\mu$, then select a feasible $\nu_\mu^{[\mc{P}]}$ from $\mb{P}(\beta_\mu^\mc{P})$, and finally construct $e^{[\Omega]}(\nu_\mu^{[\mc{P}]})$, which, according to Definition~\ref{D:HMSS}, is confined to a hypercuboid that can be easily determined from $\nu_\mu^{[\mc{P}]}$, $\bs{p}^{[\Omega]}$ and $q^{[\Omega]}$ alone. This makes possible intermediate tradeoffs between flexibility and efficiency. Given $m = |\mc{P}|$, to fully determine $\mb{P}(\beta_\mu^\mc{P})$, according to (\ref{E:betaP}), $2^m$ evaluations of $\beta_\mu$ are required. As $m$ decreases, $\mb{P}(\beta_\mu^\mc{P})$ becomes a coarser projection of $\mb{P}(\beta_\mu)$ and easier to determine, but we lose flexibility. In contrast, as $m$ increases, $\mb{P}(\beta_\mu^\mc{P})$ becomes a finer projection of $\mb{P}(\beta_\mu)$ and harder to determine, but we gain flexibility.

Recall from Section~\ref{SS:how} that, by selecting a vertex of $\mb{P}(\beta_\mu)$, we can serve flows according to a priority order. Similarly, by selecting vertex $v_\sigma^{[\mc{P}]}(\beta_\mu^\mc{P})$, and thus
\begin{equation}\label{E:HMSSvertex}
e_\sigma^{[\Omega]}(\beta_\mu^\mc{P}) := e^{[\Omega]}(v_\sigma^{[\mc{P}]}(\beta_\mu^\mc{P})),
\end{equation}
an inter-class priority order is enforced such that the larger $\sigma(\Gamma)$ is, the higher the priority that class $\Gamma$ enjoys.

\begin{example}
{\it In the case that $\mc{P} = \{\Gamma,\overline\Gamma\}$, $\mb{P}(\beta_\mu^\mc{P})$ is a line segment with a vertex at each end. At one vertex,
\[
(\nu_\mu^\Gamma, \nu_\mu^{\overline\Gamma}) = (\mu-\beta_\mu(\overline\Gamma), \beta_\mu(\overline\Gamma)),
\]
which assigns to class $\Gamma$ the highest priority possible relative to class $\overline\Gamma$. At the other,
\[
(\nu_\mu^\Gamma, \nu_\mu^{\overline\Gamma}) = (\beta_\mu(\Gamma), \mu-\beta_\mu(\Gamma)),
\]
which reverses the priority assignment to favor $\overline\Gamma$. Of course, to fine-tune the balance, any integral point along the line segment could be selected.}
\end{example}

Recall also from Section~\ref{SS:how} that, by selecting the vertex centroid of $\mb{P}(\beta_\mu)$, we can serve flows according to a fairness criterion. Similarly, by selecting the vertex centroid,
\begin{equation}\label{E:u-f}
v_\text{F}^{[\mc{P}]}(\beta_\mu^\mc{P}) = \frac{1}{m!} \sum_{\sigma\in\Pi^\mc{P}} v_\sigma^{[\mc{P}]}(\beta_\mu^\mc{P}),
\end{equation}
and thus
\begin{equation}\label{E:v-f}
e_\text{F}^{[\Omega]}(\beta_\mu^\mc{P}) := e^{[\Omega]}(v_\text{F}^{[\mc{P}]}(\beta_\mu^\mc{P})),
\end{equation}
where $v_\text{F}^{[\mc{P}]}(\beta_\mu^\mc{P})$ needs to be rounded if it is not an integral point, an inter-class fairness criterion is enforced. But this fairness is not intra-class. In fact, a flow can still be starved by other flows in the same class. To address this concern, an alternative approach is inspired by the multi-level feedback queue widely used in operating systems \cite{ArpaciDusseau:2018} (ch. 8). The idea is to fix an inter-class priority order, but allow each flow's priority to be adjusted dynamically. In particular, if a flow has been starved, it will be moved to a higher-priority class, while if a flow has been served more than adequately, it will be moved to a lower-priority class.

\subsection{Worst-Case Deadlines}\label{SS:deadline}

So far all of our discussion has been flow-centric in the sense that it has focused on how many tasks from each flow should be served. In the case of max-slack schedules, it is also enlightening to take a task-centric view.

Consider the $h$th task of a generic flow. According to (\ref{E:tau}), $\tau_h(\bs{p})$ is non-decreasing with respect to $h$. Using (\ref{E:tau}) and (\ref{E:MSSDef}), it is also easy to verify that
\begin{equation}\label{E:tauMSS}
\forall\, 0 < h \leq q, \blank\blank
\left\{
\begin{IEEEeqnarraybox}[][c]{ll}
h \leq e_\mu & \blank\blank\text{if } \tau_h(\bs{p}) < j_\mu\\
h > e_\mu    & \blank\blank\text{if } \tau_h(\bs{p}) > j_\mu
\end{IEEEeqnarraybox}\right. ~.
\end{equation}
That is to say, if $e_\mu$ tasks are served, the $h$th task is served if $\tau_h(\bs{p}) < j_\mu$, but is not served if $\tau_h(\bs{p}) > j_\mu$.\footnote{If $\tau_h(\bs{p}) = j_\mu$, the $h$th task can either be served or not served.} As $j_\mu$ is constant across all flows, this implies that max-slack schedules must serve tasks in non-decreasing order of their respective $\tau_h(\bs{p})$ values. Conversely, it can be verified that serving tasks in this order always results in a max-slack schedule.

To see why max-slack scheduled tasks' $\tau_h(\bs{p})$ values dictate their service order, it is helpful to relate $\tau_h(\bs{p})$ to $\bs\psi(\bs{q})$. For each $0 < h \leq q$, using (\ref{E:tau}) and (\ref{E:p++}), we have
\begin{IEEEeqnarray}{rCl}
\tau_h(\bs{p}) &=& \max\{j \in \mb{N}^+ |{p_j = \textstyle \max_{\bs{q} \in \mb{U}|q}} \min\{\psi_j(\bs{q}), q\} < h\} \IEEEnonumber\\
               &=& \max\{j \in \mb{N}^+ |\psi_j(\bs{q}) < h  ~\forall\, \bs{q} \in \mb{U}|q\}            \IEEEnonumber\\
               &=& \max\{j \in \mb{N}^+ |j \leq \tau_h(\bs\psi(\bs{q})) ~\forall\, \bs{q} \in \mb{U}|q\} \IEEEnonumber\\
               &=& \min_{\bs{q} \in \mb{U}|q} \tau_h(\bs\psi(\bs{q})),                           \label{E:deadline}
\end{IEEEeqnarray}
where the third equality holds because, by definition, $\psi_j(\bs{q}) < h$ if and only if $j \leq \tau_h(\bs\psi(\bs{q}))$. According to (\ref{E:WCSDef+}) and (\ref{E:tau}), to guarantee $\bs{d} \geq \bs\psi(\bs{q})$, the $h$th task must be served no later than slot $t+\tau_h(\bs\psi(\bs{q}))$. So, to guarantee $\bs{d} \geq \bs\psi(\bs{q})$ no matter which $\bs{q} \in \mb{U}|q$ is realized, (\ref{E:deadline}) implies that it must be served no later than slot $t+\tau_h(\bs{p})$. In view of this observation, we call $t+\tau_h(\bs{p})$ the {\it \textbf{worst-case deadline}} for the $h$th task, and (\ref{E:tauMSS}) says that max-slack schedules must serve tasks in non-decreasing order of their respective worst-case deadlines.

Notice that, in general, worst-case deadlines are dynamic. But, unless a task's service guarantee is upgraded during its time in the queue, its worst-case deadline can only be relaxed because, well, the worst case cannot get worse. To see this, suppose that $q \geq h > d \geq p$. In this case, the $h$th task in slot~$t$ becomes the $(h-d)$th task in $t+1$. Accordingly, its worst-case deadline becomes $t+1+\tau_{h-d}(\bs{\dot{p}})$. Using (\ref{E:deadline}), (\ref{E:tau}) and (\ref{E:WCSUpdateB}), we have
\[
\begin{IEEEeqnarraybox}[][c]{Cl}
  & \tau_{h-d}(\bs{\dot{p}}) \\
= & \min_{\bs{\dot{q}} \in \mb{U}|\dot{q}} \tau_{h-d}(\bs{\dot\psi}(\bs{\dot{q}})) \\
= & \min_{\bs{\dot{q}} \in \mb{U}|\dot{q}} \max\{j\in\mb{N}^+|\dot\psi_j(\bs{\dot{q}}) < h-d\} \\
= & \min_{\bs{q} \in \mb{U}|q, q_2=\dot{q}+d} \!\!\!\max\{j\in\mb{N}^+|[\mc{R}^{-1}(\bs\psi(\bs{q})-d\bs\delta)^+]_j < h-d\},
\end{IEEEeqnarraybox}
\]
where the final equality holds because, according to (\ref{E:vecqUpdate}) and (\ref{E:qUpdate++}), $\bs{\dot{q}} \in \mb{U}|\dot{q}$ is equivalent to $\bs{q} \in \mb{U}|q$ and $q_2=\dot{q}+d$. Using (\ref{E:LShift}) and (\ref{E:tau}), we then have
\begin{IEEEeqnarray}{Cl}
  & 1+\tau_{h-d}(\bs{\dot{p}}) \IEEEnonumber\\
= & 1+\min_{\bs{q} \in \mb{U}|q, q_2=\dot{q}+d} \max\{j\in\mb{N}^+|\psi_{j+1}(\bs{q})-d < h-d\}                 \IEEEnonumber\\
= & \min_{\bs{q} \in \mb{U}|q, q_2=\dot{q}+d} \max\{j\in\mb{N}^+\setminus\{0\}|\psi_{j}(\bs{q}) < h\} \IEEEnonumber\\
= & \min_{\bs{q} \in \mb{U}|q, q_2=\dot{q}+d} \tau_h(\bs\psi(\bs{q})),           \label{E:deadline+}
\end{IEEEeqnarray}
where the final equality holds because, according to (\ref{E:WCSImmediate}), \mbox{$h > p$} implies that $\psi_1(\bs{q}) < h$ for all $\bs{q} \in \mb{U}|q$, which in turn implies that the same maximum is obtained when searching over $\mb{N}^+$ as $\mb{N}^+\setminus\{0\}$. Comparing (\ref{E:deadline+}) to (\ref{E:deadline}), it is immediate that $1+\tau_{h-d}(\bs{\dot{p}}) \geq \tau_h(\bs{p})$. So, unless its service guarantee is upgraded, the task's worst-case deadline can only be relaxed.

\subsection{Deadline-Rigid Services}\label{SS:deadline+}

The task-centric interpretation of max-slack schedules suggests a strong connection to EDF schedules, that is, schedules that serve tasks in non-decreasing order of static deadlines assigned to each task upon arrival. The difference is that worst-case deadlines are dynamic. There are, however, worst-case services for which the worst-case deadlines are static.

\begin{definition}\label{D:DDS}
{\it A worst-case service, $\bs\psi$, is \textbf{deadline-rigid} if
\begin{equation}\label{E:DDS}
\forall\, \bs{q}, \bs{q'} \in \mb{U}\ua b, h \in \mb{N},\blank\blank \bs\psi(\bs{q}) \stackrel{[h]}{=} \bs\psi(\bs{q'}) \textup{ if } \bs{q} \stackrel{[h]}{=} \bs{q'},
\end{equation}
where $\bs{q} \stackrel{[h]}{=} \bs{q'}$ denotes $\min\{\bs{q}, h\bs\delta\} = \min\{\bs{q'}, h\bs\delta\}$.}\footnote{This relation is distinct from $\bs{q} \stackrel{j}{=} \bs{q'}$, which holds if $q_i=q'_i$ for all $i \leq j$. It is easy to verify that $\bs{q} \stackrel{[q_j]}{=} \bs{q'}$ if $\bs{q} \stackrel{j}{=} \bs{q'}$, and  $\bs{q} \stackrel{j}{=} \bs{q'}$ if $\bs{q} \stackrel{[q_j+1]}{=} \bs{q'}$. So, roughly speaking, $\bs{q} \stackrel{j}{=} \bs{q'}$ lies somewhere between $\bs{q} \stackrel{[q_j]}{=} \bs{q'}$ and $\bs{q} \stackrel{[q_j+1]}{=} \bs{q'}$.}
\end{definition}

\begin{figure}[t]
\centering \scalebox{0.875}{\includegraphics{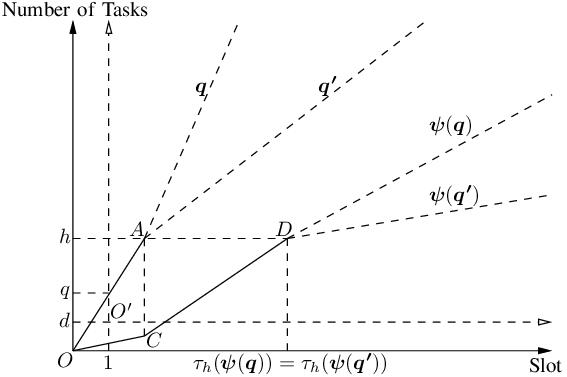}} \caption{Properties of a deadline-rigid service.}\label{F:Deadline}
\end{figure}

The intuition underlying this definition is illustrated in Fig.~\ref{F:Deadline}. If $\bs{q}$ and $\bs{q'}$ are identical up to point~$A$, deadline-rigidness requires that $\bs\psi(\bs{q})$ and $\bs\psi(\bs{q'})$ be identical up to point~$D$. Since causality would only require them to be identical up to point~$C$, an immediate corollary is that deadline-rigid services are causal, but not vice versa. For instance, it is easy to verify that the uniform-delay services in Example~\ref{Ex:CDS} are deadline-rigid while the uniform-backlog services in Example~\ref{Ex:CBS} are not, although both are causal. In contrast, since single-task services are non-causal, they are not deadline-rigid. In fact, even their causal counterparts are not deadline-rigid, which is easy to verify using (\ref{E:STSCausal}).

If $\bs\psi$ is deadline-rigid, on the one hand, using (\ref{E:tau}) and (\ref{E:DDS}), it is easy to verify that $\tau_h(\bs\psi(\bs{q})) = \tau_h(\bs\psi(\bs{q'}))$ if $\bs{q} \stackrel{[h]}{=} \bs{q'}$, as illustrated in Fig.~\ref{F:Deadline}. On the other hand, for all $q \geq h > d$, by definition, $\bs{q} \stackrel{[h]}{=} q\bs\delta$ for all $\bs{q} \in \mb{U}|q$. Therefore, according to (\ref{E:deadline}) and (\ref{E:deadline+}),
\[
1+\tau_{h-d}(\bs{\dot{p}}) = \tau_h(\bs{p}) = \tau_h(\bs\psi(q\bs\delta)).
\]
That is to say, deadline-rigid services' worst-case deadlines do remain static from slot~$t$ to $t+1$. Moreover, this stasis will persist because deadline-rigid services are update invariant. To see this, recall, from Section~\ref{SS:WCSUpdate}, that the effect of the update rule can be explained via a translated coordinate frame, as is illustrated in Fig.~\ref{F:Deadline}, where $O$ is translated to $O'$. As this translation does not change the relative positions of points $A$ and $D$, it must preserve the deadline-rigidness of $\bs\psi$.

\subsubsection*{\textup{[}The EDF Scheduler\textup{]}} When each task, upon arrival, is assigned a static deadline, an EDF scheduler can be used to generate EDF schedules. It sorts all tasks in a single, system queue according to their static deadlines, with the task possessing the earliest static deadline placed at the head of the queue and served first. When a new task arrives, it is inserted into the queue according to its newly assigned static deadline, without disturbing the relative order of any older tasks. A comprehensive survey of EDF scheduling can be found in \cite{Stankovic:1998}.

As we have seen, worst-case deadlines may be dynamic. So, when worst-case systems are EDF scheduled, unless every worst-case service is deadline-rigid, for the tasks in the single, system queue, their worst-case deadlines may be relaxed during their stays, in which case, it is impossible to ensure that the tasks are always sorted, and thus served, according to their worst-case deadlines.

A {\it \textbf{deadline-rigid system}} is a worst-case system in which every worst-case service is deadline-rigid. It is update invariant because its constituent deadline-rigid services are update invariant. In such a system, all worst-case deadlines are static, so the EDF scheduler, according to the task-centric interpretation, automatically generates max-slack schedules.\footnote{However, this implementation, besides sacrificing scheduling flexibility, also compromises the dynamic nature of state-based scheduling because, if a flow's deadline-rigid service is renegotiated on the fly, each of its older tasks in the single, system queue may be assigned a new static deadline, thereby triggering a re-sorting of the entire queue.} According to Theorem~\ref{T:MSS}, to ensure that the schedules so generated are feasible, we need only ensure that $\mu$, the total service to be allocated, is feasible. In light of this observation, an additional corollary of Theorem~\ref{T:MSS} is that, if any scheduler can meet {\it all} static deadlines for {\it all} tasks {\it simultaneously}, so can the EDF scheduler, which is a well-known fact.\footnote{An independent proof of this fact can be found in \cite{Georgiadis:1997}. The key idea is that when two tasks are not served in the order dictated by their respective static deadlines, the order can always be reversed without violating any deadline. For a deadline-rigid system, this idea also provides the basis for an alternative proof of Theorem~\ref{T:MSS}.}

It is easy to extend the task-centric interpretation to per-class max-slack schedules, which, intra-class, must serve tasks in non-decreasing order of their respective worst-case deadlines. Accordingly, for a deadline-rigid system, we can also extend the EDF scheduler to generate per-class max-slack schedules. Such an extended scheduler keeps a separate queue for each class and, intra-class, serves tasks earliest-deadline-first. Of course, to ensure that the schedules so generated are feasible, we must first select a feasible inter-class schedule.

\section{Min-Plus Services}\label{S:MPS}

A second source of complexity in our framework is the fact that worst-case services, as uncountably infinite, full-blown maps between cumulative vectors, are difficult to specify and update. On the surface, it may seem that to admit new service requests, and to identify feasible schedules, state-based scheduling only requires that we keep track of the services' spectra and conditional spectra, both of which are countably infinite. But these spectra cannot be updated independently from their underlying services because different worst-case services may share the same spectrum.

For instance, worst-case service $\bs\psi$ and its causal counterpart, $\bs\psi^\text{C}$, defined in (\ref{E:causalService}), share the same spectrum. To see this, notice that, as $\bs\psi^\text{C} \geq \bs\psi$, $\lambda_{ij}(\bs\psi^\text{C}) \geq \lambda_{ij}(\bs\psi)$ for all $i,j\in\mb{N}$. Now suppose that $\bs\psi$ and $\bs\psi^\text{C}$ do not share the same spectrum. Then $\lambda_{ij}(\bs\psi^\text{C}) > \lambda_{ij}(\bs\psi)$ for some $i,j \in \mb{N}$. So it may not preserve schedulability to replace $\bs\psi$ in a schedulable system by $\bs\psi^\text{C}$, contradicting the fact that, to guarantee $\bs\psi$, the causal scheduler must guarantee $\bs\psi^\text{C}$.\footnote{Technically, since $\lambda_{ij}(\bs\psi^\text{C}) \geq \lambda_{ij}(\bs\psi)$, we need only show that $\lambda_{ij}(\bs\psi^\text{C}) \leq \lambda_{ij}(\bs\psi)$. Using (\ref{E:signature}) and (\ref{E:causalService}), we have
\[
\begin{IEEEeqnarraybox}[][c]{rCl}
\lambda_{ij}(\bs\psi^\text{C}) &=& \max_{\bs{q} \in \mb{U} \ua b} \left(\max_{i' \leq j} \max_{\bs{q'} \in \mb{U} \ua b, \bs{q'} \stackrel{i'}{=} \bs{q}} \psi_{i'}(\bs{q'})-q_i\right)^+\\
                               &=& (\psi_{i_*'}(\bs{q_*'})-[\bs{q_*}]_i)^+,
\end{IEEEeqnarraybox}
\]
where $i_*'$, $\bs{q_*'}$, and $\bs{q_*}$ maximize $\psi_{i'}(\bs{q'})-q_i$. Then, on the one hand, since $\bs{q_*'} \stackrel{i_*'}{=} \bs{q_*}$, if $i_*' \leq i$, according to (\ref{E:WCSDef}),
\[
\psi_{i_*'}(\bs{q_*'})-[\bs{q_*}]_i \leq [\bs{q_*'}]_{i_*'} - [\bs{q_*}]_i = [\bs{q_*}]_{i_*'} - [\bs{q_*}]_i \leq 0,
\]
and on the other, since $i_*' \leq j$ and $\bs{q_*'} \stackrel{i_*'}{=} \bs{q_*}$, if $i_*' > i$,
\[
\psi_{i_*'}(\bs{q_*'})-[\bs{q_*}]_i \leq \psi_j(\bs{q_*'})-[\bs{q_*}]_i = \psi_j(\bs{q_*'})-[\bs{q_*'}]_i.
\]
So, according to (\ref{E:signature}), $\lambda_{ij}(\bs\psi^\text{C}) \leq \lambda_{ij}(\bs\psi)$.} Of course, due precisely to this fact, upgrading $\bs\psi$ to $\bs\psi^\text{C}$ is but a formality.

Can $\bs\psi^\text{C}$ be further upgraded without changing its spectrum? In fact, it can. Consider, for instance, a $\bs\psi^\text{D}$ such that
\begin{equation}
\psi_j^\text{D}(\bs{q}) := \!\!\max_{\bs{q'} \in \mb{U} \ua b, h \leq q_j, \bs{q'} \stackrel{[h]}{=} \bs{q}} \!\!\!\min\{\psi_j(\bs{q'}), h\} \blank\blank\forall\, \bs{q}\in\mb{U} \ua b, j\in\mb{N}.
\end{equation}
Although we will not do so here, it can be shown that $\bs\psi^\text{D}$ is deadline-rigid, $\bs\psi^\text{D} \geq \bs\psi^\text{C}$, and $\lambda_{ij}(\bs\psi^\text{D}) = \lambda_{ij}(\bs\psi)$ for all $i,j\in\mb{N}$. This raises two questions: Is there a limit to such upgrades? and if so, What is the limit?

What we will show in this section is that, starting from any worst-case service, $\bs\psi$, if we make an arbitrary series of spectrum-preserving upgrades, resulting in $\bs\psi \leq \bs\psi' \leq \bs\psi'' \leq \ldots$, until no further upgrade can be made without changing the spectrum, we will always end up with the same worst-case service. That is to say, among all worst-case services that share the same spectrum, there exists a maximum. Notice that, as $\leq$ only defines a partial order among worst-case services, the existence of this maximum is not self-evident.

In the remainder of this section, we use the min-plus algebra to characterize the maximum worst-case service with a given spectrum. For this reason, we call these services min-plus services. That min-plus services are maximums, on the one hand, underlies their efficiency because it necessarily implies that they can be uniquely identified by their spectra, and on the other, allows each schedulable system of non-min-plus services to be upgraded to a schedulable system of min-plus services. We conclude this section by providing an alternative definition for min-plus services, which, among other things, helps us establish the composition rule for these services.

\subsection{Definition}\label{SS:MPSDef}

As in the case of $\bs\psi$, $\bs\psi^\text{C}$, and $\bs\psi^\text{D}$, in general, worst-case services cannot be uniquely identified by their spectra. But they can be upper bounded by their spectra. Notice that, according to (\ref{E:signature}), $\psi_j(\bs{q}) -q_i \leq \lambda_{ij}(\bs\psi)$ for all $i,j\in\mb{N}$, so
\begin{equation}\label{E:psiBound}
\psi_j(\bs{q}) \leq \min_{i \in \mb{N}} (q_i+\lambda_{ij}(\bs\psi)) \blank\blank\forall\, j\in\mb{N}.
\end{equation}
This bound can be rewritten concisely using the {\it \textbf{min-plus algebra}}, an algebra in which operators $\min$ and $+$ replace, respectively, operators $+$ and $\times$ in the standard algebra. First, construct the matrix of $\lambda_{ij}(\bs\psi)$'s,
\[
\setlength\arraycolsep{3pt}
\Lambda(\bs\psi) = [\lambda_{ij}(\bs\psi)]_{i,j\in\mb{N}} = \left[
\begin{array}{ccccc}
\lambda_{00}(\bs\psi) & \lambda_{01}(\bs\psi) & \lambda_{02}(\bs\psi) & \cdots\\
\lambda_{10}(\bs\psi) & \lambda_{11}(\bs\psi) & \lambda_{12}(\bs\psi) & \cdots\\
\lambda_{20}(\bs\psi) & \lambda_{21}(\bs\psi) & \lambda_{22}(\bs\psi) & \cdots\\
\vdots                & \vdots                & \vdots                & \ddots
\end{array}
\right].
\setlength\arraycolsep{5pt}
\]
Next, using $\otimes$ to denote min-plus matrix multiplication, and applying the matrix multiplication rule, rewrite (\ref{E:psiBound}) in matrix form as
\begin{equation}\label{E:psiBound+}
\bs\psi(\bs{q}) \leq \bs{q} \otimes \Lambda(\bs\psi) \blank\blank \forall\, \bs{q} \in \mb{U} \ua b.
\end{equation}
This version of the bound motivates the following definition.

\begin{definition}\label{D:SM-MPS}
{\it For a flow with $b$ tasks left unserved in its buffer, a semi-infinite matrix, $S=[s_{ij}]_{i,j\in\mb{N}}$, is a \textbf{spectral matrix} if, for all $i,j\in\mb{N}$,
\begin{equation}\label{E:SMDefA}
s_{ij} = 0 \blank\textup{if } i \geq j,
\end{equation}
\begin{equation}\label{E:SMDefB}
s_{ij} \leq s_{i, j+1},
\end{equation}
\begin{equation}\label{E:SMDefC}
s_{ij} \geq s_{i+1, j},
\end{equation}
and
\begin{equation}\label{E:SMDefD}
s_{ij} \leq (s_{0j}-b\delta_i)^+.
\end{equation}
We call $\bs\psi^S$ the \textbf{min-plus service} identified by $S$ if
\begin{equation}\label{E:psiSVec}
\bs\psi^S(\bs{q}):= \bs{q} \otimes S \blank\blank \forall\, \bs{q} \in \mb{U}\ua b.
\end{equation}}
\end{definition}

According to this definition, spectral matrices are non-negative, strictly upper triangular, and have non-decreasing rows and non-increasing columns. Notice that, for all $j \in \mb{N}$, using (\ref{E:psiSVec}) and (\ref{E:SMDefA}), we have
\begin{IEEEeqnarray}{rCl}
\psi_j^S(\bs{q}) = \bs{q} \otimes S_{\bs\cdot j} &=& \min_{i\in\mb{N}} (q_i+s_{ij})                                   \label{E:psiS}\\
                                        &=& \min \left\{\min_{i<j} (q_i+s_{ij}), \min_{i \geq j} q_i\right\} \IEEEnonumber \\
                                        &=& \min \left\{\min_{i<j} (q_i+s_{ij}), q_j\right\}                 \IEEEnonumber \\
                                        &=& \min_{i \leq j} (q_i+s_{ij}),                                    \label{E:psiS+}
\end{IEEEeqnarray}
where $S_{\bs\cdot j}$ denotes the $j$th column of $S$. According to (\ref{E:psiS}) and (\ref{E:SMDefB}), $\psi_j^S(\bs{q}) \leq \psi_{j+1}^S(\bs{q})$. So $\bs\psi^S(\bs{q})$ is a cumulative vector. Moreover, according to (\ref{E:psiS+})'s derivation, $\psi_j^S(\bs{q}) \leq q_j$. So $\bs\psi^S(\bs{q})\leq \bs{q}$. Therefore, according to Definition~\ref{D:WCS}, $\bs\psi^S$ is indeed a worst-case service.

Min-plus services are deadline-rigid because, using (\ref{E:psiS}), we have
\[
\begin{IEEEeqnarraybox}[][c]{rCl}
\min\{\psi_j^S(\bs{q}), h\} &=& \min_{i\in\mb{N}} \min\{q_i+s_{ij}, h\}              \\
                            &=& \min_{i\in\mb{N}} \min\{\min\{q_i, h\}+s_{ij}, h\},
\end{IEEEeqnarraybox}
\]
implying that $\min\{\bs\psi^S(\bs{q}), h\bs\delta\}$ depends on $\min\{\bs{q}, h\bs\delta\}$ alone. Min-plus services are also update invariant, as is made clear by the next theorem, the proof of which is omitted because an easier-to-prove alternative will be proved in Section~\ref{SS:MPSDef+}.

\begin{theorem}\label{T:SUpdate}
{\it In Theorem~\ref{T:WCSUpdate}, if $\bs\psi$ is a min-plus service that can be identified by a spectral matrix, $S$, that is, $\bs\psi = \bs\psi^S$, then, if $q \geq d \geq p$, $\bs{\dot\psi}$ is also a min-plus service that can be identified by a spectral matrix, $\dot{S}=[\dot{s}_{ij}]_{i,j\in\mb{N}}$, that is, \mbox{$\bs{\dot\psi} = \bs\psi^{\dot{S}}$}, where
\begin{equation}\label{E:SUpdate}
\dot{s}_{ij}=\left\{
\begin{IEEEeqnarraybox}[][c]{ll}
(\min \{s_{0, j+1}, q+s_{1, j+1}\}-d)^+ & \blank\blank\textup{if } i=0\\
\min \{(s_{0, j+1}-q)^+, s_{i+1, j+1}\} & \blank\blank\textup{if } i>0
\end{IEEEeqnarraybox}\right. ~.
\end{equation}}
\end{theorem}

\subsection{The Spectrum}\label{SS:MPSSig}

By construction, (\ref{E:SMDefA}), (\ref{E:SMDefB}), (\ref{E:SMDefC}), and (\ref{E:SMDefD}) parallel (\ref{E:SigPropA}), (\ref{E:SigPropB}), (\ref{E:SigPropC}), and (\ref{E:SigPropD}), respectively. So, not surprisingly, $s_{ij}$ behaves very much like $\lambda_{ij}$, the spectral value of {\it some} worst-case service. It is fundamental that {\it this} worst-case service can be $\bs\psi^S$ itself.

\begin{theorem}\label{T:SSig}
{\it For all $i,j\in\mb{N}$,
\begin{equation}\label{E:SSig}
\lambda_{ij}(\bs\psi^S) = s_{ij},
\end{equation}
or in matrix form
\begin{equation}\label{E:SSigMat}
\Lambda(\bs\psi^S) = S.
\end{equation}}
\end{theorem}

To prove this theorem, it is convenient to introduce $\bs\varepsilon = [\varepsilon_j]_{j\in\mb{N}}:=[0, \infty, \infty, \ldots] \in \mb{U}$, that is,
\begin{equation}\label{E:epsilon+}
\varepsilon_j:= \left\{
\begin{IEEEeqnarraybox}[][c]{ll}
0      & \blank\blank\text{if } j=0\\
\infty & \blank\blank\text{if } j>0
\end{IEEEeqnarraybox}\right. ~.
\end{equation}
Intuitively $\bs\varepsilon$ models an infinite traffic burst in slot~$t$. Then, according to (\ref{E:psiS}) and (\ref{E:SMDefC}),
\begin{IEEEeqnarray}{rCl}
\psi_j^S(\mc{R}^i\bs\varepsilon+b\bs\delta) &=& \min_{k\in\mb{N}} ([\mc{R}^i\bs\varepsilon]_k+b\delta_k+s_{kj}) \IEEEnonumber\\
                                            &=& \min_{k \leq i} (b\delta_k+s_{kj})                   \IEEEnonumber\\
                                            &=& \min \{s_{0j}, b\delta_i+s_{ij}\},                   \label{E:epsilon++}
\end{IEEEeqnarray}
where the second equality holds because, by definition,
\begin{equation}\label{E:epsilon}
[\mc{R}^i\bs\varepsilon]_k= \left\{
\begin{IEEEeqnarraybox}[][c]{ll}
0      & \blank\blank\text{if } k \leq i\\
\infty & \blank\blank\text{if } k>i
\end{IEEEeqnarraybox}\right. ~.
\end{equation}

\begin{IEEEproof}[Proof of Theorem~\ref{T:SSig}]
On the one hand, since by default, $\mc{R}^i\bs\varepsilon+b\bs\delta \in \mb{U}\ua b$, using (\ref{E:signature}), (\ref{E:epsilon++}), (\ref{E:epsilon}), and (\ref{E:SMDefD}), we have
\[
\begin{IEEEeqnarraybox}[][c]{rCl}
\lambda_{ij}(\bs\psi^S) & \geq & (\psi_j^S(\bs{q})-q_i)^+|_{\bs{q}=\mc{R}^i\bs\varepsilon+b\bs\delta} \\
                        & =    & (\psi_j^S(\mc{R}^i\bs\varepsilon+b\bs\delta)-[\mc{R}^i\bs\varepsilon]_i-b\delta_i)^+ \\
                        & =    & (\min\{s_{0j}, b\delta_i+s_{ij}\}-b\delta_i)^+ \\
                        & =    & \min\{(s_{0j}-b\delta_i)^+, s_{ij}\} = s_{ij}.
\end{IEEEeqnarraybox}
\]
On the other hand, using (\ref{E:signature}) and (\ref{E:psiS}), we have
\[
\begin{IEEEeqnarraybox}[][c]{rCl}
\lambda_{ij}(\bs\psi^S) & =    & \max_{\bs{q}\in\mb{U}\ua b} \left(\min_{k\in\mb{N}} (q_k+s_{kj})-q_i\right)^+ \\
                        & \leq & \max_{\bs{q}\in\mb{U}\ua b} \left(q_i+s_{ij}-q_i\right)^+ =  s_{ij}.
\end{IEEEeqnarraybox}
\]
It follows that $\lambda_{ij}(\bs\psi^S) = s_{ij}$.
\end{IEEEproof}

According to Theorem~\ref{T:SSig}, each unique spectral matrix identifies a unique min-plus service with a unique spectrum, which explains why we term such matrices spectral matrices. One obvious corollary is that each min-plus service is uniquely identified by its spectrum. Continuing our Section~\ref{SS:signatures} analogy between worst-case services and normal matrices, min-plus services loosely correspond to diagonal matrices. Both are uniquely identified by their respective spectra, on which, however, neither may lay exclusive claim.

A second corollary is that given a collection of $\lambda_{ij}$'s, properties (\ref{E:SigPropA}), (\ref{E:SigPropB}), (\ref{E:SigPropC}), and (\ref{E:SigPropD}) suffice to ensure that the collection is some worst-case service's spectrum, because the spectral matrix $S$, constructed such that $s_{ij} = \lambda_{ij}$ for all $i,j\in\mb{N}$, identifies such a service. This implies that among all worst-case services that share the same spectrum, there is always {\it one} min-plus service. Moreover, according to (\ref{E:psiSVec}) and (\ref{E:SSigMat}), $\bs\psi^S(\bs{q}) = \bs{q} \otimes \Lambda(\bs\psi^S)$. Comparing this to (\ref{E:psiBound+}), it is immediate that if $\Lambda(\bs\psi^S) = \Lambda(\bs\psi)$, $\bs\psi^S \geq \bs\psi$, that is, $\bs\psi^S(\bs{q}) \geq \bs\psi(\bs{q})$ for all $\bs{q} \in \mb{U}\ua b$. So, among all worst-case services that share the same spectrum, {\it the one} min-plus service identified by this spectrum is always the maximum.

\subsubsection*{\textup{[}Conditional Spectra\textup{]}} Since a min-plus service is uniquely identified by its spectrum, so is its conditional spectrum. We denote its conditional spectral value, $\lambda_{ij}(\bs\psi^S|q)$, by $\hat{s}_{ij}$ when no confusion can be introduced. The next theorem details how $\hat{s}_{ij}$ is identified by $s_{ij}$.

\begin{theorem}\label{T:SCondSig}
{\it For all $i,j\in\mb{N}$,
\begin{equation}\label{E:SCondSig}
\hat{s}_{ij}=\left\{
\begin{IEEEeqnarraybox}[][c]{ll}
\min \{s_{0j}, q+s_{1j}\}     & \blank\blank\textup{if } i=0\\
\min \{(s_{0j}-q)^+, s_{ij}\} & \blank\blank\textup{if } i>0
\end{IEEEeqnarraybox}\right. ~.
\end{equation}}
\end{theorem}

\begin{IEEEproof}
In the case that $i=0$, on the one hand, since by default, $\mc{R}\bs\varepsilon+q\bs\delta \in \mb{U}|q$, using (\ref{E:CondSig}) and (\ref{E:epsilon++}), we have
\[
\hat{s}_{0j} \geq \psi_j^S(\bs{q})|_{\bs{q}=\mc{R}\bs\varepsilon+q\bs\delta} = \psi_j^S(\mc{R}\bs\varepsilon+q\bs\delta) = \min \{s_{0j}, q+s_{1j}\}.
\]
On the other hand, using (\ref{E:CondSig}) and (\ref{E:psiS}), we have
\[
\begin{IEEEeqnarraybox}[][c]{rCl}
\hat{s}_{0j} = \max_{\bs{q}\in\mb{U}|q} \min_{k\in\mb{N}} (q_k+s_{kj}) & \leq & \max_{\bs{q}\in\mb{U}|q} \min \{s_{0j}, q_1+s_{1j}\}\\
                                                                       & =    & \min \{s_{0j}, q+s_{1j}\}.
\end{IEEEeqnarraybox}
\]
So (\ref{E:SCondSig}) holds when $i=0$.

In the case that $i>0$, on the one hand, since by default, $\mc{R}^i\bs\varepsilon+q\bs\delta \in \mb{U}|q$, using (\ref{E:CondSig}), (\ref{E:epsilon++}), and (\ref{E:epsilon}), we have
\[
\begin{IEEEeqnarraybox}[][c]{rCl}
\hat{s}_{ij} & \geq & (\psi_j^S(\bs{q})-q_i)^+|_{\bs{q}=\mc{R}^i\bs\varepsilon+q\bs\delta} \\
             & =    & (\psi_j^S(\mc{R}^i\bs\varepsilon+q\bs\delta)-[\mc{R}^i\bs\varepsilon]_i-q)^+ \\
             & =    & (\min \{s_{0j}, q+s_{ij}\}-q)^+ \\
             & =    & \min \{(s_{0j}-q)^+, s_{ij}\}.
\end{IEEEeqnarraybox}
\]
On the other hand, using (\ref{E:CondSig}) and (\ref{E:psiS}), we have
\[
\begin{IEEEeqnarraybox}[][c]{rCl}
\hat{s}_{ij}  & =    & \max_{\bs{q}\in\mb{U}|q} \left(\min_{k\in\mb{N}} (q_k+s_{kj})-q_i\right)^+ \\
              & \leq & \max_{\bs{q}\in\mb{U}|q} (\min \{s_{0j}, q_i+s_{ij}\}-q_i)^+ \\
              & =    & \max_{\bs{q}\in\mb{U}|q} \min \{(s_{0j}-q_i)^+, s_{ij}\} \\
              & \leq & \min \{(s_{0j}-q)^+, s_{ij}\},
\end{IEEEeqnarraybox}
\]
where the final inequality holds because, by definition, $q_i \geq q_1 = q$ for all $\bs{q}\in\mb{U}|q$. So (\ref{E:SCondSig}) also holds when $i>0$.
\end{IEEEproof}

\subsection{Min-Plus Systems}\label{SS:MPHull}

A {\it \textbf{min-plus system}} is a worst-case system in which every worst-case service is a min-plus service. It is update invariant because its constituent min-plus services are update invariant. For all $\omega\in\Omega$, let the min-plus service guaranteed to flow $\omega$ be identified by spectral matrix $S^\omega$, and denote the resulting min-plus system by $S^{[\Omega]}$. Compared to a general worst-case system, $S^{[\Omega]}$'s efficiency is impressive. Besides being more efficient to specify, according to Theorem~\ref{T:SUpdate}, $S^{[\Omega]}$ can be more efficiently updated. Moreover, using Theorems~\ref{T:SSig} and \ref{T:SCondSig}, it is straightforward to track $S^{[\Omega]}$'s spectra and conditional spectra. In particular, according to Theorems~\ref{T:schedulability} and \ref{T:SSig}, $S^{[\Omega]}$ is schedulable if and only if
\begin{equation}\label{E:schedulabilityS}
S^{\l\Omega\r} \leq C = [c_{ij}]_{i,j\in\mb{N}} = [(j-i)^+c]_{i,j\in\mb{N}}.
\end{equation}

A worst-case system, $\bs\psi^{[\Omega]}$, is said to be {\it \textbf{dominated}} by another worst-case system, $\bs{\bar\psi}^{[\Omega]}$, if for all $\omega \in \Omega$, $\bs\psi^\omega \leq \bs{\bar\psi}^\omega$. If both systems are also schedulable, we can then, by upgrading $\bs\psi^{[\Omega]}$ to $\bs{\bar\psi}^{[\Omega]}$, improve the service guaranteed to each flow, and yet, preserve schedulability. Within this broader context, the efficiency of min-plus systems is no longer only of isolated interest but, as we will see, assumes a broader significance in that any schedulable non-min-plus system can be upgraded to a schedulable min-plus system.

\subsubsection*{\textup{[}The Min-Plus Hull\textup{]}} Given a worst-case system, $\bs\psi^{[\Omega]}$, construct a min-plus system, $S^{[\Omega]}$, such that $S^\omega = \Lambda(\bs\psi^\omega)$ for all $\omega \in \Omega$. We call $S^{[\Omega]}$ the {\it \textbf{min-plus hull}} of $\bs\psi^{[\Omega]}$. It is immediate from (\ref{E:SSigMat}) that $\Lambda(\bs\psi^{S^\omega}) = \Lambda(\bs\psi^\omega)$, and from (\ref{E:psiBound+}) and (\ref{E:psiSVec}), that $\bs\psi^{S^\omega} \geq \bs\psi^\omega$, so $S^{[\Omega]}$ dominates $\bs\psi^{[\Omega]}$. Moreover, as $S^{[\Omega]}$ and $\bs\psi^{[\Omega]}$ share the same spectrum system, if $\bs\psi^{[\Omega]}$ is schedulable, so is $S^{[\Omega]}$. So we can always upgrade a schedulable non-min-plus system to its min-plus hull. A corollary is that the {\it Pareto} frontier bounding the schedulable region in Fig.~\ref{F:StatePath} must consist solely of min-plus systems.

\begin{example}\label{Ex:hull}
{\it Let $\bs\psi^{[\Omega]}$ be a single-task system. According to (\ref{E:specSTS}), $S^{[\Omega]}$ is its min-plus hull if for all $\omega\in\Omega$ and $i,j\in\mb{N}$,
\begin{equation}\label{E:hullEx}
s_{ij}^\omega=\left\{
\begin{IEEEeqnarraybox}[][c]{ll}
1 & \blank\blank\textup{if } j-i > \bar\theta^\omega\\
0 & \blank\blank\textup{if } j-i \leq \bar\theta^\omega
\end{IEEEeqnarraybox}\right. ~.
\end{equation}
Using (\ref{E:psiS}), it can be verified that for all $\bs{q}^\omega \in \mb{U}$,
\begin{equation}\label{E:hullEx+}
\bs\psi^{S^\omega}(\bs{q}^\omega)=\left\{
\begin{IEEEeqnarraybox}[][c]{ll}
\mc{R}^{i+\bar\theta^\omega}\bs\delta & \blank\blank\textup{if } \bs{q}^\omega \stackrel{[1]}{=} \mc{R}^i\bs\delta \textup{ for some } i\in\mb{N}\\
\bs{0}                                & \blank\blank\textup{if } \bs{q}^\omega = \bs{0}
\end{IEEEeqnarraybox}\right. ~,
\end{equation}
which specifies that the first task of flow $\omega$ is to be served and that the task's delay will not exceed $\bar\theta^\omega$. Comparing (\ref{E:hullEx+}) to (\ref{E:STS}), it is immediate that $\bs\psi^{S^\omega} \geq \bs\psi^\omega$ because $\bs{q}^\omega = \mc{R}^i\bs\delta$ implies that $\bs{q}^\omega \stackrel{[1]}{=} \mc{R}^i\bs\delta$, but not vice versa.\footnote{\textup{Recall, from Example~\ref{Ex:STS}, that $\bs\psi^\omega$, as a single-task service, is non-causal. So, to guarantee $\bs\psi^\omega$, the causal scheduler must guarantee its causal counterpart, $\bs\psi^\text{C}$, in (\ref{E:STSCausal}). Yet, comparing (\ref{E:hullEx+}) to (\ref{E:STSCausal}), we still have $\bs\psi^{S^\omega} \geq \bs\psi^\text{C}$ because $\bs{q}^\omega \stackrel{i+\bar\theta}{=} \mc{R}^i\bs\delta$ implies that $\bs{q}^\omega \stackrel{[1]}{=} \mc{R}^i\bs\delta$, but not vice versa.}} So $S^{[\Omega]}$ guarantees better services than $\bs\psi^{[\Omega]}$, although they share the same spectrum system.
}
\end{example}

Since, when upgrading a non-min-plus system to its min-plus hull, the spectrum system remains the same, the server's capacity slack in the current slot also remains the same. This upgrade, however, is not cost-free. Imagine running schedulers that guarantee $\bs\psi^{[\Omega]}$ and its min-plus hull, $S^{[\Omega]}$, side by side. Using (\ref{E:CondSigPropF}), (\ref{E:CondSigPropE}), and (\ref{E:CondSigPropD}), it is easy to verify that
\begin{equation}\label{E:CondSigBound}
\forall\, i,j\in\mb{N}, \blank\blank
\hat\lambda_{ij} \leq
\left\{
\begin{IEEEeqnarraybox}[][c]{ll}
\min \{\lambda_{0j}, q+\lambda_{1j}\}     & \blank\blank\text{if } i=0\\
\min \{(\lambda_{0j}-q)^+, \lambda_{ij}\} & \blank\blank\text{if } i>0
\end{IEEEeqnarraybox}\right. ~.
\end{equation}
Comparing this to (\ref{E:SCondSig}), it is immediate that $\hat\lambda_{ij}^\omega \leq \hat{s}_{ij}^\omega$ for all $\omega\in\Omega$ because, by construction, $\lambda_{ij}^\omega = s_{ij}^\omega$. So, according to (\ref{E:beta}), $\beta^S \geq \beta^{\bs\psi}$, that is, $\beta^S(\Gamma) \geq \beta^{\bs\psi}(\Gamma)$ for all $\Gamma \subseteq \Omega$, where we use $\beta^S$ and $\beta^{\bs\psi}$ to denote the baseline functions for $S^{[\Omega]}$ and $\bs\psi^{[\Omega]}$, respectively. According to Theorem~\ref{T:polytope}, $\beta^S \geq \beta^{\bs\psi}$ in turn implies that $\mb{F}^S \subseteq \mb{F}^{\bs\psi}$, where we use $\mb{F}^S$ and $\mb{F}^{\bs\psi}$ to denote the feasible polytopes for $S^{[\Omega]}$ and $\bs\psi^{[\Omega]}$, respectively. Therefore, by upgrading $\bs\psi^{[\Omega]}$ to $S^{[\Omega]}$, we may reduce the feasible polytope, and thus reduce our flexibility when selecting feasible schedules from it.

Furthermore, imagine selecting the same feasible schedule, $d^{[\Omega]} \in \mb{F}^S \subseteq \mb{F}^{\bs\psi}$, for both $\bs\psi^{[\Omega]}$ and $S^{[\Omega]}$. Comparing (\ref{E:SUpdate}) to (\ref{E:SCondSig}), we have
\begin{equation}\label{E:SUpdate+}
\forall\, i,j\in\mb{N}, \blank\blank
\dot{s}_{ij}=\left\{
\begin{IEEEeqnarraybox}[][c]{ll}
(\hat{s}_{0, j+1}-d)^+ & \blank\blank\text{if }i=0\\
\hat{s}_{i+1, j+1}     & \blank\blank\text{if }i>0
\end{IEEEeqnarraybox}\right. ~.
\end{equation}
This relation between $\dot{s}_{ij}$ and $\hat{s}_{ij}$ replicates exactly that between $\dot\lambda_{ij}$ and $\hat\lambda_{ij}$ in (\ref{E:SigUpdate}), which is not a coincidence because, according to Theorem~\ref{T:SSig}, $\lambda_{ij}(\bs\psi^{\dot{S}}) = \dot{s}_{ij}$. So, comparing (\ref{E:SigUpdate}) to (\ref{E:SUpdate+}), it is immediate that $\dot\lambda_{ij}^\omega \leq \dot{s}_{ij}^\omega$ for all $\omega\in\Omega$ because, as we have seen, $\hat\lambda_{ij}^\omega \leq \hat{s}_{ij}^\omega$. That is to say, although $\dot{S}^{[\Omega]}$ still dominates $\bs{\dot\psi}^{\raisebox{-0.9mm}{\scriptsize$[\Omega]$}}$, they may no longer share the same spectrum system. Therefore, by upgrading $\bs\psi^{[\Omega]}$ to $S^{[\Omega]}$, we may reduce the server's capacity slack in the future.

\begin{example}\label{Ex:hull+}
{\it Let $\bs\psi^{[\Omega]}$ be a single-task system. Assume that $\bs{q}^\omega = 2\bs\delta$ for all $\omega\in\Omega$. Then, according to (\ref{E:STS}), no task need ever be served, implying that, in this case, $\beta^{\bs\psi}(\Gamma) = 0$ for all $\Gamma\subseteq\Omega$. In contrast, as in Example~\ref{Ex:hull}, let $S^{[\Omega]}$ be the min-plus hull of $\bs\psi^{[\Omega]}$. Then, according to (\ref{E:hullEx+}), the first task of each flow needs to be served either immediately or later, thus reducing the current feasible polytope, the future capacity slack, or both. For instance, if $\bar\theta^\omega = 0$ for any $\omega\in\Omega$, it is easy to verify that $\beta^S(\{\omega\}) = 1 > 0 = \beta^{\bs\psi}(\{\omega\})$, thereby reducing the current feasible polytope. Or, if any first task is not served immediately, it needs to be served later, thereby reducing the future capacity slack.}
\end{example}

\subsection{An Alternative Definition}\label{SS:MPSDef+}

In Definition~\ref{D:SM-MPS}, we imposed conditions (\ref{E:SMDefA}), (\ref{E:SMDefB}), (\ref{E:SMDefC}), and (\ref{E:SMDefD}) on $S$ to mirror spectral properties (\ref{E:SigPropA}), (\ref{E:SigPropB}), (\ref{E:SigPropC}), and (\ref{E:SigPropD}), respectively. But, as nowhere in the proof that $\bs\psi^S$ is a worst-case service are (\ref{E:SMDefC}) and (\ref{E:SMDefD}) used, neither is necessary for $\bs\psi^S$ to be a worst-case service. This observation motivates the following alternative definition of min-plus services.

\begin{definition}\label{D:CM-MPS}
{\it A semi-infinite matrix, $M=[m_{ij}]_{i,j\in\mb{N}}$, is a \textbf{cumulative matrix} if, for all $i,j\in\mb{N}$,
\begin{equation}\label{E:CMDefA}
m_{ij} = 0 \blank\textup{if } i \geq j,
\end{equation}
and
\begin{equation}\label{E:CMDefB}
m_{ij} \leq m_{i, j+1}.
\end{equation}
For a flow with $b$ tasks left unserved in its buffer, we call $\bs\psi^M$ the \textbf{min-plus service} identified by $M$ if
\begin{equation}\label{E:psiMVec}
\bs\psi^M(\bs{q}):= \bs{q} \otimes M \blank\blank\forall\, \bs{q} \in \mb{U}\ua b.
\end{equation}}
\end{definition}

As the omissions of (\ref{E:SMDefC}) and (\ref{E:SMDefD}) from this definition do not affect (\ref{E:psiS}) and (\ref{E:psiS+})'s derivations, for all $j \in \mb{N}$, we can replicate (\ref{E:psiS}) and (\ref{E:psiS+}) in terms of $M$ as
\begin{IEEEeqnarray}{rCl}
\psi_j^M(\bs{q}) = \bs{q} \otimes M_{\bs\cdot j} &=& \min_{i\in\mb{N}} (q_i+m_{ij}) \label{E:psiM}\\
                                              &=& \min_{i \leq j} (q_i+m_{ij}).  \label{E:psiM+}
\end{IEEEeqnarray}
But, by omitting (\ref{E:SMDefC}) and (\ref{E:SMDefD}), it would seem that we unduly enlarge the set of min-plus services. According to the next theorem, this concern is unwarranted.

\begin{theorem}\label{T:MtoS}
{\it For a flow with $b$ tasks left unserved in its buffer, given cumulative matrix $M$, construct another semi-infinite matrix, $S=[s_{ij}]_{i,j\in\mb{N}}$, such that
\begin{equation}\label{E:MtoS}
s_{ij} = \min \left\{(m_{0j}-b\delta_i)^+, \min_{k \leq i} m_{kj}\right\}.
\end{equation}
Then $S$ is a spectral matrix and $\bs\psi^S = \bs\psi^M$.}
\end{theorem}

\begin{IEEEproof}
It is easy to verify that $S$ is indeed a spectral matrix. Then, according to (\ref{E:psiS}) and (\ref{E:MtoS}),
\[
\psi_j^S(\bs{q}) \!=\! \min \!\left\{\!\min_{i \in \mb{N}} [q_i+(m_{0j}-b\delta_i)^+], \min_{i \in \mb{N}} \min_{k \leq i} (q_i+m_{kj})\!\right\}\!.
\]
On the one hand, for the left term of the outer minimization, using (\ref{E:vecArrival}) and (\ref{E:psiM}), we have
\[
\begin{IEEEeqnarraybox}[][c]{rCl}
\min_{i \in \mb{N}} [q_i+(m_{0j}-b\delta_i)^+] &\geq& \min_{i \in \mb{N}} (q_i-b\delta_i)+m_{0j} \\
                                               & =  & \min_{i \in \mb{N}} a_i+m_{0j} = m_{0j} \geq \psi_j^M(\bs{q}),
\end{IEEEeqnarraybox}
\]
where the final inequality holds because $m_{0j} = q_0+m_{0j} \geq \psi_j^M(\bs{q})$. On the other hand, for the right term, changing the order of the two $\min$ operators, and using (\ref{E:psiM}), we have
\[
\begin{IEEEeqnarraybox}[][c]{rCl}
\min_{i \in \mb{N}} \min_{k \leq i} (q_i+m_{kj}) &=& \min_{k \in \mb{N}} \min_{i \geq k} (q_i+m_{kj}) \\
                                                 &=& \min_{k \in \mb{N}} (q_k+m_{kj}) = \psi_j^M(\bs{q}),
\end{IEEEeqnarraybox}
\]
where the second equality holds because $q_i \geq q_k$ for all $i \geq k$. It follows that $\psi_j^S(\bs{q}) = \psi_j^M(\bs{q})$ for all $\bs{q}\in\mb{U}\ua b$ and $j\in\mb{N}$, so $\bs\psi^S = \bs\psi^M$.
\end{IEEEproof}

According to Theorem~\ref{T:SSig}, each unique spectral matrix identifies a unique min-plus service with a unique spectrum. In contrast, according to Theorem~\ref{T:MtoS}, the same min-plus service can be identified by multiple cumulative matrices. While this is redundant, it is precisely this redundancy that enables us to simplify our treatment of several topics.

\subsubsection*{\textup{[}Reformulating Theorem~\ref{T:SUpdate}\textup{]}} The first simplification arises from an alternative formulation of Theorem~\ref{T:SUpdate}.

\begin{theorem}\label{T:MUpdate}
{\it In Theorem~\ref{T:WCSUpdate}, if $\bs\psi$ is a min-plus service that can be identified by a cumulative matrix, $M$, that is, $\bs\psi = \bs\psi^M$, then, if $q \geq d \geq p$, $\bs{\dot\psi}$ is also a min-plus service that can be identified by a cumulative matrix, $\dot{M}=[\dot{m}_{ij}]_{i,j\in\mb{N}}$, that is, $\bs{\dot\psi} = \bs\psi^{\dot{M}}$, where
\begin{equation}\label{E:MUpdate}
\dot{m}_{ij}=\left\{
\begin{IEEEeqnarraybox}[][c]{ll}
(\min \{m_{0, j+1}, q+m_{1, j+1}\}-d)^+ & \blank\blank\textup{if } i=0\\
m_{i+1, j+1}                            & \blank\blank\textup{if } i>0
\end{IEEEeqnarraybox}\right. ~.
\end{equation}}
\end{theorem}

Once this theorem is proved, Theorem~\ref{T:SUpdate} is but a corollary. In particular, to derive (\ref{E:SUpdate}), we first treat $S$ as a general cumulative matrix and then use (\ref{E:MUpdate}) to find $\dot{M}$. Although the resulting $\dot{M}$ need not be a spectral matrix, we can always use (\ref{E:MtoS}) to turn it into $\dot{S}$. This is straightforward conceptually, but does require a bit of derivation. There is, however, a shortcut. Once update invariance has been established using (\ref{E:MUpdate}) and (\ref{E:MtoS}), according to Theorem~\ref{T:SSig}, $\dot{s}_{ij} = \lambda_{ij}(\bs\psi^{\dot{S}})$. So (\ref{E:SUpdate+}) follows from (\ref{E:SigUpdate}), which, via (\ref{E:SCondSig}), yields (\ref{E:SUpdate}) directly. Notice that, when $i > 0$, (\ref{E:MUpdate}) is much simpler than (\ref{E:SUpdate}). This motivates a second simplification, the introduction of dual-curve services in Section~\ref{S:DVS}.

\begin{IEEEproof}[Proof of Theorem~\ref{T:MUpdate}]
We first show that $\dot{M}$ is a cumulative matrix. Using (\ref{E:MUpdate}), (\ref{E:CMDefB}), and (\ref{E:CMDefA}), it is easy to verify that $\dot{m}_{ij} \leq \dot{m}_{i, j+1}$ for all $i,j \in \mb{N}$, and that $\dot{m}_{ij} = 0$ if $i \geq j > 0$. To show that $\dot{m}_{00} = 0$, note that, since $d \geq p$, using (\ref{E:WCSImmediate}), (\ref{E:psiM+}), and (\ref{E:CMDefA}), we have
\[
\begin{IEEEeqnarraybox}[][c]{rCl}
d \geq \max_{\bs{q} \in \mb{U}|q} \psi_1^M(\bs{q}) & = & \max_{\bs{q} \in \mb{U}|q} \min \{m_{01}, q_1+m_{11}\} \\
                                                   & = & \min \{m_{01}, q\}.
\end{IEEEeqnarraybox}
\]
So, according to (\ref{E:MUpdate}) and (\ref{E:CMDefA}),
\[
\dot{m}_{00} = (\min \{m_{01}, q+m_{11}\}-d)^+= (\min \{m_{01}, q\}-d)^+ = 0.
\]

It remains to show that $\bs{\dot\psi}$ can be identified by $\dot{M}$. For all $\bs{\dot{q}} \in \mb{U}\ua\dot{b}$ and $j \in \mb{N}$, using (\ref{E:WCSUpdateB}), (\ref{E:LShift}), and (\ref{E:psiM}), we have
\[
\begin{IEEEeqnarraybox}[][c]{rCl}
\dot\psi_j(\bs{\dot{q}}) &=& [\mc{R}^{-1}(\bs\psi^M(\bs{q})-d\bs\delta)^+]_j\\
                         &=& (\psi_{j+1}^M(\bs{q})-d)^+
                          =  \left(\min_{i \in \mb{N}} (q_i+m_{i, j+1})-d\right)^+,
\end{IEEEeqnarraybox}
\]
so
\[
\begin{IEEEeqnarraybox}[][c]{Cl}
 & \dot\psi_j(\bs{\dot{q}}) \\
=& \!\left(\!\min \left\{\!m_{0, j+1}, q_1\!+\!m_{1, j+1}, \min_{i>1} (q_i\!+\!m_{i, j+1})\!\right\}\!-\!d\!\right)^+ \\
=& \!\left(\!\min \left\{\!m_{0, j+1}, q_1\!+\!m_{1, j+1}, \min_{i>0} (q_{i+1}\!+\!m_{i+1, j+1})\!\right\}\!-\!d\!\right)^+\!\!.
\end{IEEEeqnarraybox}
\]
Then, using (\ref{E:qUpdate++}) and (\ref{E:MUpdate}), we have
\[
\begin{IEEEeqnarraybox}[][c]{Cl}
 & \dot\psi_j(\bs{\dot{q}}) \\
=& \!\!\left(\!\min \!\left\{m_{0, j+1}, q\!+\!m_{1, j+1}, \min_{i>0} (\dot{q}_i\!+\!d\!+m_{i+1, j+1})\!\right\}\!-\!d\right)^+ \\
=& \!\min \!\left\{\!(\min\{m_{0, j+1}, q\!+\!m_{1, j+1}\}\!-\!d)^+\!\!, \min_{i>0} (\dot{q}_i\!+\!m_{i+1, j+1})\!\right\} \\
=& \!\min_{i \in \mb{N}}(\dot{q}_i+\dot{m}_{ij}).
\end{IEEEeqnarraybox}
\]
So $\bs{\dot\psi}(\bs{\dot{q}}) = \bs{\dot{q}} \otimes \dot{M}$ for all $\bs{\dot{q}} \in \mb{U}\ua\dot{b}$, and thus $\bs{\dot\psi} = \bs\psi^{\dot{M}}$.
\end{IEEEproof}

\subsection{Monotonicity and Composability}\label{SS:monotone}

A worst-case service, $\bs\psi$, is {\it \textbf{monotone}} if $\bs{q}\geq\bs{q'}$ implies that $\bs\psi(\bs{q}) \geq \bs\psi(\bs{q'})$. Intuitively monotonicity ensures that the service never punishes increases in flow traffic. The next theorem establishes that monotone services are composable.

\begin{figure}[t]
\centering \scalebox{1.000}{\includegraphics{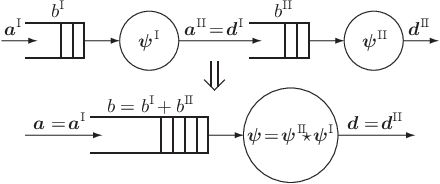}} \caption{Composing two monotone services into one.}\label{F:Composability}
\end{figure}

\begin{theorem}\label{T:composability}
{\it As illustrated in Fig.~\ref{F:Composability}, when a flow is guaranteed monotone services $\bs\psi^\textup{I}$ and $\bs\psi^\textup{II}$ by two servers operating in sequence, the effective service can be modeled by a single server, with $\bs{a}= \bs{a}^\textup{I}$, $b = b^\textup{I}+b^\textup{II}$, and $\bs{d}= \bs{d}^\textup{II}$, that guarantees monotone service $\bs\psi= \bs\psi^\textup{II} \star \bs\psi^\textup{I}$, where
\begin{equation}\label{E:composability}
\bs\psi^\textup{II} \star \bs\psi^\textup{I}(\bs{q}) := \bs\psi^\textup{II}(\bs\psi^\textup{I}(\bs{q}-b^{\textup{II}}\bs\delta)+b^{\textup{II}}\bs\delta) \blank\blank\forall\, \bs{q} \in \mb{U} \ua b.
\end{equation}}
\end{theorem}

\begin{IEEEproof}
Since $\bs\psi^\text{II}$ is monotone, using (\ref{E:WCSDef}) twice, we have
\[
\bs\psi(\bs{q}) = \bs\psi^\text{II}(\bs\psi^\text{I}(\bs{q}-b^{\text{II}}\bs\delta)+b^{\text{II}}\bs\delta) \leq \bs\psi^\text{II}(\bs{q}-b^{\text{II}}\bs\delta+b^{\text{II}}\bs\delta) \leq \bs{q}.
\]
So, according to Definition~\ref{D:WCS}, $\bs\psi$ is a worst-case service. In addition, since both $\bs\psi^\text{I}$ and $\bs\psi^\text{II}$ are monotone, if $\bs{q}\geq\bs{q'}$, it is immediate that
\[
\begin{IEEEeqnarraybox}[][c]{rCl}
\bs\psi(\bs{q}) & =    & \bs\psi^\text{II}(\bs\psi^\text{I}(\bs{q}-b^{\text{II}}\bs\delta)+b^{\text{II}}\bs\delta) \\
                & \geq & \bs\psi^\text{II}(\bs\psi^\text{I}(\bs{q'}-b^{\text{II}}\bs\delta)+b^{\text{II}}\bs\delta)=\bs\psi(\bs{q'}).
\end{IEEEeqnarraybox}
\]
Hence, $\bs\psi$ is also monotone.

To show that the flow is guaranteed $\bs\psi$, it is convenient to recast worst-case services as functions of $\bs{a}$ instead of $\bs{q}$. Specifically let
\[
\bs\varphi^\text{I}(\bs{a}^\text{I}) := \bs\psi^\text{I}(\bs{a}^\text{I}+b^\text{I}\bs\delta) = \bs\psi^\text{I}(\bs{q}^\text{I}) \leq \bs{d}^\text{I} \blank\blank\forall\, \bs{a}^\text{I} \in \mb{U},
\]
and
\[
\bs\varphi^\text{II}(\bs{a}^\text{II}) := \bs\psi^\text{II}(\bs{a}^\text{II}+b^\text{II}\bs\delta) = \bs\psi^\text{II}(\bs{q}^\text{II}) \leq \bs{d}^\text{II} \blank\blank\forall\, \bs{a}^\text{II} \in \mb{U}.
\]
As $\bs\psi^\text{II}$ is monotone, $\bs\varphi^\text{II}$ is also monotone, so
\[
\bs{d} = \bs{d}^\text{II} \geq \bs\varphi^\text{II}(\bs{a}^\text{II}) = \bs\varphi^\text{II}(\bs{d}^\text{I}) \geq \bs\varphi^\text{II}(\bs\varphi^\text{I}(\bs{a}^\text{I})) = \bs\varphi^\text{II}(\bs\varphi^\text{I}(\bs{a})).
\]
But, by definition,
\[
\bs\varphi^\text{II}(\bs\varphi^\text{I}(\bs{a})) \!=\! \bs\psi^\text{II}(\bs\psi^\text{I}(\bs{a}^\text{I}+b^\text{I}\bs\delta)+b^\text{II}\bs\delta) \!=\! \bs\psi^\text{II}(\bs\psi^\text{I}(\bs{q}-b^{\text{II}}\bs\delta)+b^{\text{II}}\bs\delta),
\]
where the final equality holds because, according to (\ref{E:vecArrival}),
\[
\bs{q} = \bs{a}+b\bs\delta = \bs{a}^\text{I} + b^\textup{I}\bs\delta + b^\textup{II}\bs\delta.
\]
Therefore, according to (\ref{E:composability}), $\bs{d} \geq \bs\psi(\bs{q})$ for all $\bs{q}\in\mb{U}\ua b$, and $\bs\psi$ is guaranteed.
\end{IEEEproof}

Using (\ref{E:psiM}), it is easy to verify that min-plus services are monotone. They are also composable as shown by the next theorem.

\begin{theorem}\label{T:MPSComposability}
{\it In Theorem~\ref{T:composability}, if $\bs\psi^\textup{I}$ and $\bs\psi^\textup{II}$ are both min-plus services that can be identified by cumulative matrices $M^\textup{I}$ and $M^\textup{II}$, respectively, that is, $\bs\psi^\textup{I} = \bs\psi^{M^\textup{I}}$ and $\bs\psi^\textup{II} = \bs\psi^{M^\textup{II}}$, then $\bs\psi=\bs\psi^\textup{II}\star\bs\psi^\textup{I}$ is also a min-plus service that can be identified by a cumulative matrix,
\begin{equation}\label{E:MPSComposability}
M = (M^\text{I}+b^\text{II} \Delta) \otimes M^\text{II},
\end{equation}
where $\Delta = [\delta_{ij}]_{i,j\in\mb{N}}$ is a cumulative matrix, with
\begin{equation}\label{E:Delta}
\delta_{ij} := (\delta_j-\delta_i)^+ = \left\{
\begin{IEEEeqnarraybox}[][c]{ll}
1 & \blank\blank\textup{if } i=0 \textup{ and } j>0\\
0 & \blank\blank\textup{otherwise}
\end{IEEEeqnarraybox}\right. ~.
\end{equation}}
\end{theorem}

According to this theorem, if a flow passes through a network of servers and is guaranteed a min-plus service by each server along its path, its end-to-end service can be modeled by a single min-plus service. This makes min-plus services amenable to network performance analysis. Notice that, in (\ref{E:MPSComposability}), even if $M^\text{I}$ and $M^\text{II}$ are both spectral matrices, $M$ is not guaranteed to be one. Of course, we can always use (\ref{E:MtoS}) to turn $M$ into $S$, but the simplicity of (\ref{E:MPSComposability}) is then lost.

\begin{IEEEproof}[Proof of Theorem~\ref{T:MPSComposability}]
For all $\bs{q}\in\mb{U}\ua b$, if $j>0$, application of the matrix multiplication rule, yields
\[
\begin{IEEEeqnarraybox}[][c]{rCl}
[(\bs{q}-b^\text{II}\bs\delta) \otimes M^\text{I}+b^\text{II}\bs\delta]_j &=& \min_{i\in\mb{N}} (q_i-b^\text{II}\delta_i+m^\text{I}_{ij})+b^\text{II}\delta_{j}\\
                                                                          &=& \min_{i\in\mb{N}} (q_i+m^\text{I}_{ij}+b^\text{II}\delta_{ij})\\
                                                                          &=& [\bs{q} \otimes (M^\text{I}+b^\text{II}\Delta)]_j,
\end{IEEEeqnarraybox}
\]
where the second equality holds because, according to (\ref{E:Delta}), $\delta_{ij} = \delta_j-\delta_i$ if $j>0$. By default, both sides of this expression equal $0$ if $j=0$. So
\[
(\bs{q}-b^\text{II}\bs\delta) \otimes M^\text{I}+b^\text{II}\bs\delta = \bs{q} \otimes (M^\text{I}+b^\text{II}\Delta).
\]
Using (\ref{E:composability}), (\ref{E:psiMVec}), (\ref{E:MPSComposability}), and $\otimes$'s associativity, we then have
\[
\begin{IEEEeqnarraybox}[][c]{rCl}
\bs\psi(\bs{q}) &=& \bs\psi^{M^\textup{II}} \star \bs\psi^{M^\textup{I}}(\bs{q})                                \\
                &=& [(\bs{q}-b^\text{II}\bs\delta) \otimes M^\text{I}+b^\text{II}\bs\delta] \otimes M^\text{II} \\
                &=& \bs{q} \otimes (M^\text{I}+b^\text{II}\Delta) \otimes M^\text{II}                           = \bs{q} \otimes M.
\end{IEEEeqnarraybox}
\]
So $\bs\psi = \bs\psi^M$.
\end{IEEEproof}

\section{Dual-Curve Services}\label{S:DVS}

\begin{figure}[t]
\centering \scalebox{0.875}{\includegraphics{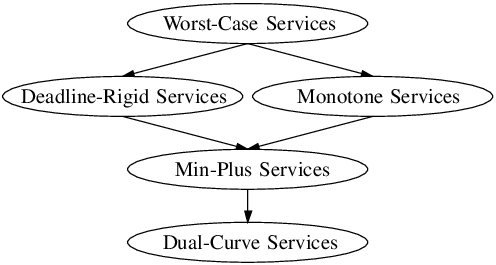}} \caption{Dual-curve services' relation to other worst-case services.}\label{F:ServiceAtlas}
\end{figure}

According to $M$'s update rule, (\ref{E:MUpdate}), if $m_{ij} = m_{i+1, j+1}$ for all $i > 0$, then $\dot{m}_{ij} = m_{ij}$ for all $i > 0$. In this case, while the $0$th row of $M$ remains a dynamic cumulative vector, all subsequent rows are static. As the elements in these rows are constant along $M$'s diagonal and off-diagonals, efficiency can be further improved by compressing them into a single, static cumulative vector. Such is the motivation for dual-curve services. Their relation to other worst-case services is shown in Fig.~\ref{F:ServiceAtlas}. Being the most specialized of these services, dual-curve services are also the most efficient, approaching near practical viability while maintaining all essential features of our general framework.

In the remainder of this section, we first summarize key properties of dual-curve services and then explore their relation to service curves. We also comment on some practical issues.

\subsection{Key Properties}\label{SS:VPS}

\begin{definition}\label{D:DVS}
{\it Given a pair of cumulative vectors, \mbox{$\bs{u}, \bs{v} \in \mb{U}$}, construct a cumulative matrix, $M^{(\bs{u}, \bs{v})} = [m_{ij}^{(\bs{u}, \bs{v})}]_{i,j\in\mb{N}}$, such~that}
\begin{equation}\label{E:rs}
m_{ij}^{(\bs{u}, \bs{v})}:=\left\{
\begin{IEEEeqnarraybox}[][c]{ll}
u_j             & \blank\blank\text{if }i=0\\
v_{(j-i)^+}     & \blank\blank\text{if }i>0
\end{IEEEeqnarraybox}\right. ~.
\end{equation}
{\it The min-plus service identified by $M^{(\bs{u}, \bs{v})}$, $\bs\psi^{M^{(\bs{u}, \bs{v})}}$, is called the \textbf{dual-curve service} identified by $(\bs{u}, \bs{v})$ and denoted by~$\bs\psi^{(\bs{u}, \bs{v})}$.}
\end{definition}

Most dual-curve service properties are straightforward specializations of the analogous min-plus service properties. We list some important ones here without derivation. First, using (\ref{E:psiM+}) and (\ref{E:rs}), it is easy to verify that
\begin{equation}\label{E:psiuv}
\psi_j^{(\bs{u}, \bs{v})}(\bs{q}) = \min \left\{u_j, \min_{0 < i \leq j} (q_i\!+\!v_{j-i}) \right\} \blank\blank\forall\, \bs{q} \in \mb{U} \ua b, j\in\mb{N}.
\end{equation}
Second, dual-curve services are update invariant. In Theorem~\ref{T:MUpdate}, if $M$ can be identified by $(\bs{u}, \bs{v})$, using (\ref{E:MUpdate}) and (\ref{E:rs}), it can be shown that $\dot{M}$ can be identified by $(\bs{\dot{u}}, \bs{v})$, so that $\bs{v}$, being static, remains the same, and $\bs{u}$, being dynamic, is updated to $\bs{\dot{u}}=[\dot{u}_j]_{j\in\mb{N}} \in \mb{U}$, where
\begin{equation}\label{E:rsUpdate}
\dot{u}_j = (\hat{u}_{j+1}-d)^+, \blank\text{with}\blank \hat{u}_{j+1} := \min\{u_{j+1}, q+v_j\},
\end{equation}
or in vector form,
\begin{equation}\label{E:rsUpdate+}
\bs{\dot{u}} = \mc{R}^{-1}(\bs{\hat{u}}-d\bs\delta)^+, \blank\text{with}\blank \bs{\hat{u}} := \min\{\bs{u}, q\bs\delta+\mc{R}\bs{v}\}.
\end{equation}
Finally, dual-curve services are composable. In Theorem~\ref{T:MPSComposability}, if $M^\text{I}$ and $M^\text{II}$ can be identified by $(\bs{u}^\text{I}, \bs{v}^\text{I})$ and $(\bs{u}^\text{II}, \bs{v}^\text{II})$, respectively, using (\ref{E:MPSComposability}) and (\ref{E:rs}), it can be shown that $M$ can be identified by $(\bs{u}, \bs{v})$, where
\begin{equation}\label{E:uvComposing}
\forall\, j\in\mb{N}, \blank\blank
\left\{
\begin{IEEEeqnarraybox}[][c]{l}
u_j = \min \left\{u_j^\text{II}, \min_{0 < i \leq j} (u_i^\text{I} + b^\text{II} + v_{j-i}^\text{II})\right\}\\
v_j = \min_{i \leq j} (v_i^\text{I} + v_{j-i}^\text{II})
\end{IEEEeqnarraybox}\right. ~.
\end{equation}

Consider next the spectra and conditional spectra of dual-curve services. For all $i, j \in \mb{N}$, according to (\ref{E:rs}), and Theorems~\ref{T:SSig} and \ref{T:MtoS},
\begin{equation}\label{E:rsMtoS}
\lambda_{ij}^{(\bs{u}, \bs{v})} = s_{ij}^{(\bs{u}, \bs{v})} = \left\{
\begin{IEEEeqnarraybox}[][c]{ll}
u_j                                        & \blank\blank\text{if }i=0\\
\min \left\{(u_j-b)^+, v_{(j-i)^+}\right\} & \blank\blank\text{if }i>0
\end{IEEEeqnarraybox}\right. ~,
\end{equation}
where $\lambda_{ij}^{(\bs{u}, \bs{v})}$ is short-hand for $\lambda_{ij}(\bs\psi^{(\bs{u}, \bs{v})})$. Similarly, according to (\ref{E:rsMtoS}) and (\ref{E:rsUpdate}), and Theorem~\ref{T:SCondSig},
\begin{equation}\label{E:rsMCondSig}
\hat\lambda_{ij}^{(\bs{u}, \bs{v})} = \hat{s}_{ij}^{(\bs{u}, \bs{v})} = \left\{
\begin{IEEEeqnarraybox}[][c]{ll}
\hat{u}_j     & \blank\blank\text{if } i=0\\
\min \{(\hat{u}_j-q)^+, v_{(j-i)^+}\} & \blank\blank\text{if } i>0
\end{IEEEeqnarraybox}\right. ~,
\end{equation}
where $\hat\lambda_{ij}^{(\bs{u}, \bs{v})}$ is short-hand for $\lambda_{ij}(\bs\psi^{(\bs{u}, \bs{v})}|q)$.

As dual-curve services are min-plus services, and min-plus services are uniquely identified by their spectra, an immediate corollary of (\ref{E:rsMtoS}) is that $\bs\psi^{(\bs{u}, \bs{v})}=\bs\psi^{(\bs{u'}, \bs{v'})}$ if and only if $\bs{u}=\bs{u'}$ and
\begin{equation}\label{E:uvEqCond}
\min\{(u_\infty-b)^+\bs\delta, \bs{v}\} = \min\{(u_\infty-b)^+\bs\delta, \bs{v'}\}.
\end{equation}
On the one hand, for instance, if $u_\infty \leq b$, each unique $\bs{u}$ identifies a unique $\bs\psi^{(\bs{u}, \bs{v})}$ irrespective of $\bs{v}$. On the other hand, if $u_\infty = \infty$, each unique $(\bs{u}, \bs{v})$ identifies a unique $\bs\psi^{(\bs{u}, \bs{v})}$, and if, additionally, $v_\infty = \infty$, according to (\ref{E:rsUpdate}), this property of unique identification is update invariant. So we call $(\bs{u}, \bs{v})$ {\it \textbf{non-degenerate}} if $u_\infty = \infty$ and $v_\infty = \infty$.

\subsubsection*{\textup{[}Dual-Curve Systems\textup{]}} A {\it \textbf{dual-curve system}} is a min-plus system in which every min-plus service is a dual-curve service. It is update invariant because its constituent dual-curve services are update invariant. For all $\omega\in\Omega$, let the dual-curve service guaranteed to flow $\omega$ be identified by $(\bs{u}^\omega, \bs{v}^\omega)$, and denote the resulting dual-curve system by $(\bs{u}^{[\Omega]}, \bs{v}^{[\Omega]})$. Using (\ref{E:rsMtoS}), and Theorem~\ref{T:schedulability}, it is easy to verify that $(\bs{u}^{[\Omega]}, \bs{v}^{[\Omega]})$ is schedulable if and only if
\begin{equation}\label{E:uvschedulability}
\max\left\{u_j^{\l\Omega\r}, \sum_{\omega\in\Omega} \min \left\{(u_\infty^\omega-b^\omega)^+, v_j^\omega\right\}\right\} \leq jc \blank\blank\forall\, j\in\mb{N}.
\end{equation}
If $(\bs{u}^{[\Omega]}, \bs{v}^{[\Omega]})$ is {\it \textbf{non-degenerate}}, that is, if, for all $\omega\in\Omega$, $(u^\omega, v^\omega)$ is non-degenerate, it is immediate that (\ref{E:uvschedulability}) can be rewritten in vector form as
\begin{equation}\label{E:uvschedulability+}
\max\{\bs{u}^{\l\Omega\r}, \bs{v}^{\l\Omega\r}\} \leq \bs{c} = [c_j]_{j\in\mb{N}} := [jc]_{j\in\mb{N}}.
\end{equation}

For all $\Gamma \subseteq \Omega$, using (\ref{E:beta}),(\ref{E:rsMCondSig}), and (\ref{E:rsUpdate}), we have
\begin{IEEEeqnarray}{rCl}
\beta(\Gamma) &=& \max_{j\in\mb{N}} \left(\sum_{\omega\in\Gamma}\hat\lambda_{0, j+1}^{(\bs{u}^\omega, \bs{v}^\omega)}+\sum_{\omega\in\overline\Gamma}\hat\lambda_{1, j+1}^{(\bs{u}^\omega, \bs{v}^\omega)}-jc\right)\IEEEnonumber\\
              &=& \max_{j \in \mb{N}} \left(
\hat{u}_{j+1}^{\l\Gamma\r} + \sum_{\omega\in\overline\Gamma} (\hat{u}_{j+1}^\omega-q^\omega)^+ - jc
\right).\label{E:betaDVS}
\end{IEEEeqnarray}
According to (\ref{E:p+}), (\ref{E:rsMCondSig}), and (\ref{E:rsUpdate}), we also have
\begin{equation}\label{E:puv}
p_j^{(\bs{u}, \bs{v})} = \min \{\hat{u}_j, q\} = \min \{u_j, q\} \blank\blank\forall\, j \in \mb{N},
\end{equation}
or in vector form,
\begin{equation}\label{E:vecpuv}
\bs{p}^{(\bs{u}, \bs{v})} = \min \{\bs{\hat{u}}, q\bs\delta\} = \min \{\bs{u}, q\bs\delta\}.
\end{equation}
These specializations enable us to use Theorems~\ref{T:polytope} and \ref{T:FP} to determine the feasible polytopes and permutohedra of dual-curve systems, and to use Definitions~\ref{D:MSS} and \ref{D:HMSS} to identify the systems' max-slack schedules and per-class extensions.

As we have seen in Section~\ref{SS:MPHull}, a schedulable worst-case system, $\bs\psi^{[\Omega]}$, can always be upgraded to its schedulable min-plus hull, $S^{[\Omega]}$. This hull is, in turn, dominated by the dual-curve system, $(\bs{u}^{[\Omega]}, \bs{v}^{[\Omega]})$, if
\begin{equation}\label{E:dvHull}
\forall\, \omega\in\Omega, j\in\mb{N}, \blank\blank
\left\{
\begin{IEEEeqnarraybox}[][c]{ll}
u_j^\omega = s_{0j}^\omega\\
v_j^\omega = \max_{i>0} s_{i, i+j}^\omega
\end{IEEEeqnarraybox}\right. ~,
\end{equation}
because then, according to (\ref{E:rs}), $S^\omega \leq M^{(\bs{u}^\omega, \bs{v}^\omega)}$. We call $(\bs{u}^{[\Omega]}, \bs{v}^{[\Omega]})$ the {\it \textbf{dual-curve hull}} of $\bs\psi^{[\Omega]}$. Unlike $S^{[\Omega]}$, however, there is no guarantee that $(\bs{u}^{[\Omega]}, \bs{v}^{[\Omega]})$ will be schedulable even if $\bs\psi^{[\Omega]}$ is schedulable. But when it is schedulable, $\bs\psi^{[\Omega]}$ can be further upgraded to $(\bs{u}^{[\Omega]}, \bs{v}^{[\Omega]})$, endowing the efficiency of dual-curve systems with a broader significance.

\begin{example}\label{Ex:dvHull}
{\it Let $\bs\psi^{[\Omega]}$ be a single-task system. Its min-plus hull, $S^{[\Omega]}$, is identified by (\ref{E:hullEx}), so, according to (\ref{E:dvHull}), its dual-curve hull, $(\bs{u}^{[\Omega]}, \bs{v}^{[\Omega]})$, must set $\bs{u}^\omega = \bs{v}^\omega = \mc{R}^{\bar\theta^\omega} \bs\delta$ for all $\omega\in\Omega$, implying that $(\bs{u}^\omega, \bs{v}^\omega)$ must be degenerate. Notice that, in this case, $S^{[\Omega]}$ is equivalent to $(\bs{u}^{[\Omega]}, \bs{v}^{[\Omega]})$ because, according to (\ref{E:hullEx}) and (\ref{E:rs}), $S^\omega = M^{(\bs{u}^\omega, \bs{v}^\omega)}$ for all $\omega\in\Omega$. Therefore, if the single-task system is schedulable, so is its dual-curve hull.
}
\end{example}

\subsection{Relation to Service Curves}\label{SS:literature}

If $\bs{u} = \bs{v}$, according to (\ref{E:psiuv}),
\begin{equation}\label{E:psivv}
\psi_j^{(\bs{v}, \bs{v})}(\bs{q}) = \min_{i \leq j} (q_i+v_{j-i}) \blank\blank\forall\, \bs{q} \in \mb{U} \ua b, j\in\mb{N}.
\end{equation}
This special case recovers the well-known min-plus convolution defining service curves. But, according to \cite{Cruz:1995}, to specify service using service curves, it is necessary that $b=0$. So, only in the case that $b=0$ is $\bs\psi^{(\bs{v}, \bs{v})}$ equivalent to the service curve specified by $\bs{v}$. For instance, since the definition of single-task services requires that $b=0$, it follows, from Example~\ref{Ex:dvHull}, that the dual-curve hull of a single-task system is equivalent to a system of degenerate service curves.

As the relations $\bs{u} = \bs{v}$ and $b=0$ are not preserved~by, respectively, (\ref{E:rsUpdate+}) and (\ref{E:bUpdate}), service curves are not update invariant. Accordingly, by introducing dual-curve services, we are only extending service curves to their dynamic closure, that is, we are only allowing for cases where $\bs{u} \neq \bs{v}$ or $b>0$. But this extension is essential to state-based scheduling because we cannot update a state to something yet undefined.

Notice, in (\ref{E:uvComposing}), that if $\bs{u}^\text{I} = \bs{v}^\text{I}$, $\bs{u}^\text{II} = \bs{v}^\text{II}$, and $b^\text{I} = b^\text{II} =0$,
\begin{equation}\label{E:scComposing}
u_j = v_j = \min_{i \leq j} (v_i^\text{I} + v_{j-i}^\text{II}) \blank\blank\forall\, j\in\mb{N}.
\end{equation}
Clearly, after composition, $\bs{u}=\bs{v}$ and $b = b^\text{I}+b^\text{II} = 0$. So (\ref{E:scComposing}) recovers the well-known convolutional service curve composition rule.

Comprehensive introductions to service curves can be found in \cite{Chang:2000, Boudec:2001, Bouillard:2018}. As special cases of dual-curve services, their well-documented versatility makes clear the versatility of dual-curve services, and by extension, min-plus, and general worst-case, services.

In \cite{Chang:1998}, it was observed that service curves can be viewed as time-invariant min-plus filters, whose time-invariant nature is reflected in the fact that, according to (\ref{E:psivv}), if $b = 0$, implying that $\mc{R}^i\bs\varepsilon \in \mb{U} \ua b$,
\begin{equation}\label{E:ti}
\bs\psi^{(\bs{v}, \bs{v})}(\mc{R}^i\bs\varepsilon) = \mc{R}^i\bs{v} \blank\blank\forall\, i\in\mb{N}.
\end{equation}
This observation led to the introduction of time-varying min-plus filters in \cite{Chang:2002}. The relation of these filters to min-plus services exactly parallels that of service curves to dual-curve services. In particular, only in the case that $b=0$, are min-plus services equivalent to these filters, whose time-varying nature is reflected in the fact that, according to (\ref{E:psiS}), if $b=0$,
\begin{equation}\label{E:tv}
\bs\psi^S(\mc{R}^i\bs\varepsilon) = S_{i\bs\cdot} \blank\blank\forall\, i\in\mb{N},
\end{equation}
where $S_{i\bs\cdot}$ denotes the $i$th row of $S$.

\subsubsection*{\textup{[}The EDF Scheduler\textup{]}} Since min-plus services are deadline-rigid, so are dual-curve services, and thus service curves. Taking advantage of this fact, the service-curve-earliest-deadline scheduler was proposed in \cite{Sariowan:1999} to guarantee service curves. At its core, it is but a work-conserving EDF scheduler that automatically generates max-slack schedules. Interestingly, it was realized in \cite{Stoica:2000} that, from time to time, the EDF scheduler need not be work-conserving, and that in these instances, the unused capacity can be put to good use.

Recall, from Section~\ref{SS:MSSProp}, that work-conserving schedules set $\mu = \min\{c, q^{\l\Omega\r}\}$. In our terminology, \cite{Sariowan:1999} proposed to set the EDF scheduler's $\mu = \min\{c, q^{\l\Omega\r}\}$. In contrast, \cite{Stoica:2000} proposed to set the EDF scheduler's $\mu = \beta(\Omega)$ so that it automatically generates $e_{\beta(\Omega)}^{[\Omega]} \in \mb{P}(\beta)$, where $\mb{P}(\beta)$ is the baseline permutohedron. Then, as suggested in Section~\ref{SS:how}, excess capacity, $c-\beta(\Omega)$, can be freely allocated using, for instance, a GPS policy. Both \cite{Sariowan:1999}, and \cite{Stoica:2000}, implicitly explored additional dynamic features of dual-curve services, but still required that service specifications be limited to the case that $b=0$.

Of course, by restricting attention to EDF schedulers, we lose almost all scheduling flexibility. In particular, the remedy proposed in \cite{Stoica:2000} is only partial and opportunistic. This is showcased in the next example by comparing the scheduling flexibility of an EDF scheduler to that of a state-\linebreak based scheduler.

\begin{example}\label{Ex:EDF}
{\it Consider a two-flow dual-curve system identified by $(b^1\mc{R}^{\bar\theta}\bs\delta, \bs{0})$ and $(b^2\mc{R}^{\bar\theta+1}\bs\delta, \bs{0})$. This specification requires that the $b^1$ tasks in flow $1$'s buffer and the $b^2$ tasks in flow $2$'s buffer be served no later than slot $t+\bar\theta$ and $t+\bar\theta+1$, respectively, but nothing more. Suppose, also, that
\[
b^1 \leq \bar\theta c, \blank b^2 \leq (\bar\theta + 1) c \blank\textup{and}\blank b^1+b^2 = (\bar\theta+2)c,
\]
implying that $b^1 \geq c$, $b^2 \geq 2 c$ and $\bar\theta \geq 1$. According to (\ref{E:uvschedulability}) and (\ref{E:betaDVS}), these conditions not only guarantee schedulability, but also ensure that
\[
\beta(\{1\}) = \beta(\{2\}) = 0 \blank\textup{and}\blank \beta(\{1, 2\}) = c.
\]
So, in slot~$t$, although both schedulers must be work-conserving, the state-based scheduler can freely serve flow $1$, flow $2$, or both in any proportion, while the EDF scheduler has no choice but to serve flow $1$ only. Notice that, in this case, \cite{Stoica:2000}'s proposal is of no help because there is no excess capacity to allocate.

Furthermore, if $b^1$, $b^2$, and $\bar\theta$ are large enough, this sharp contrast in scheduling flexibility can persist for a long time. For instance, let $b^1 = b^2 = 50 c$ and $\bar\theta = 98$. In this case, although both schedulers must be work-conserving during interval $[t, t+100)$, the state-based scheduler can freely apportion service to both flows during $[t, t+99)$ as long as only flow $2$ is served in slot $t+99$, while the EDF scheduler has no choice but to first serve flow $1$ during $[t, t+50)$ and then serve flow $2$ during $[t+50, t+100)$.
}
\end{example}

\subsection{Practical Issues}\label{SS:practical}

We expect state-based schedulers, when implemented, will first be implemented as dual-curve systems because of their efficiency. We further speculate that, initially, applications in logistic systems and data centers will be most promising. In the former case, since the natural scheduling slot may last for hours, days, or even weeks, there is time for significant computation during each slot. In the latter case, since the cost of maintaining huge data storage, processing and distribution capacities is immense, there exist strong incentives for improving utilization, even if only by a tiny fraction.\footnote{Abstractly, a data center can be viewed as a collection of independent servers that run in parallel. Ideally, it should allocate flows to servers to both maintain all flows' service guarantees and balance all servers' workloads. Suppose that each server is equipped with a non-degenerate dual-curve system. Then, according to (\ref{E:uvschedulability+}), its capacity utilization can be measured by $\bs{u}^{\l\Omega\r}/c$ and $\bs{v}^{\l\Omega\r}/c$. As both are cumulative vectors, achieving exact balance across all servers is unlikely, but improvement may still be possible.}

Despite their efficiency, it is still impossible to implement state-based schedulers for dual-curve systems in their most general form because, well, countably infinite is still infinite. Somehow, we need to further reduce the curves' dimensionality. Two approaches are outlined below.

\subsubsection*{\textup{[}The Piecewise-Linear-Curve Approach\textup{]}} We call a dual-curve service, $\bs\psi^{(\bs{u}, \bs{v})}$, {\it \textbf{piecewise-linear}} if both $\bs{u}$ and $\bs{v}$ are piecewise-linear. In this case, both $\bs{u}$ and $\bs{v}$ can be represented by a finite number of linear segments or turning points, and thus all computation can be carried out on a finite basis. For instance, to check schedulability using (\ref{E:uvschedulability}) or (\ref{E:uvschedulability+}), or to calculate $\beta$ using (\ref{E:betaDVS}), we need only account for those turning points that are relevant. An insightful discussion of piecewise-linear functions, their representation, their classification, and their efficient manipulation can be found in \cite{Bouillard:2018} (ch. 4).

The piecewise-linear property is update invariant because, according to (\ref{E:rsUpdate+}), if both $\bs{u}$ and $\bs{v}$ are piecewise-linear, so is $\bs{\dot{u}}$. As time goes by, however, $\bs{u}$ may grow unwieldy, that is, the number of linear segments needed to represent $\bs{u}$ may grow without bound. How can this growth be contained? In \cite{Sariowan:1999, Stoica:2000}, problems of a similar nature were encountered. To overcome them, additional restrictions were imposed on piecewise-linear service curves to cap the number of segments. The same techniques can be applied to piecewise-linear dual-curve services, but we will not pursue them here. Instead, we outline two new techniques that are unique to our framework and hold great promise.\footnote{These techniques, we believe, make piecewise-linear dual-curve systems manageable in most practical cases. Of course, the final arbiter of this can only be implementation, which is beyond the scope of this paper.}

As state-based scheduling is dynamic, in each slot, the scheduler is free to upgrade $(\bs{u}^{[\Omega]}, \bs{v}^{[\Omega]})$ to any schedulable $(\bs{\bar{u}}^{[\Omega]}, \bs{v}^{[\Omega]})$ such that $\bs{\bar{u}}^{[\Omega]} \geq \bs{u}^{[\Omega]}$, or to select any feasible $d^{[\Omega]}$, each of which will induce a different $(\bs{\dot{u}}^{\raisebox{-0.3mm}{\scriptsize$[\Omega]$}}, \bs{v}^{[\Omega]})$. The idea is to use this freedom, when possible, to simplify $\bs{u}^{[\Omega]}$ or $\bs{\dot{u}}^{\raisebox{-0.3mm}{\scriptsize$[\Omega]$}}$. On the one hand, given that $(\bs{u}^{[\Omega]}, \bs{v}^{[\Omega]})$ is schedulable, if there exists a less unwieldy $\bs{\bar{u}}^{[\Omega]}$ such that $(\bs{\bar{u}}^{[\Omega]}, \bs{v}^{[\Omega]})$ is also schedulable, we can upgrade $(\bs{u}^{[\Omega]}, \bs{v}^{[\Omega]})$ to $(\bs{\bar{u}}^{[\Omega]}, \bs{v}^{[\Omega]})$ to simplify $\bs{u}^{[\Omega]}$. If $(\bs{u}^{[\Omega]}, \bs{v}^{[\Omega]})$ is also non-degenerate, according to (\ref{E:uvschedulability+}), finding a less unwieldy $\bs{\bar{u}}^{[\Omega]}$ such that $\bs{\bar{u}}^{\l\Omega\r} \leq \bs{c}$ suffices. Of course, this upgrade may reduce the server's capacity slack, but it may be worth it if the number of linear segments needed to represent $\bs{u}^{[\Omega]}$ can be significantly reduced. On the other hand, according to (\ref{E:rsUpdate+}), the larger $d^\omega$ is, the simpler $\bs{\dot{u}}^{\raisebox{-0.3mm}{\scriptsize$\omega$}}$ tends to become, because more of $\bs{u}^\omega$'s segments tend to be buried below $0$ when $\bs{u}^\omega$ is updated. So, to simplify $\bs{\dot{u}}^{\raisebox{-0.3mm}{\scriptsize$[\Omega]$}}$, the more unwieldy $\bs{u}^\omega$ becomes, the higher the priority that should be given to flow $\omega$. Notice that these techniques can be applied concurrently, and to all dual-curve systems, not just those that are piecewise-linear.

\begin{example}\label{Ex:dv}
{\it Given that $(\bs{u}^{[\Omega]}, \bs{v}^{[\Omega]})$ is schedulable, when $d^{[\Omega]}=q^{[\Omega]}$, it is immediate from (\ref{E:rsUpdate+}) that $\bs{\dot{u}}^{\raisebox{-0.3mm}{\scriptsize$[\Omega]$}} \leq \bs{v}^{[\Omega]}$. So, if $(\bs{u}^{[\Omega]}, \bs{v}^{[\Omega]})$ is non-degenerate, implying, according to (\ref{E:uvschedulability+}), that $\bs{v}^{\l\Omega\r} \leq \bs{c}$, $(\bs{\dot{u}}^{\raisebox{-0.3mm}{\scriptsize$[\Omega]$}}, \bs{v}^{[\Omega]})$ can be upgraded to $(\bs{v}^{[\Omega]}, \bs{v}^{[\Omega]})$. Since $d^{[\Omega]}=q^{[\Omega]}$ implies that $\dot{b}^\omega = 0$ for all $\omega\in\Omega$, it follows that, whenever all flows' buffers are emptied, a non-degenerate dual-curve system can be upgraded to a system of non-degenerate service curves. A corollary is that, starting from a system of non-degenerate service curves, we can always return to the same system of service curves whenever all flows' buffers are emptied, and then these renewal epochs allow us to activate the scheduler only during busy periods, as is well known \cite{Sariowan:1999}.}
\end{example}

\subsubsection*{\textup{[}The Transient-Curve Approach\textup{]}} Although in reality all service guarantees are transient, in most cases, the guarantees are either too long or too indefinite to be taken as transient. That said, there are still cases where the guarantees become both short enough and definite enough that they can be considered transient, for instance, the service guaranteed to cargoes conveyed via a transportation line, or files transferred via a communication link. In these cases, flows' service specifications can be further simplified.

We call a worst-case service, $\bs\psi$, {\it \textbf{transient}} if there exists a finite $g > 0$ such that $\psi_j(\bs{q}) = \psi_g(\bs{q})$ for all $\bs{q} \in \mb{U} \ua b$ and $j \geq g$. Transiency limits $\bs\psi$'s guarantee to interval \mbox{$[t, t+g)$}. Additionally, we call a cumulative matrix, $M$, {\it \textbf{transient}} if $m_{ij} = m_{ig}$ for all $j \geq g$. According to (\ref{E:CMDefA}), this implies that $m_{ij} = 0$ for all $i \geq g$. So $M$ is solely determined by $M_{[0:g] \times [0:g]}$, the $(g+1) \times (g+1)$ submatrix in the upper-left corner of $M$.

Certainly $\bs\psi^M$ is transient if $M$ is transient. But, unless $u_j = u_g$ for all $j \geq g$ and $\bs{v} = \bs{0}$, $M^{(\bs{u}, \bs{v})}$ cannot be transient because, according to (\ref{E:rs}), if $u_j > u_g$ for some $j > g$, $m_{0j}^{(\bs{u}, \bs{v})} > m_{0g}^{(\bs{u}, \bs{v})}$ and, if $v_{j-g} > 0$ for some $j > g$, \mbox{$m_{gj}^{(\bs{u}, \bs{v})} > 0$}. So typical dual-curve services are non-transient.\footnote{Our argument here is not airtight. The {\it caveat} is that $M$'s transiency is sufficient but not necessary for $\bs\psi^M$'s. To see this, recall, from Section~\ref{SS:MPSDef+}, that the same min-plus service may be identified by multiple cumulative matrices. So $\bs\psi^M$ is transient if {\it any} of these matrices is transient. According to Theorem~\ref{T:MtoS}, we can always construct an $S$ that identifies $\bs\psi^M$. Using (\ref{E:MtoS}), it is easy to verify that this $S$ is transient if $M$ is transient. As, according to Theorem~\ref{T:SSig}, this $S$ must be the same for all cumulative matrices that identify $\bs\psi^M$, it is transient if {\it any} of these matrices is transient. This suggests that $\bs\psi^M = \bs\psi^S$ is transient if and only if $S$ is transient, the necessity of which is confirmed by the fact that, for all $i,j\in\mb{N}$, according to (\ref{E:SMDefD}) and (\ref{E:epsilon++}),
\[
s_{ij} = (\min\{s_{0j}, b\delta_i+s_{ij}\}-b\delta_i)^+ = (\psi_j^S(\mc{R}^i\bs\varepsilon+b\bs\delta)-b\delta_i)^+.
\]
It follows that $\bs\psi^{(\bs{u}, \bs{v})}$ is transient if and only if $S^{(\bs{u}, \bs{v})}$ is transient. So, according to (\ref{E:rsMtoS}), $\bs\psi^{(\bs{u}, \bs{v})}$ is transient, not only if $u_j = u_g$ for all $j \geq g$ and $\bs{v} = \bs{0}$, but also if $u_j = u_g$ for all $j \geq g$ and $u_\infty \leq b$. This, however, does not contradict our claim that typical dual-curve services are non-transient.} In fact, no non-degenerate dual-curve service can be transient. Even in degenerate cases, dual-curve services need not be transient. Although both dual-curve services in Example~\ref{Ex:EDF} are transient, the dual-curve services in the dual-curve hull of the single-task system in Example~\ref{Ex:dvHull}, like the single-task services themselves, are non-transient. These observations motivate the following definition.

\begin{definition}\label{D:DVS+}
{\it Given a pair of cumulative vectors, $\bs{u}, \bs{v} \in \mb{U}$, and a finite $g > 0$, construct a cumulative matrix, $M^{(\bs{u}, \bs{v}, g)} = [m_{ij}^{(\bs{u}, \bs{v}, g)}]_{i,j\in\mb{N}}$, such that}
\begin{equation}\label{E:rs++}
m_{ij}^{(\bs{u}, \bs{v}, g)} := m_{\min\{i, g\}, \min\{j, g\}}^{(\bs{u}, \bs{v})},
\end{equation}
or, according to (\ref{E:rs}),
\begin{equation}\label{E:rs+}
m_{ij}^{(\bs{u}, \bs{v}, g)} := \left\{
\begin{IEEEeqnarraybox}[][c]{ll}
u_{\min\{j, g\}}                      & \blank\blank\text{if }i=0\\
v_{(\min\{j, g\}-\min\{i, g\})^+}     & \blank\blank\text{if }i>0
\end{IEEEeqnarraybox}\right. ~.
\end{equation}
{\it The min-plus service identified by $M^{(\bs{u}, \bs{v}, g)}$, $\bs\psi^{M^{(\bs{u}, \bs{v}, g)}}$, is called the \textbf{dual-transient-curve service} identified by $(\bs{u}, \bs{v}, g)$ and denoted by $\bs\psi^{(\bs{u}, \bs{v}, g)}$.}
\end{definition}

By definition, $M^{(\bs{u}, \bs{v}, g)}$ is transient, so $\bs\psi^{(\bs{u}, \bs{v}, g)}$ is transient. Instead of $\bs{u}$ and $\bs{v}$, only $\bs{u}_{[0:g]}$ and $\bs{v}_{[0:g-1]}$ matter in $\bs\psi^{(\bs{u}, \bs{v}, g)}$'s specification, which explains why we call these services dual-transient-curve services.\footnote{We call them dual-transient-curve services, instead of transient-dual-curve services, to distinguish them from dual-curve services that are transient, which, according to the preceding footnote, can arise when $u_j = u_g$ for all $j \geq g$, and either  $u_\infty \leq b$ or $\bs{v} = \bs{0}$.}  Analogous to the dual-curve service properties surveyed in Section~\ref{SS:VPS}, we can easily develop dual-transient-curve counterparts. For instance, we can still use (\ref{E:rsUpdate+}) to update $\bs{u}$. But, since $\dot{g} = g-1$ by default, $(\bs{u}, \bs{v}, g)$ needs to be updated to $(\bs{\dot{u}}, \bs{v}, g-1)$. Moreover, we can even make dual-transient-curve services piecewise-linear, thereby combining the two approaches.

\section{Concluding Remarks}\label{S:remarks}

In this paper, we have shed new light on the classical problem of using short-run scheduling decisions to provide long-run service guarantees. We have drawn our inspiration from a variety of abstractions, including: cumulative vectors, the state-space approach, polymatroid theory, EDF scheduling, the min-plus algebra, and service curves. Although, individually, none of these abstractions is new, it has been our contribution to weave them all into a general framework for guaranteeing worst-case services using state-based scheduling that provides novel solutions to the classical problem.

Along the way we have noted several topics for future work. For instance, we have outlined how to analyze performance bounds when $\bs{q}$ is uncertain in Section~\ref{SS:WCSDef}, how to partition $\Omega$ to trade flexibility for efficiency in Section~\ref{SS:howHMSS}, and how to reduce the dimensionality of dual-curve systems in Section~\ref{SS:practical}. Of course, there are many other topics that we have not noted. Among these, some of the most interesting are suggested by generalizing our capacity model.

For instance, comparing the definition of $c_{ij}$ in (\ref{E:schedulability}) to that of $\bs{c}$ in (\ref{E:uvschedulability+}), it is immediate that
\begin{equation}\label{E:cij}
c_{ij}= (c_j-c_i)^+ \blank\blank\forall\, i,j\in\mb{N}.
\end{equation}
Given this formulation, if we allow $\bs{c}$ to be an arbitrary cumulative vector, then, the server's capacity is $c_{j+1}-c_j$ in slot $t+j$, and we obtain a {\it \textbf{time-varying}} capacity model. It is straightforward to extend our results, with minimal modifications, to this model.

Alternatively, the capacity constraint, (\ref{E:capacity}), can be generalized to $d^{[\Omega]} \in \mb{C}$. To sidestep integrality issues, let us also switch to a continuous traffic model so that $d^{[\Omega]}$ takes its value in $\mb{R}^n$ and $\mb{C}$ is a subset of $\mb{R}^n$.\footnote{As discussed following the proof of Theorem~\ref{T:HMSSCond}, although such integrality issues may be subtle, they are not insurmountable.} Now, for simplicity, let $\mb{C}$ be determined by
\begin{equation}\label{E:capacity+}
\sum_{\omega\in\Omega} \frac{d^\omega}{c^\omega} \leq 1.
\end{equation}
Then we obtain a {\it \textbf{linear}} capacity model. It is easy to extend our results to this model by scaling the worst-case service guaranteed to each flow accordingly. For instance, in this case, the schedulability condition, (\ref{E:schedulability}), becomes
\begin{equation}\label{E:schedulability++}
\sum_{\omega\in\Omega} \frac{\lambda_{ij}^\omega}{c^\omega} \leq (j-i)^+ \blank\blank\forall\, i, j \in \mb{N}.
\end{equation}

\begin{figure}[t]
\centering \scalebox{0.875}{\includegraphics{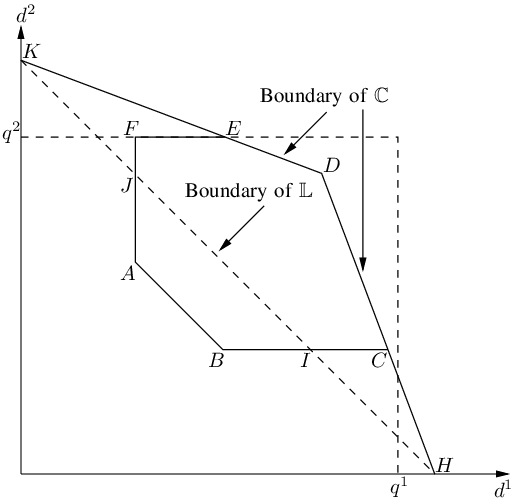}} \caption{The non-linear capacity model determined by two linear constraints in a two-flow case.}\label{F:ConvexCapacity}
\end{figure}

More generally, the specification of $\mb{C}$ can be non-linear. For instance, multiuser capacity in wireless systems is typically non-linear \cite{Tse:2005} (ch. 6). However, accounting for even simple non-linearities in $\mb{C}$'s specification poses a great challenge. Consider the two-flow case illustrated in Fig.~\ref{F:ConvexCapacity}, in which two linear constraints, the line segments $\overline{HD}$ and $\overline{KD}$, bound $\mb{C}$, modeling, for instance, a server with two linear bottlenecks internally, but no internal cache. Even in this simple case, it is not clear how to find the schedulability condition.\footnote{This case can be viewed as the intersection of two linear models. A naive guess is that it is schedulable if (\ref{E:schedulability++}) holds separately for each model. This is certainly a necessary condition, but it is not sustainable because there is no guarantee that the two models' feasible polytopes will always intersect.}

The heart of the problem lies in the fact that in the non-linear case, there is no obvious counterpart to the concept of spectrum, which, as we have seen in the linear case, helps us distill all essential information inherent in worst-case services that is relevant to schedulability. Therefore, a key bridge to schedulability is lost. This situation is not unlike that in system theory, where frequency-domain analysis, although powerful in the linear case, has no obvious counterpart in the non-linear case. We leave further study of this issue to future work.

Of course, the fact that it is hard to find the schedulability condition does not mean that it is also hard to find a sustainable condition. Given any $\mb{L} \subset \mb{C}$, where $\mb{L}$ is linear, the schedulability condition for $\mb{L}$, (\ref{E:schedulability++}), is sustainable for $\mb{C}$. So, once this condition is met, in each slot, we can first, as if the capacity is modeled by $\mb{L}$, determine the feasible polytope for $\mb{L}$, $\mb{F}_\mb{L}$, and then, as if $\mb{C}\setminus\mb{L}$ is some windfall capacity, expand $\mb{F}_\mb{L}$ accordingly. For instance, in Fig.~\ref{F:ConvexCapacity}, as the boundary of $\mb{L}$ is line segment $\overline{HK}$, $\mb{F}_\mb{L}$ is the trapezoid, $ABIJ$, which can be expanded to the hexagon, $ABCDEF$, from which $d^{[\Omega]}$ can be selected. In this instance, throughput, that is, $d^{\l\Omega\r} = d^1+d^2$, is maximized at point~$D$.

\appendix

In this appendix, we prove two lemmas from Section~\ref{SS:polymatroid}.

\begin{IEEEproof}[Proof of Lemma~\ref{L:non-empty}]
Denote $\mb{P}_\mc{S}(\chi)$ by $\mb{P}_\mc{S}$ for simplicity. If $\mb{P}_\mc{S}$ is non-empty, given any $d^{[\Omega]} \in \mb{P}_\mc{S}$, (\ref{E:P-S}) and (\ref{E:permutohedron}) ensure that, for all $\Gamma,\Gamma' \in \mc{S}$,
\[
\begin{IEEEeqnarraybox}[][c]{rCl}
\chi(\Gamma)+\chi(\Gamma') = d^{\l\Gamma\r}+d^{\l\Gamma'\r} &=& d^{\l\Gamma+\Gamma'\r}+d^{\l\Gamma\Gamma'\r}\\
& \geq & \chi(\Gamma+\Gamma')+\chi(\Gamma\Gamma').
\end{IEEEeqnarraybox}
\]
According to (\ref{E:supermodular}) and (\ref{E:permutohedron}), this implies that:
\begin{itemize}
\item[I1] if $\mb{P}_\mc{S}$ is non-empty, (\ref{E:supermodular}) must hold with equality for all $\Gamma,\Gamma' \in \mc{S}$; and
\item[I2] given any $d^{[\Omega]} \in \mb{P}_\mc{S}$, $d^{\l\Gamma+\Gamma'\r} = \chi(\Gamma+\Gamma')$ and $d^{\l\Gamma\Gamma'\r} = \chi(\Gamma\Gamma')$ for all $\Gamma,\Gamma' \in \mc{S}$;
\end{itemize}

When (\ref{E:supermodular}) holds with equality for specific $\Gamma,\Gamma'\subseteq\Omega$, we denote the relation imposed by this equality by $\Gamma \simeq \Gamma'$. By default, $\Gamma \simeq \Gamma'$ if $\Gamma\subseteq\Gamma'$ or $\Gamma'\subseteq\Gamma$. In the case that, for all $\Gamma,\Gamma'\subseteq\Omega$, $\Gamma\subseteq\Gamma'$ or $\Gamma'\subseteq\Gamma$ if $\Gamma \simeq \Gamma'$, we call $\chi$ {\it \textbf{strictly supermodular}}. According to I1, $\mb{P}_\mc{S}$'s non-emptiness implies that $\Gamma \simeq \Gamma'$ for all $\Gamma,\Gamma' \in \mc{S}$. So, if $\chi$ is strictly supermodular, $\mc{S}$ is a chain.

If $\chi$ is not strictly supermodular, there may exist $\Gamma,\Gamma'\in\mc{S}$ such that $\Gamma \simeq \Gamma'$ but $\Gamma\nsubseteq\Gamma'$ and $\Gamma'\nsubseteq\Gamma$, a relation that we denote by $\Gamma \sim \Gamma'$.\footnote{Although we will not do so here, it can be shown that $\mb{P}(\chi)$ is degenerate if and only if $\chi$ is not strictly supermodular.} In this case, we can replace $\Gamma$ and $\Gamma'$ by $\Gamma+\Gamma'$ and $\Gamma\Gamma'$ to form $\mc{S}'$ such that $\mb{P}_\mc{S'}=\mb{P}_\mc{S}$. To see this, notice that according to I2, $\mb{P}_\mc{S}\subseteq\mb{P}_{\mc{S}'}$, so we need only show $\mb{P}_{\mc{S}'}\subseteq\mb{P}_\mc{S}$. Given any $d^{[\Omega]}\in\mb{P}_{\mc{S}'}$, (\ref{E:P-S}), and the fact that $\Gamma\sim\Gamma'$, ensure that
\[
\begin{IEEEeqnarraybox}[][c]{rCl}
d^{\l\Gamma\r}+d^{\l\Gamma'\r} = d^{\l\Gamma+\Gamma'\r}+d^{\l\Gamma\Gamma'\r} &=& \chi(\Gamma+\Gamma')+\chi(\Gamma\Gamma')\\
&=& \chi(\Gamma)+\chi(\Gamma').
\end{IEEEeqnarraybox}
\]
According to (\ref{E:permutohedron}), this implies that $d^{\l\Gamma\r} = \chi(\Gamma)$ and $d^{\l\Gamma'\r} = \chi(\Gamma')$, so $d^{[\Omega]}\in\mb{P}_\mc{S}$.

If $\Gamma,\Gamma'\in\mc{S}'$ satisfying the $\sim$ relation remain, $\mc{S}'$ can be replaced by $\mc{S}''$ in the same manner that $\mc{S}$ was replaced by $\mc{S}'$, and so on. The question is whether through this process, all such $\sim$ pairs can be eliminated to arrive at a chain. The answer is yes, because $\mc{S}'$ is guaranteed to contain less $\sim$ pairs than $\mc{S}$. To see this, notice that, on the one hand, at least one $\sim$ pair, $\Gamma\sim\Gamma'$, is eliminated when $\mc{S}$ is replaced by $\mc{S}'$. On the other hand, for all $\Gamma''\in\mc{S}$ with $\Gamma''\neq\Gamma,\Gamma'$, we need only consider four cases:
\begin{itemize}
\item[C1] if $\Gamma''\subset\Gamma$ and $\Gamma''\subset\Gamma'$, then $\Gamma''\subset\Gamma+\Gamma'$ and $\Gamma''\subseteq\Gamma\Gamma'$;
\item[C2] if $\Gamma\subset\Gamma''$ and $\Gamma'\subset\Gamma''$, then $\Gamma+\Gamma'\subseteq\Gamma''$ and $\Gamma\Gamma'\subset\Gamma''$;
\item[C3] if $\Gamma''\subset\Gamma$ and $\Gamma''\sim\Gamma'$, or if $\Gamma''\sim\Gamma$ and $\Gamma''\subset\Gamma'$, then $\Gamma''\subset\Gamma+\Gamma'$; and finally,
\item[C4] if $\Gamma\subset\Gamma''$ and $\Gamma'\sim\Gamma''$, or if $\Gamma\sim\Gamma''$ and $\Gamma'\subset\Gamma''$, then $\Gamma\Gamma'\subset\Gamma''$.
\end{itemize}
In all cases, the number of $\sim$ pairs cannot increase. So $\mc{S}'$ must contain at least one less $\sim$ pair than $\mc{S}$.
\end{IEEEproof}

\begin{IEEEproof}[Proof of Lemma~\ref{L:vertex}]
For all $\Gamma\subseteq\Omega$ and $1 \leq i \leq n$, it is immediate from (\ref{E:GammapiA}) that
\[
\Gamma\Gamma_\pi^i\Gamma_\pi^{i-1}=\Gamma\Gamma_\pi^{i-1}.
\]
If $\omega_\pi^i \in \Gamma$, using (\ref{E:GammapiB}), it is also easy to verify that
\[
\Gamma\Gamma_\pi^i+\Gamma_\pi^{i-1}=\Gamma_\pi^i.
\]
so according to (\ref{E:supermodular}),
\[
\chi(\Gamma\Gamma_\pi^i)+\chi(\Gamma_\pi^{i-1}) \leq \chi(\Gamma_\pi^i)+\chi(\Gamma\Gamma_\pi^{i-1}).
\]
Let $\Gamma=\{\omega_\pi^{i_1},\omega_\pi^{i_2},\ldots,\omega_\pi^{i_l}\}$, with $i_1 < i_2 < \cdots < i_l$. Then, on the one hand, using (\ref{E:vertex}) and the previous inequality, we have
\[
\begin{IEEEeqnarraybox}[][c]{Cl}
     & v_\pi^{\l\Gamma\r}(\chi)\\
=    & \sum_{k=1}^l (\chi(\Gamma_\pi^{i_k})-\chi(\Gamma_\pi^{i_k-1}))\\
\geq & \sum_{k=1}^l (\chi(\Gamma\Gamma_\pi^{i_k})-\chi(\Gamma\Gamma_\pi^{i_k-1}))\\
=    & \chi(\Gamma\Gamma_\pi^{i_l}) +\sum_{k=1}^{l-1} (\chi(\Gamma\Gamma_\pi^{i_k})-\chi(\Gamma\Gamma_\pi^{i_{k+1}-1})) -\chi(\Gamma\Gamma_\pi^{i_1-1}).
\end{IEEEeqnarraybox}
\]
On the other hand, using (\ref{E:GammapiB}), it is also easy to verify that
\[
\Gamma\Gamma_\pi^{i_l}=\{\omega_\pi^{i_1},\omega_\pi^{i_2},\ldots,\omega_\pi^{i_l}\}=\Gamma,
\]
\[
\Gamma\Gamma_\pi^{i_k} = \{\omega_\pi^{i_1},\omega_\pi^{i_2},\ldots,\omega_\pi^{i_k}\} = \Gamma\Gamma_\pi^{i_{k+1}-1},
\]
and
\[
\Gamma\Gamma_\pi^{i_1-1} =\phi.
\]
It follows that $v_\pi^{\l\Gamma\r}(\chi) \geq \chi(\Gamma)$.
\end{IEEEproof}

\bibliographystyle{IEEEtran}
\bibliography{IEEEabrv,References}

\end{document}